\providecommand{\forArxiv}[1]{}

\providecommand{\forThesis}[1]{}

\documentclass[12pt]{article}

\usepackage{tikz}
\usetikzlibrary{arrows.meta}
\usetikzlibrary{positioning,shapes,shadows,arrows,calc}
\usetikzlibrary{shapes.geometric}
\tikzset{%
  >={Latex[width=2mm,length=2mm]}
}
\makeatletter
\global\let\tikz@ensure@dollar@catcode=\relax
\makeatother

\usepackage{amsmath}
\usepackage{amssymb}
\usepackage{array}
\usepackage{authblk}
\usepackage{blindtext}
\usepackage{booktabs}
\usepackage{caption}
\usepackage{dsfont}
\usepackage{enumitem}
\usepackage[T1]{fontenc}
\usepackage{hyperref}
\usepackage{ifthen}
\usepackage{latexsym}
\usepackage{letltxmacro}
\usepackage{mathtools}
\usepackage{multirow}
\usepackage{setspace}
\usepackage{scalefnt}
\usepackage{tla2}
\usepackage{verbatim}
\usepackage{xcolor}
\usepackage{xspace}
\usepackage[square,numbers]{natbib}
\usepackage{fullpage}
\usepackage{fontawesome}

\newcommand{\capcite}[1]{\protect\cite{#1}}
\newcommand{\capref}[1]{\protect\ref{#1}}
\newcommand{\msg}[1]{\ifmmode\text{\textcolor{blue}{\textsf{#1}}}\else\textcolor{blue}{\textsf{#1}}\fi}
\newcommand{\msgtype}[1]{\ifmmode\text{\textcolor{blue}{\textsf{``#1''}}}\else\textcolor{blue}{\textsf{``#1''}}\fi}
\newcommand{\agg}[1]{\ifmmode \text{\texttt{#1}}\else \texttt{#1}\fi}
\newcommand{\secref}[1]{Section~\ref{#1}}
\newcommand{\figref}[1]{Fig.~\ref{#1}}
\newcommand{\mysec}[1]{\section{#1}}
\newcommand{\mypar}[1]{\vspace{\baselineskip}\noindent\textbf{#1.}}
\newcommand{\tlaplus}{TLA\textsuperscript{+}}
\newcommand\m[1]{\mbox{$#1$}} 
\newenvironment{myeqn}{\begin{equation}\begin{aligned}}{\end{aligned}\end{equation}}
\definecolor{commentGreen}{rgb}{0,0.5,0}
\newcommand{\tlaComment}[1]{\textcolor{commentGreen}{\textbackslash* #1}}
\newcommand{\tlaif}{\mbox{\textcolor{purple}{\sc if}}}
\newcommand{\tlathen}{\mbox{\textcolor{purple}{\sc then}}}
\newcommand{\tlaelse}{\mbox{\textcolor{purple}{\sc else}}}
\newcommand{\Ballots}{\mathcal{B}}
\newcommand{\Slots}{\mathcal{S}}
\newcommand{\Values}{\mathcal{V}}
\newcommand{\Acceptors}{\mathcal{A}}
\makeatletter
\newcommand{\showfontsize}{\f@size{} pt}
\makeatother

\providecommand{\notes}[1]{}

\hypersetup{ 
    colorlinks=true,
    urlcolor=cyan,
}

\begin{document}

\title{Simpler Specifications and Easier Proofs of Distributed Algorithms Using History Variables}

\author{Saksham Chand}
\author{Yanhong A. Liu}
\affil{%
	\{schand, liu\}@cs.stonybrook.edu\\
	Computer Science Department, Stony Brook University,\\
	Stony Brook, New York, USA, 11794
}

\maketitle
\begin{abstract}
This paper studies specifications and proofs of distributed algorithms when only message history variables are used, using the Basic Paxos and Multi-Paxos algorithms for distributed consensus as precise case studies.
We show that not using and maintaining other state variables yields simpler specifications that are more declarative and easier to understand.  It also allows easier proofs to be developed by needing fewer invariants and facilitating proof derivations.  Furthermore, the proofs are mechanically checked more efficiently.

We show that specifications in \tlaplus{}, Lamport's temporal logic of actions, and proofs in TLAPS, the \tlaplus{} Proof System (TLAPS) are reduced by a quarter or more for single-value Paxos and by about half or more for multi-value Paxos.
Overall we need about half as many manually written invariants and proof obligations. Our proof for Basic Paxos takes about 25\% less time 
for TLAPS to check, and our proofs for Multi-Paxos are checked 
within 1.5 minutes whereas prior proofs 
fail to be checked by TLAPS.


\end{abstract}



\mysec{Introduction}
Reasoning about correctness of distributed algorithms is notoriously difficult due to a number of reasons including concurrency, asynchronous networks, 
and failures. Emerging technologies such as autonomous cars are bringing vehicular clouds closer to reality~\cite{gerla2014internet}; decentralized digital currencies are gathering more attention from academia and industry than ever~\cite{tschorsch2016bitcoin}; and with the explosion in the number of nano- and pico- satellites being launched, a similar trend is expected in the field of space exploration~\cite{schilling2017perspectives}. All of these systems deal with critical resources such as human life, currency, and intricate machinery. This only amplifies the need for employing formal methods to guarantee their correctness.


Verification of distributed algorithms continues to pose a demanding challenge to computer scientists, exacerbated by the fact that paper proofs of these algorithms cannot be trusted~\cite{zave2012using}. 
The usual line of reasoning in static analysis of such systems involves manually writing invariants and then using theorem provers to verify that the invariants follow from the specification and that they imply correctness.

\mypar{History variables and derived variables} A distributed system comprises a set of processes communicating with each other by message passing while performing local actions that may be triggered upon receiving a set of messages and may conclude with sending a set of messages~\cite{lamport1978implementation,lamport1994temporal}. As such, data processed by any distributed process fall into two categories: (i) message history variables, or in short \textit{history variables}: Sets of all messages sent and received\footnote{This is different from some other references of the term history variables which include sequences of local actions, i.e., execution history~\cite{clint1973program}} and (ii) \textit{derived variables:} 
Local data maintained for efficient computation. Derived variables are often used to maintain results of aggregate queries over sent and received messages.

Derived variables are helpful for efficient implementation because instead of computing expensive queries from scratch as messages are sent and received, the variable is incrementally updated to maintain the query result.
While this approach works well for efficient implementation, the same is not true for reasoning. For specifications written with derived variables, invariants have to be added to their proofs which, at the very least, establish that the derived variable holds the query result.

One reason to use derived variables in formal specifications is their existence in pseudocode and implementations. Another reason is the lack of high-level languages that provide elegant support for quantifications, history variables, and automatic optimal maintenance of aggregate queries over history variables. The barrier of lack of executable language support for such expressiveness is overcome by high-level languages like DistAlgo~\cite{liu2017clarity}, which provides native support for history variables, quantifications, and aggregate queries. This motivated us to dispense with derived variables, and study specifications written with only history variables and the impact of this change on the proofs.

Note that uses of history variables provide higher-level specifications of systems in terms of what to compute, as opposed to how to compute with employing and updating derived variables.  It makes proofs easier, independent of the logics used for doing the proofs, because important invariants are captured directly in the specifications, rather than hidden under all the incremental updates.  On the other hand, it can make model checking much less efficient, just as it can make straightforward execution much less efficient.  This is not only because high-level queries are time consuming, but also because maintaining history variables can blow up the state space.  This is why automatic incrementalization~\cite{PaiKoe82,RotLiu08OSQ-GPCE,Gor11thesis,Liu13book} is essential for efficient implementations, including implementations of distributed algorithms~\cite{Liu+12DistPL-OOPSLA,Liu+16IncOQ-PPDP}.  The same transformations for incrementalization can drastically speed up both program execution and model checking.


\mypar{This paper} We first describe a systematic style to write specifications of distributed algorithms 
using message history variables. The only variables in these specifications are the sets of sent and/or received messages. We show (i) how these are different from the usual pseudocode, (ii) why these are sufficient for specifying all distributed algorithms, and (iii) when these are better for the provers than other specifications. A method is then explained which, given such specifications, leads us to systematically derive many important invariants needed for correctness proofs. This method exploits the fact that the sets of sent and received messages grow monotonically --- messages can only be added or read from these sets, not modified or deleted.

We use three algorithm variants already specified in \tlaplus{}, Lamport's temporal logic of actions~\cite{lamport1994temporal}, and proved using TLAPS, the \tlaplus{} proof system as our case studies: (i) Basic Paxos for single-valued consensus by Lamport et al.~\cite{basicpaxos2014}, (ii) Multi-Paxos for multi-value consensus by Chand et al.~\cite{chand2016formal}, and (iii) Multi-Paxos with Preemption~\cite{chand2016formal}. Paxos is chosen because it is famous for being a difficult algorithm to grasp, while at the same time it is the core algorithm for distributed consensus---the most fundamental problem in distributed computing.
We show that our approach led to significantly reduced sizes of specifications and proofs, numbers of needed manually written invariants, and proof checking times.
Our specifications and proofs are available at \url{https://github.com/Distalgo/proofs}.

This paper is an extended and revised version of~\cite{chand2018simpler}. Besides overall revision and improvement, the main extensions are as follows:
\begin{enumerate}

    \item \secref{sec-pre} is extended with new Sections~\ref{secTLA+} and~\ref{secTLAPS} introducing \tlaplus{} and TLAPS, respectively.
    
    \item \secref{sec-impact} is extended with new \secref{sec-hist-basic-proof}, describing the complete proof for Basic Paxos.
        
    \item \secref{sec-multi}, on specifications and proofs for Multi-Paxos, is almost entirely new, instead of only a specification for preemption and a paragraph about verification in~\cite{chand2018simpler}. \secref{sec-multi-spec} explains and compares two approaches to developing Multi-Paxos specifications that use only history variables.
    \secref{sec-multi-preempt-spec} presents a new specification of preemption to use only history variable $sent$, not both $sent$ and $received$.  \secref{sec-multi-inv} and \secref{sec-multi-proof} describe all key invariants and changes for the proofs of Multi-Paxos and Multi-Paxos with Preemption.
    
    \item Complete simplified specification, invariants, and proof for Basic Paxos are given in new Appendices~\ref{appendix:spec},~\ref{appendix:prop}, and~\ref{appendix:proof}, respectively.
    
    \item Complete simplified specification of Multi-Paxos with Preemption is given in new Appendix~\ref{appendix:mppspec}. Complete simplified invariants used in the proofs for Multi-Paxos and Multi-Paxos with Preemption are given in new Appendix~\ref{appendix:mpinvcomp}.
    
    \item New Appendix~\ref{appendix-p2a} shows the need of a specific condition in Lamport's specification for Basic Paxos for it to be safe, even though the condition is missing in the English description in~\cite{lamport2001paxos}.
\end{enumerate}

The rest of the paper is organized as follows.
\secref{sec-pre} covers preliminaries: distributed consensus, Paxos, \tlaplus{}, and TLAPS.
\secref{sec-basic-paxos} details our style of writing specifications using Basic Paxos as an example. \secref{sec-impact} describes our strategy to systematically derive invariants and how using history variables leads to needing fewer invariants. \secref{sec-multi} explains approaches, specifications, invariants, and proofs for verifying Multi-Paxos and Multi-Paxos with Preemption. \secref{sec-results} compares our specifications and proofs with those that do not use history variables. \secref{sec-related} discusses related work and concludes.

\mysec{Preliminaries}
\label{sec-pre}

\subsection{Distributed consensus}
\label{secDistCons}
A distributed system is a set of processes that process data locally and communicate with each other by sending and receiving messages. 
The processes may crash and may later recover, and
the messages may be lost, delayed, reordered, and duplicated.

The basic consensus problem, called single-value consensus, is for a set of processes to agree on a single value.  An algorithm for single-value consensus is said to be \textit{safe} if it satisfies the following conditions~\cite{lamport2001paxos}:
\begin{enumerate}
    \item[C1.] Only a value that has been proposed may be chosen,
    \item[C2.] Only a single value is chosen, and
    \item[C3.] A process never learns that a value has been chosen unless it actually has been.
\end{enumerate}
Following Lamport et al.~\cite{basicpaxos2014} and Chand et al.~\cite{chand2016formal}, we consider only C2, also called $Agreement$. 
Conditions C1 and C3 
are straightforward and easy to prove.
For $Agreement$, we formally specify the following:
\begin{myeqn}
Agree == \A v1, v2 \in \mathcal{V} : Chosen(v1) /\ Chosen(v2) => v1 = v2
\end{myeqn}%
where $\mathcal{V}$ is the set of possible proposed values, and
$Chosen$ is a predicate that given a value $v$ evaluates to true iff $v$ was chosen by the algorithm. The specification of $Chosen$ is part of the algorithm. The complete $Agreement$ property is formally specified in \eqref{cons}.

The more general consensus problem, called multi-value consensus, is to agree on a sequence of values, instead of a single value. Here we have
\begin{myeqn}\label{eqnsafe}
Agree_{multi} == \A v1, v2 \in \mathcal{V}, s \in \mathcal{S} : Chosen(s, v1) /\ Chosen(s, v2) => v1 = v2
\end{myeqn}%
where $\mathcal{V}$ is as above, $\mathcal{S}$ is a set of \textit{slots} used to index the sequence of chosen values, and
$Chosen(s,v)$ is true iff for slot $s$, value $v$ was chosen by the algorithm.

\subsection{Basic Paxos and Multi-Paxos}
\label{secPaxosEng}
Paxos solves the problem of consensus. Two main roles of the algorithm are performed by two kinds of processes:
\begin{itemize}

\item $\mathcal{P}$, the set of proposers that propose values that can be chosen.

\item $\mathcal{A}$, the set of acceptors that vote for proposed values.  A value is chosen when there are enough votes for it.

\end{itemize}
These roles can be co-located, that is, a single process can take on more than one role.

A set $\mathcal{Q}$ of subsets of the acceptors, that is, $\mathcal{Q} \subseteq 2^{\mathcal{A}}$, is used as a quorum system.  It must satisfy the property that any two quorums in $\mathcal{Q}$ overlap, that is, $\forall Q1, Q2 \in \mathcal{Q} : Q1 \cap Q2 \neq \emptyset$. The most commonly used quorum system $\mathcal{Q}$ takes any majority of acceptors as an element in $\mathcal{Q}$. For example, if $\mathcal{A} = \{1, 2, 3\}$, then the majority based quorum set is $\mathcal{Q} = \{\{1, 2\}, \{2, 3\}, \{1, 3\},$ $\{1, 2, 3\}\}$. Quorums are needed because the system can have failures. If a process waits for replies from all other processes, the system will hang in the presence of even one failed process. For example, in the system defined above, the system will continue to work even if acceptor 3 fails because at least one quorum, which is $\{1, 2\}$, is alive.

Basic Paxos solves the problem of single-value consensus.  It defines predicate $Chosen$ as
\begin{myeqn}
Chosen(v) == \E Q \in \mathcal{Q} : \A a \in Q : \E b \in \mathcal{B} : sent(\textcolor{blue}{"2b"}, a, b, v)
\end{myeqn}%
where $\mathcal{B}$ is the set of proposal numbers, also called ballot numbers, which is any set that can be totally ordered. $sent(\textcolor{blue}{"2b"}, a, b, v)$ means that a message of type $\textcolor{blue}{\textsf{2b}}$ with ballot number $b$ and value $v$ was sent by acceptor $a$.  An acceptor votes (for value $v$) by sending such a message.

Multi-Paxos solves the problem of multi-value consensus.  It extends predicate $Chosen$ to decide a value for each slot $s$ in $\mathcal{S}$:
\begin{myeqn}\label{voting}
Chosen(s, v) == \E Q \in \mathcal{Q} : \A a \in Q : \E b \in \mathcal{B} : sent(\textcolor{blue}{"2b"}, a, b, s, v)
\end{myeqn}%
To satisfy the $Agree_{multi}$ property, $\mathcal{S}$ can be any set.  In practice, $\mathcal{S}$ is usually the set of natural numbers. Multi-Paxos can be built from Basic Paxos by carefully adding slots as described in~\cite{chand2016formal}.

\subsection{\texorpdfstring{\tlaplus{}}{TLA+}}
\label{secTLA+}
The specifications presented in this article are written in the language \tlaplus{}, Lamport's temporal logic of actions~\cite{lamport1994temporal,lamport2002specifying,merz2008spec,merz2003logic}, a logic for specifying concurrent and distributed systems and reasoning about their properties. In \tlaplus{}, a \textit{state} is an assignment of values to the variables. An \textit{action} is a relation between a current state and a new state, specifying the effect of executing a sequence of instructions. For example, the instruction $x := x + 1$ is specified in \tlaplus{} by the action $x' = x + 1$. An action is specified as a formula over unprimed and primed variables, where unprimed variables refer to the values of the variables in the current state, and primed variables refer to the values of the variables in the new state.

A system is specified by its actions and initial states. Formally, a system is specified as $Spec == Init /\ [][Next]_{vars}$ where $Init$ is a predicate that holds for initial states of the system, $Next$ is a disjunction of all actions of the system, and $vars$ is the tuple of all variables. The expression $[Next]_{vars}$ is true if either $Next$ is true, implying some action is true and therefore executed, or $vars$ stutters, that is, the values of the variables are same in the current and next states. $[]$ is the temporal operator \textit{always}. Thus, $Spec$ defines a set of infinite sequences of steps where in each step either an action is executed 
or $vars$ stutters. Such a sequence is called a \textit{behavior}.
\def\laclkexample {}

\def\clkH {\relax\ifmmode H0\else $H0$\fi}
\def\clkh {\relax\ifmmode h\else $h$\fi}
\def\clkmdm {\relax\ifmmode meridiem\else $meridiem$\fi}

\def\laclkc {\relax\ifmmode c\else $c$\fi}

\newcommand{\mytlakw}[1]{\textcolor{purple}{\footnotesize\textsc{\scalefont{1.15}{#1}}}}
\ifdefined\clkexample
    As a simple example, consider this specification of the hour hand of a clock:

    \begin{myeqn}\label{specclk}\begin{aligned}
        \begin{tabular}{@{}l@{}l@{}l@{}}
            \multicolumn{3}{@{}l@{}}{$\CONSTANT \clkH{}$}\\
            \multicolumn{3}{@{}l@{}}{$\VARIABLE \clkh{}, \clkmdm{}$}\\
            $Init$ &$==$ &$/\ \clkh{} = \clkH{}$\\
            & &$/\ \clkmdm{} \in \{0, 1\}$\\
            $Incr$ &$==$ &$/\ \clkh{} < 12$ \\
            & &$/\ \clkh{}' = \clkh{} + 1$\\
            & &$/\ \UNCHANGED <<\clkmdm{}>>$\\
            $Wrap$ &$==$ &$/\ \clkh{} = 12$ \\
            & &$/\ \clkh{}' = 1$\\
            & &$/\ \clkmdm{}' = 1 - \clkmdm{}$\\
            $Next$ &$==$ &$Incr \/ Wrap$\\
            $Spec$ &$==$ &$Init /\ [][Next]_{<<\clkh{}, \clkmdm{}>>}$\\
        \end{tabular}
    \end{aligned}\end{myeqn}
    
    The variable \clkh{} stores the current hour value and \clkmdm{} tells whether it is am/pm. Initially, the hour hand reads \clkH{}. Actions $Incr$ and $Wrap$ specify how the values of \clkh{} and \clkmdm{} change. The first predicate in both actions is the guard.
\else
    \ifdefined\laclkexample
        As a simple example, consider the following specification of a clock based on Lamport's logical clock~\cite{lamport1978time} but on a shared memory system:
        \begin{myeqn}\label{speclaclk}\begin{aligned}
            \begin{tabular}{@{}l@{}l@{}l@{}}
                \multicolumn{3}{@{}l@{}}{$\VARIABLE \laclkc{}$}\\
                $Max(S)$ &$==$ &$\CHOOSE e \in S: \A f \in S: e \geq f$\\
                $Init$ &$==$ &$\laclkc{} = [p \in \{0, 1\} |-> 0]$\\
                $LocalEvent(p)$ &$==$ &$\laclkc{}' = [\laclkc{} \EXCEPT![p] = \laclkc{}[p] + 1]$\\
                $ReceiveEvent(p)$ &$==$ &$\laclkc{}' = [\laclkc{} \EXCEPT![p] = 
                Max(\{\laclkc{}[p], \laclkc{}[1-p]\}) + 1]$\\
                $Next$ &$==$ &$\E p \in \{0, 1\}: LocalEvent(p) \/ ReceiveEvent(p)$\\
                $Spec$ &$==$ &$Init /\ [][Next]_{<<\laclkc{}>>}$\\
            \end{tabular}
        \end{aligned}\end{myeqn}
        
        The system has two processes numbered 0 and 1. Variable \laclkc{} stores their current clock values as a function from process numbers to clock values. Both processes start with clock value 0, as specified in $Init$. $LocalEvent(p)$ specifies that process $p$ has executed some local action and therefore increments its clock value. The expression $\laclkc{}' = [\laclkc{} \EXCEPT![p] = \laclkc{}[p] + 1]$ means that function $\laclkc{}'$ is the same as function \laclkc{} except that $\laclkc{}'[p]$ is $\laclkc{}[p] + 1$. $ReceiveEvent(p)$ specifies that process $p$ updates its clock value to 1 greater than the higher of its and the other process' clock value. We define operator $Max$ to obtain the highest of a set of values. \CHOOSE 
        returns an arbitrarily chosen value satisfying the body of the \CHOOSE expression if one exists, or an arbitrary value otherwise.
    \fi
\fi


\subsection{TLAPS}
\label{secTLAPS}

TLAPS, the \tlaplus{} Proof System~\cite{chaudhuri2008tlaps,cousineau2012tlaproofs,tlaps}, is a tool for mechanically checking proofs of properties of systems specified in \tlaplus{}. Proofs are written in a hierarchical style~\cite{lamport2012write}, and are transformed to individual proof obligations that are sent to backend theorem provers. An obligation is a logical formula of the form $P \protect\Rightarrow Q$. For proving an obligation, the default behaviour of TLAPS is to try three backend provers in succession: CVC3 (an SMT solver), Zenon, and Isabelle~\cite{merz2012harnessing,merz2012automatic,tlaps2019provers}. If none of them finds a proof, TLAPS reports a failure on the obligation. Other SMT solvers supported by TLAPS are Z3, veriT, and Yices. Temporal formulas are proved using LS4, a propositional temporal logic (PTL) prover. Users can specify which prover they want to use by using its name and can specify the timeout for each obligation separately.

\def\clkRange {\relax\ifmmode \{1, \ldots, 12\}\else $\{1, \ldots, 12\}$\fi}
\providecommand{\prfstep}[2]{\langle#1\rangle#2.\,}
\providecommand{\prfstepnum}[2]{\langle#1\rangle#2}

\ifdefined\clkexample
    As an example, we present the proof of a simple type invariant about the clock specification in~(\ref{specclk}) - Assuming $\clkH{} \in 
    \clkRange{}$, it is always the case that $\clkh{} \in \clkRange{}$:
    
    \begin{myeqn}\label{prfclk}\begin{aligned}
    &\ASSUME ParamAssumption == \clkH{} \in \clkRange{}\\
    &TypeOK == \clkh{} \in \clkRange{}\\
    &\THEOREM Inv == Spec => [](TypeOK)\\
    &\prfstep{1}{} \USE \DEF TypeOK\\
    &\prfstep{1}{1} Init => TypeOK\, \BY ParamAssumption\, \DEF Init\\
    &\prfstep{1}{2} TypeOK /\ [Next]_{<<\clkh{}, \clkmdm{}>>} => TypeOK'\, \BY \DEF Next, Incr, Wrap\\
    &\prfstep{1}{} \QED \BY \prfstepnum{1}{1}, \prfstepnum{1}{2}, \texttt{PTL}\, \DEF Spec
    \end{aligned}\end{myeqn}
    
    The proof of theorem $Inv$ is written in a hierarchical fashion. It is proved by two steps, named $\langle1\rangle1$ and $\langle1\rangle2$, and RuleINV1~\cite{lamport1994temporal}. Proof steps in TLAPS are typically written as:
    \begin{myeqn}
        \prfstep{x}{y}\, Assertion\,\, \BY e_1, \ldots, e_m\, \DEF d_1, \ldots, d_n
    \end{myeqn}
    which states that step number $\prfstepnum{x}{y}$ proves $Assertion$ by assuming $e_1, \ldots, e_m$, and expanding the definitions of $d_1, \ldots, d_n$. For example, step $\prfstepnum{1}{1}$ asserts that $Init => TypeOK$ by assuming $ParamAssumption$ and expanding the definition of operator $Init$. The \QED step for $\langle1\rangle$ requires us to invoke \texttt{PTL} - Propositional Temporal Logic prover because $Inv$ is a temporal formula.
\else
    \ifdefined\laclkexample
        As an example, we present the proof of a simple type invariant about the clock specification in~(\ref{speclaclk})---it is always the case that $\laclkc{} \in [\{0, 1\} -> \mathds{N}]$, where $\mathds{N}$ is the set of natural numbers:
        
        \begin{myeqn}\label{prflaclk}\begin{aligned}
        &TypeOK == \laclkc{} \in [\{0, 1\} -> \mathds{N}]\\
        &\THEOREM Inv == Spec => [](TypeOK)\\
        &\prfstep{1}{} \USE \DEF TypeOK\\
        &\prfstep{1}{1} Init => TypeOK\, \BY \DEF Init\\
        &\prfstep{1}{2} TypeOK /\ [Next]_{<<\laclkc{}>>} => TypeOK'\\
        &\phantom{\prfstep{1}{2}}\prfstep{2}{} \ASSUME TypeOK, [Next]_{<<\laclkc{}>>}\, \PROVE TypeOK'\\
        &\phantom{\prfstep{1}{2}}\prfstep{2}{1} \CASE \E p \in \{0, 1\}: LocalEvent(p)\, \BY \prfstepnum{2}{1}\, \DEF LocalEvent\\
        &\phantom{\prfstep{1}{2}}\prfstep{2}{2} \CASE \E p \in \{0, 1\}: ReceiveEvent(p)\, \BY \prfstepnum{2}{2}\, \DEF ReceiveEvent\\
        &\phantom{\prfstep{1}{2}}\prfstep{2}{3} \CASE \UNCHANGED <<\laclkc{}>>\, \BY \prfstepnum{2}{3}\\
        &\phantom{\prfstep{1}{2}}\prfstep{2}{} \QED \BY \prfstepnum{2}{1}, \prfstepnum{2}{2}, \prfstepnum{2}{3}\, \DEF Next\\
        &\prfstep{1}{} \QED \BY \prfstepnum{1}{1}, \prfstepnum{1}{2}, \texttt{PTL}\, \DEF Spec
        \end{aligned}\end{myeqn}
        
        The proof of theorem $Inv$ is written in a step-by-step fashion. It is proved by two steps, named $\prfstepnum{1}{1}$ and $\prfstepnum{1}{2}$, and the PTL solver.
        Proof steps in TLAPS are typically written as:
        \begin{myeqn}
            \prfstep{x}{y}\, Assertion\,\, \BY e_1, \ldots, e_m\, \DEF d_1, \ldots, d_n
        \end{myeqn}%
        which states that step number $\prfstepnum{x}{y}$ proves $Assertion$ by using $e_1, \ldots, e_m$, and expanding the definitions of $d_1, \ldots, d_n$. For example, step $\prfstepnum{1}{1}$ proves $Init => TypeOK$ by expanding the definition of $Init$. If TLAPS does not know if $e_i$ is true, it would try to prove $e_i$ using $e_1, \ldots, e_{i-1}$ and the current context. If TLAPS is unable to prove $e_i$, it would display both $e_i$ and $Assertion$ as failed obligations. The step ``$\prfstep{1}{}$ \USE \DEF $TypeOK$'' instructs the prover to expand the definition of $TypeOK$ in all proof steps till the \QED step for $\prfstepnum{1}{}$. The \QED step for $\langle1\rangle$ instructs TLAPS to invoke a \texttt{PTL} prover because $Inv$ is a temporal formula.
        
        To demonstrate the hierarchical proof style advocated in TLAPS, we break down the proof of step $\prfstepnum{1}{2}$. The step ``$\prfstepnum{2}{} \ASSUME \ldots \PROVE$'' specifies the assumptions and goal to be proved in the current proof level, which is level 2. The next two steps $\prfstepnum{2}{1}$ and $\prfstepnum{2}{2}$ prove the goal for the two actions specified in $Next$. Finally, $\prfstepnum{2}{3}$ proves the goal for the case of stuttering. Together, $\prfstepnum{2}{\text{1--3}}$ cover all cases of $[Next]_{<<\laclkc{}>>}$, thus concluding the proof.
    \fi
\fi

\mysec{Specifications using message history variables}
\label{sec-basic-paxos}
We demonstrate our approach by developing a specification of Basic Paxos in which we only maintain the set of sent messages. This specification is made to correspond to the specification of Basic Paxos in \tlaplus{} by Lamport et al.~\cite{basicpaxos2014}. 
This is done to better understand 
the applicability of our approach. We also simultaneously show Lamport's description of the algorithm in English~\cite{lamport2001paxos} to aid the comparison, except we rename message types and variable names to match those in Lamport et al.'s \tlaplus{} specification: \m{prepare} and \m{accept} messages are renamed \msg{1a} and \msg{2a}, respectively, their responses are renamed \msg{1b} and \msg{2b}, respectively, and variable $n$ is renamed $b$ and $bal$ in different places.

\mypar{Basic Paxos variables} Lamport et al.'s specification of Basic Paxos has four global variables.

\begin{itemize}
\item $msgs$: history variable maintaining the set of messages that have been sent. Processes read from or add to this set but cannot remove from it. We rename this to $sent$ in both ours and Lamport et al.'s specifications for clarity purposes. This is the only variable maintained in our specifications.
\item $maxBal$: per acceptor, the highest ballot seen by the acceptor.
\item $maxVBal$ and $maxVal$: per acceptor, the highest ballot in which the acceptor has voted and the value the acceptor voted for in the highest ballot, respectively.
\end{itemize}

\mypar{Basic Paxos algorithm steps}
\label{secSpecAlgo}
The algorithm consists of repeatedly executing two phases. Each phase comprises two actions, one by acceptors and one by proposers.

\begin{itemize}
\item \textbf{Phase 1a.} \figref{fig-hist-spec-p1a} shows Lamport's description in English followed by Lamport et al.'s and our specifications. $Send$ is an operator that adds its argument to $sent$, i.e., $Send(m) \mathrel{\smash{\triangleq}} sent' = sent \cup \{m\}$. 
\begin{enumerate}
    \item The first conjunct in Lamport et al.'s specification is not mentioned in the English description and is not needed. Therefore it was removed.
    \item The third conjunct is also removed because the only variable our specification maintains is $sent$, which is updated by $Send$.
\end{enumerate}

\begin{figure*}[t]
\centering
\begin{tabular}{|l|l|}
    \hline
    \multicolumn{2}{|p{0.8\textwidth}|}{\textbf{Phase 1a.} A proposer selects a proposal number \m{b} and sends a \msg{1a} request with number \m{b} to a majority of acceptors.}\\
    \hline
    Lamport et al.'s & Using $sent$ only \\
    \hline
    \begin{tabular}{@{}l@{}}
        $Phase1a(b \in \mathcal{B}) ==$\\
        $\; /\ \nexists\, m \in sent : (m.type = \msgtype{1a}) /\ (m.bal = b)$\\
        $\; /\ Send([type |-> \msgtype{1a}, bal |-> b])$\\
        $\; /\ \UNCHANGED <<maxVBal, maxBal, maxVal>>$\\
    \end{tabular}
 &
    \begin{tabular}{@{}l@{}}
        $Phase1a(b \in \mathcal{B}) ==$\\
        \\
        $\; Send([type |-> \msgtype{1a}, bal |-> b])$\\
        \\
    \end{tabular}\\
    \hline
\end{tabular}

 \caption{Specifications of Phase 1a of Basic Paxos}
 \label{fig-hist-spec-p1a}
\end{figure*}

\item \textbf{Phase 1b.} \figref{fig-hist-spec-p1b} shows the English description and the specifications of Phase 1b. The first two conjuncts in both specifications capture the precondition in the English description. The remaining conjuncts specify the action.
\begin{enumerate}
    \item The first conjunct states that message $m$ received by acceptor $a$ is of type \msg{1a}.
    
    \item The second conjunct ensures that the proposal number $bal$ in the \msg{1a} message $m$ is higher than that of any \msg{1a} request responded to by $a$. In Lamport et al.'s specification, derived variable $maxBal[a]$ maintains the highest proposal number 
    that $a$ has ever responded to, in both \msg{1b} and \msg{2b} messages, and its second conjunct uses $m.bal > maxBal[a]$. Using $sent$ only, we capture this intent 
    more directly, as $\A m2 \in sent: m2.type \in \{\msgtype{1b}, \msgtype{2b}\} /\ m2.acc = a => m.bal > m2.bal$, because those $m2$'s are the response messages that $a$ has ever sent.
   
    \item The third conjunct is the action of sending a promise (\msg{1b} message) not to accept any more proposals numbered less than $bal$ and with the highest-numbered proposal (if any) that $a$ has accepted, i.e., has sent a \msg{2b} message.   This proposal is maintained in Lamport et al.'s specification in derived variables $maxVBal$ and $maxVal$. We specify this proposal as $max\_prop(a)$, which is either the set of proposals that have the highest proposal number among all accepted by $a$ or, if $a$ has not accepted anything, $\{[bal |-> -1, val |-> \bot]\}$, where $-1 \notin \mathcal{B}$ and is smaller than all ballots and $\bot \notin \mathcal{V}$. The latter corresponds to initialization in Lamport et al.'s specification as shown in \figref{fig-hist-spec-comp}, discussed later. Note that the specification in Appendix~\ref{appendix:spec} writes $\bot$ as $None$.
    
    \item The remaining conjuncts in Lamport et al.'s specification maintain the variable $maxBal[a]$. A compiler that implements incrementalization~\cite{Liu13book} over queries would automatically generate and maintain such a derived variable to optimize the corresponding query.
\end{enumerate}

\begin{figure*}
  \centering
\begin{tabular}{|l|l|}
    \hline
    \multicolumn{2}{|p{0.95\textwidth}|}{\textbf{Phase 1b.} If an acceptor receives a \msg{1a} request with number \m{bal} greater than that of any \msg{1a} request to which it has already responded, then it responds to the request with a promise not to accept any more proposals numbered less than \m{bal} and with the highest-numbered proposal (if any) that it has accepted.}\\
    \hline
    Lamport et al.'s & Using $sent$ only\\
    \hline
    \begin{tabular}{@{}l@{}}
        $Phase1b(a \in \mathcal{A}) ==$\\
        $\E m \in sent :$\\
        $\; /\ m.type = \msgtype{1a}$\\
        $\; /\ m.bal > maxBal[a]$\\
        \\
        $\; /\ Send([type |-> \msgtype{1b},$\\
        $\quad\; acc |-> a, bal |-> m.bal,$\\
        $\quad\; maxVBal |-> maxVBal[a],$\\ 
        $\quad\; maxVal |-> maxVal[a]])$\\
        \\
        $\; /\ maxBal' =$\\
        $\quad\; [maxBal \EXCEPT ![a] = m.bal]$\\
        $\; /\ \UNCHANGED <<maxVBal, maxVal>>$\\
        \\
    \end{tabular}
 &
    \begin{tabular}{@{}l@{}}
        $Phase1b(a \in \mathcal{A}) ==$\\
        $\E m \in sent, r \in max\_prop(a):$\\
        $\; /\ m.type = \msgtype{1a}$\\
        $\; /\ \A m2 \in sent: m2.type \in \{\msgtype{1b}, \msgtype{2b}\} /\ $\\
        $\quad\; m2.acc = a => m.bal > m2.bal$\\
        $\; /\ Send([type |-> \msgtype{1b},$\\
        $\quad\; acc |-> a, bal |-> m.bal,$\\
        $\quad\; maxVBal |-> r.bal,$\\ 
        $\quad\; maxVal |-> r.val])$\\
        \\
        $2bs(a) \!\!==\!\! \{m \in sent\!:\! m.type = \msgtype{2b} /\ m.acc = a\}$\\
        $max\_prop(a) ==$\\
        $\;$\tlaif{} $2bs(a) = \emptyset$ \tlathen{} $\{[bal |-> -1, val |-> \bot]\}$\\
        $\;$\tlaelse{}\!
        $\{\!m \in 2bs(a)\!:\! \A\! m2 \in 2bs(a)\!:\! m.bal \geq m2.bal \}$
    \end{tabular}\\
    \hline
\end{tabular}

 \caption{Specifications of Phase 1b of Basic Paxos}
 \label{fig-hist-spec-p1b}
\end{figure*}

\item \textbf{Phase 2a.} \figref{fig-hist-spec-p2a} shows the specifications of Phase 2a. The specifications differ from the English description by using a set of quorums, $\mathcal{Q}$, instead of a majority. The only difference between the two specifications is the removed $\UNCHANGED$ conjunct when using $sent$ only.  It is important to note that the English description fails to mention the first conjunct, without which the specification is unsafe. Appendix~\ref{appendix-p2a} describes a run that violates $Agreement$ when the first conjunct of $Phase2a$ is removed.  That is, for Lamport's specification to be safe, every \msg{2a} message must have a unique ballot.

Note that the first conjunct in Lamport et al.'s specification (and therefore ours as well) states that none of the \msg{2a} messages sent so far has $bal$ equal to $b$. This is not directly implementable in a real system because this quantification query requires accessing message histories of all processes. We leave this query as is for two main reasons: (i) The focus of this paper is to demonstrate the use of history variables against derived variables and compare them in the light of simpler specification and verification. This removes derived variables but leaves queries on history variables unchanged even though they are not directly implementable.  (ii) There is a commonly-used, straightforward, efficient way to implement this query, namely, realizing ballot as a tuple in $\mathds{N} \times \mathcal{P}$~\cite{van2015paxos}. So a proposer only executes Phase 2a on a ballot proposed by itself (i.e., sent a \msg{1a} message with that ballot) and, for efficient implementation, only executes Phase 2a on the highest ballot that it has proposed.

\begin{figure*}[h!]
  \centering

\begin{tabular}{|l|l|}
    \hline
    \multicolumn{2}{|p{0.95\textwidth}|}{\textbf{Phase 2a.} If the proposer receives a response to its \m{\msg{1a}} requests (numbered \m{b}) from a majority of acceptors, then it sends a \m{\msg{2a}} request to each of those acceptors for a proposal numbered \m{b} with a value \m{v}, where \m{v} is the value of the highest-numbered proposal among the \m{\msg{1b}} responses, or is any value if the responses reported no proposals.}\\
    \hline
    Lamport et al.'s & Using $sent$ only \\\hline
    \begin{tabular}{@{}l@{}}
    $Phase2a(b \in \mathcal{B}) ==$\\
    $/\ \nexists m \in sent : m.type = \msgtype{2a} /\ m.bal = b$\\ 
    $/\ \E v \in \mathcal{V}, Q \in \mathcal{Q}, S \subseteq \{m \in sent :$\\
       $\quad m.type = \msgtype{1b} /\ m.bal = b\} :$\\
               $\qquad /\ \A a \in Q : \E m \in S : m.acc = a$\\
               $\qquad /\ \hspace{1pt} \/ \A m \in S : m.maxVBal = -1$\\
                  $\qquad \quad \/ \E c \in 0..(b-1) :$\\ 
                        $\qquad \qquad /\ \A m \in S : m.maxVBal \leq c$\\
                        $\qquad \qquad /\ \E m \in S : /\ m.maxVBal = c$\\
                        $\hspace{89pt} /\ m.maxVal = v$\\
       $\quad /\ Send([type |-> \msgtype{2a}, bal |-> b, val |-> v])$\\
    $/\ \UNCHANGED <<maxBal, maxVBal, maxVal>>$\\
    \end{tabular}
 &
    \begin{tabular}{@{}l@{}}
    $Phase2a(b \in \mathcal{B}) ==$\\
    $/\ \nexists m \in sent : m.type = \msgtype{2a} /\ m.bal = b$\\ 
    $/\ \E v \in \mathcal{V}, Q \in \mathcal{Q}, S \subseteq \{m \in sent :$\\
       $\quad m.type = \msgtype{1b} /\ m.bal = b\} :$\\
               $\qquad /\ \A a \in Q : \E m \in S : m.acc = a$\\
               $\qquad /\ \hspace{1pt} \/ \A m \in S : m.maxVBal = -1$\\
                  $\qquad \quad \/ \E c \in 0..(b-1) :$\\ 
                        $\qquad \qquad /\ \A m \in S : m.maxVBal \leq c$\\
                        $\qquad \qquad /\ \E m \in S : /\ m.maxVBal = c$\\
                        $\hspace{89pt} /\ m.maxVal = v$\\
       $\quad /\ Send([type |-> \msgtype{2a}, bal |-> b, val |-> v])$\\
       \\
    \end{tabular}\\
    \hline
\end{tabular}

 \caption{Specifications of Phase 2a of Basic Paxos}
 \label{fig-hist-spec-p2a}
\end{figure*}

\item \textbf{Phase 2b.} \figref{fig-hist-spec-p2b} shows specifications of Phase 2b. Like for Phase 1b, we replace the second conjunct with the corresponding query over $sent$ and remove updates to the derived variables.
\begin{figure*}[h!]
  \centering

\begin{tabular}{|l|l|}
    \hline
    \multicolumn{2}{|p{0.95\linewidth}|}{\textbf{Phase 2b.} If an acceptor receives a \m{\msg{2a}} request for a proposal numbered \m{bal}, it accepts the proposal unless it has already responded to a \m{\msg{1a}} request having a number greater than \m{bal}.}\\
    \hline
    Lamport et al.'s & Using $sent$ only \\
    \hline
    \begin{tabular}{@{}l@{}}
        $Phase2b(a \in \mathcal{A}) ==$\\
        $\E m \in sent :$\\
        $\; /\ m.type = \msgtype{2a}$\\
        $\; /\ m.bal \geq maxBal[a]$\\
        \\
        $\; /\ Send([type |-> \msgtype{2b}, acc |-> a,$\\
        $\quad\; bal |-> m.bal, val |-> m.val])$\\
        $\; /\ maxBal' = [maxBal \EXCEPT ![a] = m.bal]$\\
        $\; /\ maxVBal' = [maxVBal \EXCEPT ![a] = m.bal]$\\
        $\; /\ maxVal' = [maxVal \EXCEPT ![a] = m.val]$\\
    \end{tabular}
 &
    \begin{tabular}{@{}l@{}}
        $Phase2b(a \in \mathcal{A}) ==$\\
        $\E m \in sent :$\\
        $\; /\ m.type = \msgtype{2a}$\\
        $\; /\ \A m2 \in sent\!:\! m2.type \in \{\msgtype{1b},\! \msgtype{2b}\} /\ $\\
        $\qquad m2.acc = a => m.bal \geq m2.bal$\\
        $\; /\ Send([type |-> \msgtype{2b}, acc |-> a,$\\
        $\quad\; bal |-> m.bal, val |-> m.val])$\\
        \\
        \\
        \\
    \end{tabular}\\
    \hline
\end{tabular}

 \caption{Specifications of Phase 2b of Basic Paxos}
 \label{fig-hist-spec-p2b}
\end{figure*}
\end{itemize}

\mypar{Overall Basic Paxos algorithm}
To complete the algorithm specification, we define, and compare, $vars$, $Init$, $Next$, and $Spec$ which are typical \tlaplus{} operator names for the set of variables, the initial state, possible actions leading to the next state, and the system specification, respectively, in \figref{fig-hist-spec-comp}.

Lamport et al.'s initialization of $maxVBal$ and $maxVal$ to $-1$ and $\bot$, respectively, is moved to our definition of $max\_prop$ in \figref{fig-hist-spec-p1b}. Note that we do not need $maxBal$ at all, but instead use the universally quantified queries directly in \figref{fig-hist-spec-p1b} and \figref{fig-hist-spec-p2b}. 
Lamport et al.'s specification uses $maxBal$ and initializes it to $-1$, 
which is smaller than all ballots, and thus, the conjunct $m.bal > maxBal[a]$ in \figref{fig-hist-spec-p1b} and \figref{fig-hist-spec-p2b} holds if no \msg{1a} or \msg{2a} messages were received before.
%

The complete Basic Paxos algorithm specification is given in Appendix~\ref{appendix:spec}.

\begin{figure*}[ht!]
  \centering

\begin{tabular}{|l|l|}
    \hline
    Lamport et al.'s & Using $sent$ only \\
    \hline
    \begin{tabular}{@{}l@{}}
        $vars == <<sent, maxBal, maxVBal, maxVal>>$\\
        $Init == /\  sent = \emptyset$\\
        $\hspace{41.75pt} /\  maxVBal = [a \in \mathcal{A} |-> -1]$\\
        $\hspace{41.75pt} /\  maxBal  = [a \in \mathcal{A} |-> -1]$\\
        $\hspace{41.75pt} /\  maxVal  = [a \in \mathcal{A} |-> \bot]$\\
    \end{tabular}
  &
    \begin{tabular}{@{}l@{}}
        $vars == <<sent>>$\\
        $Init == sent = \emptyset$\\
        \\
        \\
        \\
    \end{tabular}\\
    \hline
    \multicolumn{2}{|p{0.5\textwidth}|}{$Next \mathrel{\smash{\triangleq}} \/ \E b \in \mathcal{B} : Phase1a(b) \/ Phase2a(b)$}\\
    \multicolumn{2}{|p{0.5\textwidth}|}{
        $\qquad \;\;\;\;\; \/ \E a \in \mathcal{A} : Phase1b(a) \/ Phase2b(a)$}\\
    \multicolumn{2}{|p{0.5\textwidth}|}{
        $Spec \mathrel{\smash{\triangleq}} Init \land \Box[Next]_{vars}$}\\
    \hline
\end{tabular}

 \caption{Overall algorithm specification}
 \label{fig-hist-spec-comp}
\end{figure*}

\mysec{Invariants and proofs using message history variables}
\label{sec-impact}

\subsection{Invariants}
\label{sec-basic-inv}
Invariants of a distributed algorithm can be categorized into the following three kinds:
\begin{enumerate}
    \item \textit{Type invariants.} These ensure that all data processed in the algorithm are of valid types. For example, messages of type \msg{1a} must have a field $bal \in \mathcal{B}$. If an action sends a \msg{1a} message with $bal$ missing or $bal \notin \mathcal{B}$, a type invariant is violated.
    \item \textit{Message invariants.} These are invariants defined on message history variables. For example, each message of type \msg{2a} has a unique $bal$. This is expressed by the invariant $\A m1, m2 \in sent: m1.type = \msgtype{2a} /\ m2.type = \msgtype{2a} /\ m1.bal = m2.bal => m1 = m2$.
    \item \textit{Process invariants.} These state properties about the data maintained in derived variables. For example, in Lamport et al.'s specification, one such invariant is that for any acceptor $a$, $maxBal[a] \geq maxVBal[a]$.
\end{enumerate}

\figref{fig-hist-inv} shows and compares all invariants used in Lamport et al.'s proof vs.\ ours. The following operators are used in the invariants for brevity (single-line comments start with \tlaComment{} in \tlaplus{}):
{
\begin{myeqn}
    &VotedForIn(a, v, b) \deq \E m \in sent: \land m.type = \msgtype{2b} \land m.acc = a \land m.val = v \land m.bal = b\\
    &WontVoteIn(a, b) \deq \A v \in \mathcal{V}: \qquad \mbox{\tlaComment{Lamport et al.'s}}\\
    &\quad ~VotedForIn(a, v, b) \land maxBal[a] > b \\
    &WontVoteIn(a, b) \deq \A v \in \mathcal{V}: \qquad \mbox{\tlaComment{Using $sent$ only}}\\
    &\quad ~VotedForIn(a, v, b) \land \E m \in sent: m.type \in \{\msgtype{1b}, \msgtype{2b}\} \land m.acc = a \land m.bal > b\\
    &SafeAt(v, b) \deq \A b2 \in 0..(b-1)\!:\! \E Q \in \mathcal{Q}\!:\! \A a \in Q\!:\! VotedForIn(a, v, b2) \lor WontVoteIn(a, b2)
\end{myeqn}
}

\begin{figure*}[ht!]
    \centering
    \small
    \noindent
    \begin{tabular}{|@{~}p{0.1\textwidth}|>{\raggedright\arraybackslash}p{0.45\textwidth}|@{~}p{0.3\textwidth}|}
         \hline
         & Lamport et al.'s proof & Our proof\\
         \hline
         \multirow{4}{*}{\shortstack[c]{Type\\Invariants}} & (I1) $sent \subseteq Messages$ & $sent \subseteq Messages$\\
         & (I2) $maxVBal \in [\mathcal{A} -> \mathcal{B} \cup \{-1\}]$ & \\
         & (I3) $maxBal \in  [\mathcal{A} -> \mathcal{B} \cup \{-1\}]$ & \\
         & (I4) $maxVal \in  [\mathcal{A} -> \mathcal{V} \cup \{\bot\}]$ & \\
         \hline
         \multirow{6}{*}{\shortstack[c]{Process\\Invariants\\\\\\$\A a \in \mathcal{A}$}} & (I5) $maxBal[a] \geq maxVBal[a]$ & \\
         & (I6) $maxVal[a] = \bot \Leftrightarrow maxVBal[a] = -1$ & \\
         & (I7) $maxVBal[a] \geq 0 =>$ &\\
         & $\quad VotedForIn(a, maxVal[a], maxVBal[a])$ & \\
         & (I8) $\A b \in \mathcal{B} : b > maxVBal[a] =>$ &\\
         & $\quad \nexists v \in \mathcal{V} : VotedForIn(a, v, b)$ & \\
         \hline
         \multirow{14}{*}{\shortstack[c]{Message\\Invariants\\\\\\$\A\!m\!\in\! sent$}} & (I9)  $m.type \!=\! \msgtype{2b} \!=>\! m.bal\! \leq\! maxBal[m.acc]$ & \\
         & (I10) $m.type \!=\! \msgtype{1b} \!=>\! m.bal\! \leq\! maxBal[m.acc]$ & \\
         \cline{2-3}
         
         & (I11) $m.type = \msgtype{1b} =>$ & $m.type = \msgtype{1b} =>$\\
         & $\/ /\ m.maxVal \in \mathcal{V} /\ m.maxVBal \in \mathcal{B}$ &\\
         & $ \hspace{0.9em} /\ VotedForIn(m.acc, $ & $\/ VotedForIn(m.acc,$\\
         & $\hspace{3.0em} m.maxVal, m.maxVBal)$ & $\hspace{2em} m.maxVal, m.maxVBal)$\\
         &$ \/ m.maxVBal = -1 /\ m.maxVal = \bot$ & $\/ m.maxVBal = -1$\\
         \cline{2-3}
         
         & \multicolumn{2}{p{0.78\textwidth}|}{(I12) $m.type = \msgtype{1b} \Rightarrow$}\\
         & \multicolumn{2}{p{0.78\textwidth}|}{$\quad \A b2 \in m.maxVBal+1..m.bal-1: \nexists v \in \mathcal{V} : VotedForIn(m.acc, v, b2)$}\\
         & \multicolumn{2}{p{0.78\textwidth}|}{(I13) $m.type = \msgtype{2a} \Rightarrow SafeAt(m.val, m.bal)$}\\
         & \multicolumn{2}{p{0.78\textwidth}|}{(I14) $m.type = \msgtype{2a} \Rightarrow$}\\
         & \multicolumn{2}{p{0.78\textwidth}|}{$\quad \A m2 \in sent : m2.type = \msgtype{2a} \land m2.bal = m.bal \Rightarrow m2 = m$}\\
         & \multicolumn{2}{p{0.78\textwidth}|}{(I15) $m.type = \msgtype{2b} \Rightarrow$}\\
         & \multicolumn{2}{p{0.78\textwidth}|}{$\quad \E m2 \in sent : m2.type = \msgtype{2a} \land m2.bal = m.bal \land m2.val = m.val$}\\
         \hline
    \end{tabular}
    \caption{Comparison of invariants. Our proof does not need I2-I10, and needs only I1, a simpler I11, and I12-I15.}
    \label{fig-hist-inv}
\end{figure*}

The complete invariants, auxiliary operators, and the safety property to be proved can be found in Appendix~\ref{appendix:prop}.

\subsection{Proving type invariants and process invariants}
\label{sec-hist-basic-prove-type-invs}

\mypar{Type invariants reduced to one} Lamport et al.\ define four type invariants, one for each variable they maintain. $Messages$ is the set of all possible valid messages.  We require only one, (I1). This invariant asserts that the type of all sent messages is valid. (I2) - (I4) are not applicable to our specification.

\mypar{Process invariants not needed} Lamport et al.\ define four process invariants, (I5) - (I8), regarding variables $maxVal$, $maxVBal$, and $maxBal$.  They are not applicable to our specification, and need not be given in our proof.
\begin{itemize}[wide]
    \item \textbf{(I5).} {Because $maxBal[a]$ is the highest ballot ever seen by $a$ and $maxVBal[a]$ is the highest ballot $a$ has voted for, the following invariants hold:
    {
    \notla
    \begin{myeqn}\label{eqn-hist-maxbal}
    \mathit{maxBal[a]} \equiv\; &\agg{max}(\{m.bal\!: m \in sent \land m.type\! \in\! \{\msgtype{1b}\!,\! \msgtype{2b}\} &\land m.acc = a\})\\
    maxVBal[a] \equiv\; &\agg{max}(\{m.bal\!: m \in sent \land m.type\! \in\! \{\msgtype{2b}\} &\land m.acc = a\})\\
    \end{myeqn}%
    }
    where $\agg{max}(S) \!\!==\!\! \CHOOSE e \in S \cup \{-1\}: \A f \in S: e \geq f$. Note that \agg{max} is not in \tlaplus{} and has to be user-defined. Invariant (I5) is needed in Lamport et al.'s proof but not ours because they use derived variables whereas we specify the properties directly. For example, for Lamport et al.'s Phase 1b, one cannot deduce $m.bal > maxVBal[a]$ without (I5), whereas in our Phase 1b, definitions of $2bs$ and $max\_prop$ along with the second conjunct are enough to deduce it.
    }
    
    \item \textbf{(I6).} {Lamport et al.'s proof needs this invariant to prove (I11). Because the initial values are part of $Init$ and are not explicitly present in their Phase 1b, this additional
    invariant is needed to carry this information along. We include the initial values when specifying the action in Phase 1b and therefore do not need this invariant.}
    
    \item \textbf{(I7).} {This invariant is obvious from the definition of $VotedForIn$ in Equation~(3) and property of $maxVBal$ in Equation~(4). The premise $maxVBal[a] \geq 0$ is needed by Lamport et al.'s proof to differentiate from the initial value $-1$ of $maxVBal[a]$.}
    
    \item \textbf{(I8).} {This states that $a$ has not voted for any value at a ballot higher than $maxVBal[a]$. This invariant need not be manually given in our proofs because it is implied from the definition of $maxVBal[a]$.}
\end{itemize}

\subsection{Proving message invariants}
\label{sec-hist-basic-prove-msg-invs}

With history variables, message invariants are either not needed or are more easily proved.
Message invariants (I9) and (I10) follow directly from \eqref{eqn-hist-maxbal} and need not be manually specified for our proof.
Before detailing the other message invariants, we present a systematic method that can derive 
all but one useful invariant used by Lamport et al.\ and thus make the proofs easier.

Our method is based on the following properties of our specifications and distributed algorithms:
\begin{enumerate}
    \item $sent$ grows monotonically, that is, the only operations on it are read and add.
    
    \item Message invariants hold for each sent message of some type, i.e., they are of the form $\A m \in sent: m.type = \textcolor{blue}{\tau} => \Phi(m)$, or more conveniently if we define $sent_{\tau} = \{m \in sent: m.type = \textcolor{blue}{\tau}\}$, we have $\A m \in sent_{\tau}: \Phi(m)$.

    \item $sent = \emptyset$ initially, so the message invariants are vacuously true in the initial state of the system.

    \item Distributed algorithms usually implement a logical clock for ordering two arbitrary messages. In Paxos, this is done by ballots.
\end{enumerate}

We demonstrate our method by deriving (I15).  The method is applied for each message type used in the algorithm. Invariant (I15) is about \msg{2b} messages. We first identify all actions that send \msg{2b} messages and then do the following:
\begin{enumerate}
    \item \textbf{Increment.} \msg{2b} messages are sent in Phase 2b as specified in \figref{fig-hist-spec-p2b}. We first determine the increment to $sent$, $\Delta(sent)$, the new messages sent in Phase 2b. We denote a message in $\Delta(sent)$ by $\delta$ for brevity.  We have, from \figref{fig-hist-spec-p2b},
    \begin{myeqn}
        \delta = [&type |-> \msgtype{2b}, acc |-> a, bal |-> m.bal, val |-> m.val]
    \end{myeqn}
    
    \item \textbf{Analyze.} We deduce properties about the messages in $\Delta(sent)$. 
    For \msg{2b} messages, we deduce the most straightforward property that connects the contents of messages in $\Delta(sent)$ with the message $m$, from \figref{fig-hist-spec-p2b},
    \begin{myeqn}
        \phi(\delta) = &\ \E m \in sent: m.type = \msgtype{2a} /\ \delta.bal = m.bal /\ \delta.val = m.val
    \end{myeqn}
    
    \item \textbf{Integrate.} Because (i) $sent$ monotonically increases, and (ii) $\phi$ is an existential quantification over $sent$, 
    $\phi$ holds for all increments to $sent_{2b}$. Property (i) means that once the existential quantification in $\phi$ holds, it holds forever. 
    Integrating both sides of Equation~(6) in the space of \msg{2b} messages yields (I15), that is,
    \begin{myeqn}
        \Phi(sent_{2b}) = &\ \A m2 \in sent_{2b}: \E m \in sent: m.type = \msgtype{2a} /\ m2.bal = m.bal /\ m2.val = m.val
    \end{myeqn}%
    The case for $\phi$ being universally quantified over $sent$ is discussed with invariant (I12).
\end{enumerate}

We also derive 
(I11), (I12), and (I14) as described in the following.

\begin{itemize}[wide]
	\item \textbf{(I11).} {Like (I15), (I11) can also be systematically derived, from our Phase 1b in \figref{fig-hist-spec-p1b}. 
	This invariant is less obvious and harder to prove when variables $maxVal$ and $maxVBal$ are explicitly used and updated because (i) they are not updated in the same action that uses them, requiring additional invariants to carry their meaning to the proofs involving the actions that use them, and (ii)  it is not immediately clear if these variables are being updated in Lamport et al.'s Phase 2b in \figref{fig-hist-spec-p2b} because a \msg{2b} message is being sent or because a \msg{2a} message was received.}
	
	\item \textbf{(I12).} {To derive (I11) and (I15), we focused on \textit{where} the contents of the new message come from. For (I12), we analyze \textit{why} those contents 
	were chosen.
	From our Phase 1b with definitions of $2bs$ and $max\_prop$ in \figref{fig-hist-spec-p1b}, we have
	{
	\notla
	\begin{myeqn}\label{eqn-hist-i12}
        &\phi(\delta) =\\
        &\quad \lor \land \ \E m \in sent: \, m.type = \msgtype{2b} \land m.acc = \delta.acc \\
        &\hspace{1.9em} \land \A m \in sent: \, m.type = \msgtype{2b} \land m.acc = \delta.acc \Rightarrow \delta.maxVBal \geq m.bal\\
        &\quad \lor \land \, \nexists\ m \in sent: \, m.type = \msgtype{2b} \land m.acc = \delta.acc\\
        &\hspace{1.9em} \land \delta.maxVBal = -1
    \end{myeqn}
    }

	$\phi$ has two disjuncts---the first has a universal quantification and the second has a negated existential quantification, which is universal in disguise. If $sent$ is universally quantified, integration as for (I15) is not possible because the quantification only holds \textit{at the time of the action}. As new messages are sent in the future, the universal may become violated.
	
	The key is the phrase \textit{at the time}. One way to work around the universal is to add a time field in each message and update it in every action as a message is sent, such as using a logical clock. Then, a property of the form $\phi(\delta) = \A m \in sent_{\tau}: \psi(m)$ can be integrated to obtain
	\begin{myeqn}
	    &\Phi(sent_{\tau}) = \A m2 \in sent_{\tau}: \A m \in sent: m.time < m2.time => \psi(m)
	\end{myeqn}%
	Because ballots act as the logical clock in Paxos, we do not need to specify a separate logical clock and we can perform the above integration on \eqref{eqn-hist-i12} to obtain invariant (I12).
	}
	
	\item \textbf{(I14).} {This invariant is of the form $\A m1,$ $m2 \in sent_{\tau}, t: \psi(m1, t) /\ \psi(m2, t) => m1 = m2$. In this case, $\psi(m, t) == m.bal = t$. Deriving invariants like (I14) is nontrivial unless $\psi$ is already known. In some cases, $\psi$ can be guessed. The intuition is to look for a universal quantification (or negated existential) in the specification of an action. The ideal case is when the quantification is on the message type being sent in the action. Potential candidates for $\psi$ may be hidden in such quantifications. Moreover, if message history variables are used, these quantifications are easier to identify.
	
	Starting with a guess of $\psi$, we identify the change in the counting measure (cardinality) of the set $\{t: m \in sent_{\tau} /\ \psi(m, t)\}$ along with that of $sent_{\tau}$. In the case of (I14), we look for $\Delta(|\{m.bal: m \in sent_{2a}\}|)$. From our Phase 2a in \figref{fig-hist-spec-p2a}, we have
	\begin{equation}
	    \begin{aligned}
        &\Delta(\{m.bal: m \in sent_{2a}\}) = \{b\}\\
        &\phi(b) = \nexists m \in sent: m.type = \msgtype{2a} /\ m.bal = b,\\
        &\quad \text{where}\ b \in \Delta(\{m.bal: m \in sent_{2a}\})
        \end{aligned}
	\end{equation}
	
	Rewriting $\phi$ as $\{b\} \not\subseteq \{m.bal: m \in sent_{2a}\}$, it becomes clear that $\Delta(|\{m.bal: m \in sent_{2a}\}|)$ $= 1$. Meanwhile, $\Delta(|\{m \in sent_{2a}\}|) = 1$. Because the counting measure increases by the same amount for both, (I14) can be derived safely. 
	}
\end{itemize}



\subsection{Basic Paxos proof}
\label{sec-hist-basic-proof}
The main property to prove is $Agreement$, defined as follows:
\begin{myeqn}\label{cons}\begin{aligned}
&Inv == TypeOK /\ MsgInv\\
&Agree == \A v1, v2 \in \mathcal{V} : Chosen(v1) /\ Chosen(v2) => v1 = v2\\
&\THEOREM Agreement\!\! ==\!\! Spec => []Agree
\end{aligned}\end{myeqn}%

To proceed, we first prove $Inv => Agree$. We then prove $Spec => []Inv$ and, by temporal logic, conclude $Spec => []Agree$. Note that property $Agreement$ is called $Consistent$, and invariant $Agree$ is called $Consistency$ by Lamport et al.~\cite{basicpaxos2014}.

To prove $Agree$ for the algorithm, 
we first prove the following helper lemmas for three important properties:
\begin{enumerate}
    \item Lemma $VotedInv$. If any acceptor votes any pair $\langle v, b \rangle$, then the predicate $SafeAt(v, b)$ holds:
    \begin{equation}\label{votedinv}\begin{aligned}
        &\LEMMA VotedInv == MsgInv /\ TypeOK =>\A a \in \mathcal{A}, v \in \mathcal{V}, b \in \mathcal{B}:\\
        &\qquad VotedForIn(a, v, b) => SafeAt(v, b)
    \end{aligned}\end{equation}

    \item Lemma $VotedOnce$. If acceptor $a1$ votes pair $\langle v1, b \rangle$ and acceptor $a2$ votes pair $\langle v2, b \rangle$, then $v1 = v2$:
    \begin{equation}\label{votedonce}\begin{aligned}
        &\LEMMA VotedOnce == MsgInv => \A a1, a2 \in \mathcal{A}, v1, v2 \in \mathcal{V}, b \in \mathcal{B}:\\
        &\qquad VotedForIn(a1, v1, b) /\ VotedForIn(a2, v2, b) => v1 = v2
    \end{aligned}\end{equation}
    
    \item Lemma $SafeAtStable$. If pair $\langle v, b \rangle$ is safe in the current state, it remains safe in the next state, where state transition is defined by $Next$.
    \begin{equation}\label{safeatstable}\begin{aligned}
        &\LEMMA SafeAtStable == Inv /\ Next => \A v \in \mathcal{V}, b \in \mathcal{B}:\\
        &\qquad SafeAt(v, b) => SafeAt(v, b)'
    \end{aligned}\end{equation}
\end{enumerate}

The proof of $Spec => []Inv$ follows the same strategy as used in Chand et al.~\cite{chand2016formal}. The proof is inductive, written in a hierarchical style~\cite{lamport2012write}. The base case proves $Init => Inv$. The inductive case considers each action in $Next$ individually, and proves that $Inv$ holds in the next state given that it holds in the current state.

The complete TLAPS-checked proof for Basic Paxos spans about 2 pages, and is summarized as follows:

\begin{enumerate}
\item The three lemmas and their proofs are about half a page.

\item The proof of type invariant $TypeOK$ is a quarter page, using only a 1-level proof for each action. 

\item The proof of message invariant $MsgInv$ is less than a page, using 1-level proofs for actions $Phase1a$, $Phase1b$, and $Phase2b$, together taking less than a half a page, and a 4-level proof for $Phase2a$, taking half a page.

\item The proof of theorem $Agreement$ using $Spec => []Inv$ is a quarter page, with a straightforward argument of $Inv => Agree$.
\end{enumerate}
The complete proof is given in Appendix~\ref{appendix:proof}. 

\mysec{Multi-Paxos}
\label{sec-multi}
We have developed new specifications of Multi-Paxos and Multi-Paxos with Preemption that use only message history variables. 
\secref{sec-multi-spec} outlines and compares two approaches to develop these specifications from existing specifications. 
\secref{sec-multi-preempt-spec} discusses specification of Multi-Paxos with Preemption. The complete specification of Multi-Paxos with Preemption is provided in Appendix~\ref{appendix:mppspec}. 
\secref{sec-multi-inv} describes two invariants used in our proofs. One of the invariants is new compared to Chand et al.~\cite{chand2016formal}, while the other is similar to an invariant used in~\cite{chand2016formal} but our invariant is simpler. The other invariants used by us compare to theirs similar to our invariants for Basic Paxos compared with Lamport et al.'s as described in \secref{sec-impact}, and are therefore moved to Appendix~\ref{appendix:mpinvcomp}.
\secref{sec-multi-proof} discusses the importance of these specifications to the proof, in particular how these helped in reducing proof size by 48\%.

\subsection{Specification of Multi-Paxos}
\label{sec-multi-spec}
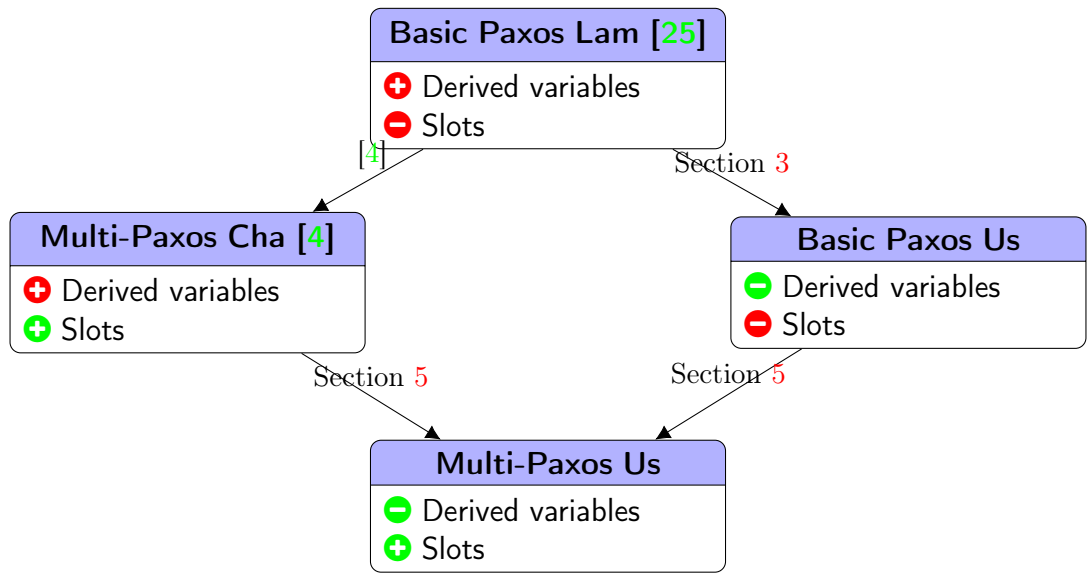
\begin{figure}[h]
    \centering
    \begin{tikzpicture}[node distance=1.5cm,
        specnode/.style={ rectangle split,
                            rectangle split parts=2,
                            rectangle split part fill={blue!30,white},
                            rounded corners,
                            draw=black,
                            minimum width={26ex},
                            text width={24ex},
                            minimum height=1cm,
                            font=\sffamily,
                            align=left},
        align=center,
        invisible/.style={opacity=0},
        ]

      \node [specnode] (bpfm)
        {
        \centerline{\textbf{Basic Paxos Lam~\cite{basicpaxos2014}}}
        \nodepart{second}
        \textcolor{red}{\faPlusCircle{}} Derived variables\\
        \textcolor{red}{\faMinusCircle{}} Slots
        };
    
      \node [invisible] (dummy3) [below = of bpfm] {extra};
      
      \node [specnode] (mpfm) [left = 10ex of dummy3]
        {
        \centerline{\textbf{Multi-Paxos Cha~\cite{chand2016formal}}}
        \nodepart{second}
        \textcolor{red}{\faPlusCircle} Derived variables\\
        \textcolor{green}{\faPlusCircle} Slots
        };

      \node [specnode] (bphist) [right = 10ex of dummy3]
        {
        \centerline{\textbf{Basic Paxos Us}}
        \nodepart{second}
        \textcolor{green}{\faMinusCircle} Derived variables\\
        \textcolor{red}{\faMinusCircle} Slots
        };

      \node [specnode] (mphist) [below = 10ex of dummy3]
        {
        \centerline{\textbf{Multi-Paxos Us}}
        \nodepart{second}
        \textcolor{green}{\faMinusCircle} Derived variables\\
        \textcolor{green}{\faPlusCircle} Slots
        };
    
      \draw[->] (bpfm) --  node [above] {\small{Section~\ref{sec-basic-paxos}}} (bphist);
      \draw[->] (bpfm) -- node [above] {\small{~\cite{chand2016formal}}} (mpfm);
      \draw[->] (mpfm) -- node [above] {\small{Section~\ref{sec-multi}}} (mphist);
      \draw[->] (bphist) -- node [above] {\small{Section~\ref{sec-multi}}} (mphist);

    \end{tikzpicture}

    \caption{Derivation showing two approaches to develop specification for Multi-Paxos with history variables.}
    \label{fig-hist-spec-diamond}
\end{figure}

\newcommand{\myleft}{left}
\newcommand{\myright}{right}
\figref{fig-hist-spec-diamond} shows two approaches to derive the specification of Multi-Paxos with history variables using existing specifications: (1) From Chand et al.'s Multi-Paxos that uses derived variables~\cite{chand2016formal}, removing derived variables to use only history variables, similar to the specification of Basic Paxos described in \secref{sec-basic-paxos} and, (2) From Basic Paxos that uses only history variables, by adding slots to obtain Multi-Paxos.

To compare these two approaches, we use Phase 1b as an example. Following is the English description of Phase 1b of Multi-Paxos, obtained from that of Basic Paxos (\figref{fig-hist-spec-p1b}) by adding ``\textit{for each slot}'' at the end:

\begin{quote}
If an acceptor receives a \msg{1a} request with number \m{bal} greater than that of any \msg{1a} request to which it has already responded, then it responds to the request with a promise not to accept any more proposals numbered less than \m{bal} and with the highest-numbered proposal (if any) that it has accepted \textit{for each slot}.
\end{quote}

This implies an obvious way to obtain Multi-Paxos---by specifying Basic Paxos per slot. In practice, this would mean executing multiple instances of Basic Paxos. This is inefficient and yields impractical specifications. For instance, in Phase 1b sending multiple \msg{1b} messages, one per slot, is worse than sending a single \msg{1b} message with the set of accepted proposals for each slot.

\mypar{From Multi-Paxos that uses derived variables}
The development of Phase 1b for this approach follows closely that of Basic Paxos. The two specifications are shown in \figref{fig-hist-spec-p1b-multiderived}---on the left we have the specification of Phase 1b from Chand et al.~\cite{chand2016formal} and on the right we have our specification of Phase 1b that uses only history variables.
The first difference is replacing $aBal$ with a query over $sent$. $aBal$ is called $maxBal$ in Basic Paxos (\figref{fig-hist-spec-p1b}) and therefore the query is the same as in our specification of Basic Paxos that uses only history variables.
The second difference is replacing $aVoted[a]$ with $PartialBmax(voteds(a))$. $voteds(a)$ is the set of all $\langle ballot, slot, value \rangle$ triples that acceptor $a$ has voted for. $PartialBmax(T)$ is a subset of votes in $T$, one vote per slot, that contains only the highest-numbered vote for that slot in $T$.

\begin{figure*}[h]
  \centering

\begin{tabular}{|l|l|}
    \hline
    Multi-Paxos that uses derived variables~\cite{chand2016formal} & After removing derived variables\\
    \hline
    \begin{tabular}{@{}l@{}}
        $Phase1b(a \in \mathcal{A}) ==$\\
        $\E m \in msgs :$\\
        $\quad /\ m.type = \textcolor{blue}{"1a"}$\\
        $\quad /\ m.bal > aBal[a]$\\
        \\
        $\quad /\ Send([type |-> \textcolor{blue}{"1b"},$\\
        $\qquad from |-> a, bal |-> m.bal,$\\
        $\qquad voted |-> aVoted[a]])$\\
        $\quad /\ aBal' = [aBal \EXCEPT ![a] = m.bal]$\\
        $\quad /\ \UNCHANGED <<pBal, aVoted>>$\\
        \\
        \\
        \\
        \\
        \\
    \end{tabular}
 &
    \begin{tabular}{@{}l@{}}
        $Phase1b(a \in \mathcal{A}) ==$\\
        $\E m \in sent :$\\
        $\quad /\ m.type = \textcolor{blue}{"1a"}$\\
        $\quad /\ \A m2 \in sent: m2.type \in \{\msgtype{1b}, \msgtype{2b}\} /\ $\\
        $\qquad m2.acc = a => m.bal > m2.bal$\\
        $\quad /\ Send([type |-> \textcolor{blue}{"1b"},$\\
        $\qquad from |-> a, bal |-> m.bal,$\\
        $\qquad voted |-> PartialBmax(voteds(a))])$\\
        \\
        \\
        $PartialBmax(T) == \{t \in T : \A t2 \in T :$\\
        $\quad t.slot = t2.slot => t.bal \geq t2.bal\}$\\
        $voteds(a) ==$\\
        $\quad \{[bal |-> m.bal, slot |-> m.slot, val |-> m.val]:$\\
        $\quad m \in 2bs(a)\}$\\
    \end{tabular}\\
    \hline
\end{tabular}

 \caption{Change to Phase 1b of Multi-Paxos that uses derived variables, to using only history variables.}
 \label{fig-hist-spec-p1b-multiderived}
\end{figure*}

\mypar{From Basic Paxos that uses only history variables}
Two modifications need to be made for this approach: (1) adding slots and, (2) replying with a set of votes as opposed to a single vote in Basic Paxos. Recalling that ``the highest-numbered proposal (if any) that it has accepted'' was specified by $max\_prop$ in \figref{fig-hist-spec-p1b} and that the only change to the English description was adding ``for each slot'' to the aforementioned phrase, so the main change to add slots would be a small change in $max\_prop$ as shown in \figref{fig-hist-add-slots1b}. Finally, adding the change to reply with a set of votes leads to the specification shown in ~\figref{fig-hist-spec-p1b-multiderived}.

\begin{figure*}[h]
  \centering
\begin{tabular}{|l|l|}
    \hline
    Basic Paxos that uses only history variables & After adding slots\\
    \hline
    \begin{tabular}{@{}l@{}}
        $max\_prop(a) ==$\\
        $\;$\tlaif{} $2bs(a) = \emptyset$ \tlathen{} $\{[bal |-> -1, val |-> \bot]\}$\\
        $\;$\tlaelse{}
        $\;\{m \in 2bs(a): \A m2 \in 2bs(a):$\\
        $m.bal \geq m2.bal \}$
    \end{tabular}
 &
    \begin{tabular}{@{}l@{}}
        $max\_prop(a) ==$\\
        $\;$\tlaif{} $2bs(a) = \emptyset$ \tlathen{} $\{[bal |-> -1, val |-> \bot]\}$\\
        $\;$\tlaelse{}
        $\;\{m \in 2bs(a): \A m2 \in 2bs(a):$\\
        ${\textcolor{commentGreen}{m.slot = m2.slot =>}} m.bal \geq m2.bal \}$
    \end{tabular}\\
    \hline
\end{tabular}

 \caption[.]{Change to Phase 1b of Basic Paxos that uses only history variables to add slots.}
 \label{fig-hist-add-slots1b}
\end{figure*}

\mypar{Lessons learned} The second approach is easier 
than the first approach because the change is smaller and more easily isolated. The change in the first approach requires replacing derived variables. which is more difficult, especially for $aVoted[a]$ which is updated in Phase 2b of Chand et al.~\cite{chand2016formal} using the following complex expression:

\begin{align}\label{eqn-aVoted}
aVoted' = &\ [aVoted \EXCEPT ![a] = \nonumber\\
          &\ \cup \{[bal \mapsto m.bal, slot \mapsto d.slot, val \mapsto d.val] : d \in m.propSV\}\\
          &\ \cup \{e \in aVoted[a] : \nexists\, r \in m.propSV : e.slot = r.slot \}]\nonumber
\end{align}

\subsection{Specification of Multi-Paxos with Preemption}
\label{sec-multi-preempt-spec}
Specification of preemption required defining a new action and changing Phase 1a as follows:
\begin{itemize}
    \item Add action Preempt. With preemption, if an acceptor receives a \msg{1a} or \msg{2a} message with a ballot smaller than the highest that it has seen, it responds with a \msg{preempt} message that contains the highest ballot it has seen. Chand et al.~\cite{chand2016formal} specify this by changing their specifications of Phase 1b and Phase 2b with repeated logic formulae as shown in \figref{fig-hist-spec-pmt} (the \tlaelse{} branch). We specify a new action leading to a cleaner specification. In our specification in \figref{fig-hist-spec-pmt}, $m$ is the \msg{1a} or \msg{2a} message that acceptor $a$ received, and $m2$ is a \msg{1b} or \msg{2b} message already sent by $a$ whose ballot is higher than $m.bal$.
    
    Most importantly, our specification led us to identify a subtlety in Chand et al.'s specification that can cause unnecessary preemptions. In their specification of Phase 1b, an acceptor sends a \msg{preempt} message upon receiving a duplicate \msg{1a} message instead of ignoring it, causing the proposer to unnecessarily preempt. While this does not affect safety, it may impact liveness if not handled correctly by the proposer.
    \begin{figure*}[h]
    \centering
    \begin{tabular}{|l|l|}
    
        \hline
        Chand et al.~\cite{chand2016formal} & Us \\
        \hline
        \begin{tabular}{@{}l@{}}
            $Phase1b(a \in \mathcal{A}) == \E m \in sent:$\\
            $\; /\ \; m.type = \msgtype{1a}$\\
            $\; /\ \; \tlaif{}\, m.bal > aBal[a]\, \tlathen{} \ldots$\\
            $\hspace{1.15em}    \tlaelse{} /\ Send([type |-> \msgtype{preempt},$\\
            $\hspace{5.15em}                    to |-> m.from, bal |-> aBal[a]])$\\
            $\hspace{3.15em}               /\ \UNCHANGED <<aVoted, aBal, pBal>>$\\
            \\
            $Phase2b(a \in \mathcal{A}) == \E m \in sent:$\\
            $\; /\ \; m.type = \msgtype{2a}$\\
            $\; /\ \; \tlaif{}\, m.bal \geq aBal[a]\, \tlathen{} \ldots$\\
            $\hspace{1.15em}    \tlaelse{} /\ Send([type |-> \msgtype{preempt},$\\
            $\hspace{5.15em}                    to |-> m.from, bal |-> aBal[a]])$\\
            $\hspace{3.15em}               /\ \UNCHANGED <<aVoted, aBal, pBal>>$\\
        \end{tabular}
      &
        \begin{tabular}{@{}l@{}}
            $Preempt(a \in \mathcal{A}) == \E m \in sent, m2 \in 1b2b(a):$\\
            $\; /\ \; m.type \in \{\msgtype{1a}, \msgtype{2a}\}$\\
            $\; /\ \; m2.bal > m.bal$\\
            $\; /\ \; \A m3 \in 1b2b(a): m2.bal \geq m3.bal$\\
            $\; /\ \; Send([type |-> \msgtype{preempt}, to |-> m.from,$\\
            $\hspace{1.55em} bal |-> m2.bal])$\\
            \\
            $1b2b(a) ==$\\
            $\; \{m \in sent: m.type \in \{\msgtype{1b}, \msgtype{2b}\} /\ m.from = a\}$\\
            \\
            \\
            \\
            \\
        \end{tabular}\\
        
    \hline
    \end{tabular}
     \caption[.]{Specifications for preemption in Chand et al. and us. The ``$\ldots$'' in Chand et al.'s specification correspond to parts of $Phase1b$ and $Phase2b$ that do not apply to preemption.}
     \label{fig-hist-spec-pmt}
    \end{figure*}

    \item Change $Phase1a$. $Phase1a$ remains unchanged between Basic Paxos and Multi-Paxos. However with preemption, upon receiving a \msg{preempt} message with a ballot higher than every ballot on which the receiving proposer has ever initiated Phase 1a, the proposer needs to pick a new ballot and initiate Phase 1a on it. \figref{fig-hist-spec-p1a-pmt} shows the specifications of Phase 1a for Multi-Paxos and Multi-Paxos with Preemption. If the proposer receives a \msg{preempt} message with a ballot higher than the ballots of all \msg{1a} messages that it has sent, it picks a new ballot higher than the ballot of the \msg{preempt} message and sends a \msg{1a} message with this new ballot.

    \begin{figure*}[h]
    \centering
    \begin{tabular}{|l|l|}
    
        \hline
        Multi-Paxos & Multi-Paxos with Preemption \\
        \hline
        \begin{tabular}{@{}l@{}}
            $Phase1a(p \in \mathcal{P}) == \E b \in \mathcal{B}:$\\
            \\
            \\
            \\
            \\
            \\
            $\; Send({[type |-> \msgtype{1a}, from |-> p, bal |-> b]})$
        \end{tabular}
      &
        \begin{tabular}{@{}l@{}}
            $Phase1a(p \in \mathcal{P}) == \E b \in \mathcal{B}:$\\
            $\; /\           \/ \nexists\, m \in sent: m.type = \msgtype{preempt} /\ m.to = p$\\
            $\hspace{1.15em} \/ \E m \in sent:$\\
            $\hspace{2.15em}      /\ m.type = \msgtype{preempt} /\ m.to = p /\ b > m.bal$\\
            $\hspace{2.15em}      /\ \A m2 \in sent: m2.type = \msgtype{1a} /\ m2.from = p$\\
            $\hspace{8.85em}                        => m.bal > m2.bal$\\
            $\; /\ \, Send([type |-> \msgtype{1a}, from |-> p, bal |-> b])$\\
        \end{tabular}\\
        
    \hline
    \end{tabular}
     \caption{Our specifications of Phase 1a for Multi-Paxos and Multi-Paxos with Preemption}
     \label{fig-hist-spec-p1a-pmt}
    \end{figure*}

\end{itemize}

\subsection{Invariants}
\label{sec-multi-inv}

Following our methodology, we were able to derive all but two invariants in the case of Multi-Paxos. One of the invariants is similar to (I13), which we also do not derive for Basic Paxos. The other is a \msg{1b} message invariant, (I26), shown in ~\figref{fig-hist-multi-1b-inv}. Additionally, one of the \msg{1b} message invariants generated by us, (I27), is slightly different than its counterpart in Chand et al.'s proof, also shown in ~\figref{fig-hist-multi-1b-inv}. Comparison of other invariants is similar to that for Basic Paxos (\figref{fig-hist-inv}) and therefore the complete list for Multi-Paxos is moved to Appendix~\ref{appendix:mpinvcomp}.
\begin{figure}[]
    \centering
    \begin{tabular}{|l|l|}
        \hline
        Chand et al.~\cite{chand2016formal} & Us \\
        \hline
        & (I26) $\A m \in msgs: m.type = \msgtype{1b} =>$\\
        & $\; \A b \in 0..m.bal-1, s \in \Slots, v \in \Values:$\\
        & $\;\; VotedForIn(m.from, b, s, v) =>$\\
        & $\;\; \E r \in m.voted: r.slot = s /\ r.bal \geq b$\\
        \hline
        (I27) $\A m \in msgs : m.type = \msgtype{1b} =>$ & $\A m \in msgs : m.type = \msgtype{1b} =>$ \\
        $\; \A b \in \mathcal{B}, s \in \mathcal{S}, v \in \mathcal{V}:$ & $\; \A r \in m.voted:$ \\
        $\;\; b \in MaxVotedBallotInSlot(m.voted, s)+1..m.bal-1 =>$ & $\;\; \A b \in r.bal+1..m.bal-1, v \in \mathcal{V}:$\\
        $\;\;\; \neg VotedForIn(m.from, b, s, v)$ & $\;\;\; \neg VotedForIn(m.from, b, r.slot, v)$\\
        \hline
    \end{tabular}
    \caption{Comparison of \msg{1b} message invariants that are new (I26) or are different (I27).}
    \label{fig-hist-multi-1b-inv}
\end{figure}

Invariant (I26) states that for every \msg{1b} message $m$, slot $s$ and ballot $b$ smaller than $m.bal$, if acceptor $m.from$ has voted in $b$ for $s$ then there must exist a vote in $m.voted$ with slot $s$ and ballot greater than or equal to $b$.

Invariant (I27) in Chand et al.~\cite{chand2016formal} states that for every \msg{1b} message $m$, slot $s$ and ballot $b$ higher than the highest ballot that $m.from$ has voted in for slot $s$, and lower than the ballot in $m$, $m.from$ has not voted any value $v$ for slot $s$ in ballot $b$. The operator $MaxVotedBallotInSlot$ is defined as

\begin{myeqn}
&MaxVotedBallotInSlot(D, s) \deq \textsf{max}(\{d.bal : d \in \{d \in D : d.slot = s\}\})
\end{myeqn}

We derive a similar invariant but without the notion of ``highest ballot''. Our invariant states that for every \msg{1b} message $m$, and vote $r$ in $m.voted$, $m.from$ has not voted in any ballot higher than the ballot in $r$ and less than the ballot in $m$ for the slot in $r$.

Because $m.from$ has voted for every triple in $m.voted$, this means that in Chand et al.'s specification, $m.voted$ may have multiple votes for the same slot, and therefore they take a max in their invariant, whereas in our specification, $m.voted$ has only one vote per slot---the one with the highest ballot as can be seen from~\figref{fig-hist-spec-p1b-multiderived} following the definitions of $PartialBmax$ and $voteds$.

Consider Chand et al.'s Phase 1b in \figref{fig-hist-spec-p1b-multiderived}, note that $m.voted$ copies the derived variable $aVoted[m.from]$ and consider Chand et al.'s update logic of $aVoted$ in~\eqref{eqn-aVoted}, we can see that for each slot that $m.from$ has voted for, $aVoted[m.from]$ only maintains the vote with the highest ballot. This means that even in Chand et al.'s proof, invariant (I27) could be relaxed to the one used by us, leading to a smaller proof for reasons described in~\secref{sec-multi-proof}. This only bolsters our claim that only using history variables leads to simpler specifications and easier proofs.

\subsection{Proof}
\label{sec-multi-proof}

While we observed a quarter decrease in proof size for Basic Paxos, for Multi-Paxos this decrease was almost a half.
Besides the simplifications described in \secref{sec-impact}, an important player in this decrease was the absence of the operator $MaxVotedBallotInSlot$ used by Chand et al.~\cite{chand2016formal}. Five lemmas were needed in Chand et al.'s proof to assert basic properties of the operator. For example, lemma $MVBISType$ stated that if $D \subseteq [bal : \mathcal{B}, slot : \mathcal{S}, val : \mathcal{V}]$, then the result of the operator is in $ \mathcal{B} \cup \{-1\}$.

Absence of this operator and not having derived variables attributed to more than two-thirds of the decrease in proof size. This includes: 
(1) removing lemmas for $MaxVotedBallotInSlot$ and their proofs, and deriving a different message invariant (I27) that does not contain the operator leading to a smaller proof,
(2) removing process invariants and their proofs and,
(3) removing message invariants with derived variables (similar to I9 and I10) and their proofs
\notes{
\begin{itemize}
    \item A \msg{1b} message invariant in Chand et al.'s proof uses the operator $MaxVotedBallotInSlot$. This invariant is similar to (I12). Not having the operator in the invariant (and therefore the proof) resulted in a 6\% decrease in the proof.
    
    \item A \msg{1b} and \msg{2b} message invariant respectively similar to (I10) and (I9) in Chand et al.'s proof involve derived variable $maxBal$. Absence of these invariants with derived variables resulted in a 3\% decrease in the proof.
\end{itemize}
}

The remaining decrease in proof size is attributed to proof improvements, for example, removing unnecessary manually written proofs for obligations that can be automatically proven by backend provers.

\mysec{Results}
\label{sec-results}
Table~\ref{tab-sum-bp} summarizes the results of our specifications and proofs that use only message history variables, compared with those by Lamport et al.~\cite{basicpaxos2014} and Chand et al.~\cite{chand2016formal}. 
We observe an improvement of about a quarter across all stats for Basic Paxos and about a half for Multi-Paxos and Multi-Paxos with Preemption. Following, we list some important results:
\begin{itemize}
    \item The specification size decreased by 13 lines (25\%) for Basic Paxos, from 52 lines for Lamport et al.'s specification to 39 lines for ours. For Multi-Paxos, the decrease is 36 lines (46\%), from 78 lines for Chand et al.'s to 42 lines for ours, and for Multi-Paxos with Preemption, the decrease is 45 lines (46\%), from 97 to 52.
    
    \item The total number of manually written invariants decreased by 54\% overall---by 9 (60\%) from 15 to 6 for Basic Paxos, by 8 (50\%) from 16 to 8 for Multi-Paxos, and by 9 (53\%) from 17 to 8 for Multi-Paxos with Preemption. This significant decrease is because we do not maintain derived variables $maxBal$, $maxVBal$, and $maxVal$ as explained in~\secref{sec-impact}. 
    
    \item The proof size for Basic Paxos decreased by 83 lines (27\%), from 310 to 227. This decrease is attributed to the fact that our specification does not use other state variables besides $sent$. For Multi-Paxos and Multi-Paxos with Preemption, this decrease is 468 lines (47\%), from 988 to 520, and 494 lines (48\%), from 1032 to 538, respectively. The proof size increases about 300 lines from Basic Paxos to Multi-Paxos because type invariants and message invariants have to be proved over sets of tuples.
     
    \item Proof by contradiction is used twice in the proof by Lamport et al.\ and thrice for the proofs in Chand et al. 
    We were able to remove all of them because our specification uses queries as opposed to derived variables. This yields easier-to-understand constructive proofs.

    \item The number of proof obligations decreased by 46\%---by 57 (24\%) from 239 to 182 for Basic Paxos, by 450 (49\%) from 918 to 468 for Multi-Paxos, and by 468 (49\%) from 959 to 491 for Multi-Paxos with Preemption.
    
    \item The proof-checking time decreased by 11 seconds (26\%), from 42 to 31 for Basic Paxos. For Multi-Paxos and Multi-Paxos with Preemption, TLAPS took over 3 minutes for the proofs in~\cite{chand2016formal} and failed (due to updates in the new version of TLAPS) to check the proofs of 5 obligations. In contrast, our proofs were checked successfully in 1.5 minutes or less.
\end{itemize}

\newcommand{\cntapp}{*}
\newcommand{\tlafail}{**}
\newcommand{\cnterr}{\textasciicircum}
\begin{table*}[t!]
  \centering
\begin{tabular}{l|r@{\hskip 6pt}rr|r@{~~}r@{~~}r|r@{~~}r@{~~}r}
    \hline
    \multirow{3}{*}{Metric} & \multicolumn{3}{c|}{Basic Paxos}&\multicolumn{3}{c|}{Multi-Paxos} & \multicolumn{3}{c}{Multi-Paxos}\\
    & & & & & & & \multicolumn{3}{c}{with Preemption}\\
    \cline{2-10}
    & Lam & Us & Decr & Cha & Us & Decr & Cha & Us & Decr\\
    \hline
    Specification size  & 52  & 39, 33\cntapp{}  &25\% & 56\cnterr{}   & 42  & 25\% & 75\cnterr{}   & 52 & 31\%\\
    \hline
    \# invariants    & 15 & 6 & 60\% & 16 & 8 & 50\% & 17 & 8 & 53\%\\
    \# type invariants    & 4 & 1 & 75\% & 4 & 1 & 75\% & 5 & 1 & 80\%\\
    \# process invariants & 4 & 0 & 100\% & 4 & 0 & 100\% & 4 & 0 & 100\%\\
    \# message invariants & 7 & 5 & 29\% & 8 & 7 & 13\% & 8 & 7 & 13\%\\
    \hline
    Proof size & 310 & 227, 115\cntapp{} & 27\% & 1010\cnterr{} & 520 & 47\% & 1054\cnterr{} & 538 & 48\%\\
    \hline
    Type invariants' proof size & 22 & 21, 12\cntapp{} & 5\% & 54 & 34 & 37\% & 75 & 38 & 49\%\\
    Process invariants' proof size & 27 & 0, 0\cntapp{} & 100\% & 136 & 0 & 100\% & 141 & 0 & 100\%\\
    \msg{1b}$^{\dagger}$ invariants' proof size & 21 & 15, 9\cntapp{} & 29\% & 133 & 70 & 47\% & 133 & 70 & 47\%\\
    \msg{2a}$^{\dagger}$ invariants' proof size & 73 & 57, 21\cntapp{} & 22\% & 264 & 120 & 55\% & 269 & 120 & 55\%\\
    \msg{2b}$^{\dagger}$ invariants' proof size & 14 & 12, 7\cntapp{} & 14\% & 94 & 73 & 22\% & 94 & 73 & 22\%\\
    \hline
    \# proofs by contradiction & 2 & 0 & 100\% & 3 & 0 & 100\% & 3 & 0 & 100\%\\
    \hline
    \# obligations in TLAPS & 239 & 182 & 24\% & 918 & 468 & 49\% & 959 & 491 & 49\%\\
    Type inv proof obligations & 17 & 17 & 0\% & 69 & 52 & 25\% & 100 & 60 & 40\%\\
    Process inv proof obligations & 39 & 0 & 100\% & 163 & 0 & 100\% & 173 & 0 & 100\%\\
    \msg{1b}$^{\dagger}$ inv proof obligations & 12 & 10 & 17\% & 160 & 80 & 50\% & 160 & 80 & 50\%\\
    \msg{2a}$^{\dagger}$ inv proof obligations & 62 & 52 & 16\% & 241 & 145 & 40\% & 249 & 145 & 42\%\\
    \msg{2b}$^{\dagger}$ inv proof obligations & 9 & 9 & 0\% & 77 & 44 & 43\% & 77 & 44 & 43\%\\
    \hline
    TLAPS check time (seconds) & 42 & 31 & 26\% & $>$191\tlafail{}$\!\!\!$ & 80 & $>$58\% & $>$208\tlafail{}$\!\!\!$ & 90 & $>$57\%\\
    \hline
\end{tabular}

\caption{Summary of results. Lam is for Lamport et al.~\capcite{basicpaxos2014}, Cha is from Chand et al.~\capcite{chand2016formal}, Us is ours
 in this paper, and Decr is percentage of decrease by Us from Lam and Cha.\\
 Specification and proof sizes are measured in lines excluding comments and empty lines.\\
 \cntapp{} indicates a number for the specification in Appendix~\ref{appendix:spec} and proof in Appendix~\ref{appendix:proof}, after removing unnecessary line breaks from default latex generated by \tlaplus{} Tools.\\
 \cnterr{} indicates a number; correcting a count oversight in~\cite{chand2018simpler}\\
 \textdagger~\msg{1b} invariants are (I10)--(I12), \msg{2a} invariants are (I13) and (I14), and \msg{2b} invariants are (I9) and (I15) for Basic Paxos in Figure~\capref{fig-hist-inv}, and corresponding invariants for Multi-Paxos and Multi-Paxos with Preemption in~\capcite{chand2016formal}.\\
 An obligation is a condition that TLAPS checks.\\
 Check time is taken on an Intel i7-4720HQ 2.6 GHz CPU with 16 GB of memory, running 64-bit Ubuntu 18.04.3 LTS and TLAPS 1.6.\\
 \tlafail{} indicates that TLAPS 1.5.4 failed to check and gave up after that number of seconds.\\
}
 \label{tab-sum-bp}
\end{table*}

\mysec{Related work and conclusion}
\label{sec-related}

\mypar{History variables} History variables have been at the center of much debate since they were introduced in the early 1970s~\cite{clint1973program,clarke1980proving,clint1981use}. Owicki and Gries~\cite{owicki1976axiomatic} use them in an effort to prove properties of parallel programs, criticized by Lamport in his writings~\cite{lamport2017writings}. Contrary to ours, their history variables are auxiliary variables introduced for the sole purpose of simpler proofs. Our history variables are $sent$ and $received$, whose contents are actually processed in all distributed system implementations. 

Recently, Lamport and Merz~\cite{2017arXiv170305121L} present rules to add history variables, among other auxiliary variables, to a low-level specification so that a refinement mapping from 
a high-level specification can be established. The idea is to prove invariants in the high-level specification that serves as an abstraction of the low-level specification. In contrast, we focus on high-level specifications because our target executable language is DistAlgo, and efficient lower-level implementations can be generated systematically from high-level code.

\mypar{Specification and verification} A number of systems~\cite{sergeyprogramming,druagoi2016psync,chaudhuri2010tla+}, models~\cite{wilcox2015verdi,charron2009heard,lynch1987hierarchical}, and methods~\cite{padon2017paxos,hawblitzel2015ironfleet,kufner2012formal,abadi1991existence} have been developed in the past to specify distributed algorithms and mechanically check proofs of safety and liveness properties of the algorithms. This work is orthogonal to them in the sense that the idea of maintaining only message history variables can be incorporated in their specifications as well. 

Closer to our work in terms of the specification is the work by Padon et al.~\cite{padon2017paxos}, which does not define any variable and instead defines predicate relations which would correspond to manipulations of our history variables. For example, $Send([type |-> \msgtype{1a}, bal |-> b])$ is denoted by $start\_round\_msg(b)$. Instead of using \tlaplus{}, they specify Paxos in first-order logic to later exploit benefits of Effectively Propositional Logic, such as satisfiability being decidable in it.

In contrast, we present a method to specify distributed algorithms using history variables, implementable in high-level executable languages like DistAlgo, and then show (i) how such specifications require fewer invariants for proofs and (ii) how many important invariants can be systematically derived. 

\mypar{Conclusion}
We have shown that using message history variables can lead to simpler specifications and easier proofs of challenging distributed algorithms.
Future work includes applying our method in specification and proofs of other complex distributed algorithms, and extending our method for proving liveness properties.

\providecommand{\grantsponsor}[3]{#2}
\providecommand{\grantnum}[2]{#2}

\mypar{Acknowledgements}
We thank Stephan Merz for his helpful comments on the proofs and explanations of TLAPS.
We thank anonymous reviewers for their helpful comments on this work.
This work was supported in part by 
\grantsponsor{sponsorid}{National Science Foundation}{https://www.nsf.gov} 
  grants \grantnum{sponsorid}{CCF-1248184}, 
  and \grantnum{sponsorid}{CCF-1414078}, 
  and \grantsponsor{sponsorid}{Office of Naval Research}{https://www.onr.navy.mil} grant \grantnum{sponsorid}{N000141512208}. 
  Any opinions, findings, and conclusions or recommendations expressed in
  this material are those of the authors and do not necessarily reflect
  the views of these agencies.

\bibliographystyle{ACM-Reference-Format}
\bibliography{mybib}{}


\begin{thebibliography}{45}


\ifx \showCODEN    \undefined \def \showCODEN     #1{\unskip}     \fi
\ifx \showDOI      \undefined \def \showDOI       #1{#1}\fi
\ifx \showISBNx    \undefined \def \showISBNx     #1{\unskip}     \fi
\ifx \showISBNxiii \undefined \def \showISBNxiii  #1{\unskip}     \fi
\ifx \showISSN     \undefined \def \showISSN      #1{\unskip}     \fi
\ifx \showLCCN     \undefined \def \showLCCN      #1{\unskip}     \fi
\ifx \shownote     \undefined \def \shownote      #1{#1}          \fi
\ifx \showarticletitle \undefined \def \showarticletitle #1{#1}   \fi
\ifx \showURL      \undefined \def \showURL       {\relax}        \fi
\providecommand\bibfield[2]{#2}
\providecommand\bibinfo[2]{#2}
\providecommand\natexlab[1]{#1}
\providecommand\showeprint[2][]{arXiv:#2}

\bibitem[\protect\citeauthoryear{Abadi and Lamport}{Abadi and Lamport}{1991}]%
        {abadi1991existence}
\bibfield{author}{\bibinfo{person}{Mart\'{\i}n Abadi} {and}
  \bibinfo{person}{Leslie Lamport}.} \bibinfo{year}{1991}\natexlab{}.
\newblock \showarticletitle{The Existence of Refinement Mappings}.
\newblock \bibinfo{journal}{\emph{Theor. Comput. Sci.}} \bibinfo{volume}{82},
  \bibinfo{number}{2} (\bibinfo{date}{May} \bibinfo{year}{1991}),
  \bibinfo{pages}{253--284}.
\newblock
\showISSN{0304-3975}
\urldef\tempurl%
\url{https://doi.org/10.1016/0304-3975(91)90224-P}
\showDOI{\tempurl}


\bibitem[\protect\citeauthoryear{Center}{Center}{2017}]%
        {tlaps}
\bibfield{author}{\bibinfo{person}{Microsoft Research-Inria~Joint Center}.}
  \bibinfo{year}{2017}\natexlab{}.
\newblock \bibinfo{title}{{TLA\textsuperscript{+} Proof System (TLAPS)}}.
\newblock
\newblock
\urldef\tempurl%
\url{http://tla.msr-inria.inria.fr/tlaps}
\showURL{%
Retrieved September 9, 2019 from \tempurl}


\bibitem[\protect\citeauthoryear{Chand and Liu}{Chand and Liu}{2018}]%
        {chand2018simpler}
\bibfield{author}{\bibinfo{person}{Saksham Chand} {and}
  \bibinfo{person}{Yanhong~A. Liu}.} \bibinfo{year}{2018}\natexlab{}.
\newblock \showarticletitle{Simpler Specifications and Easier Proofs of
  Distributed Algorithms Using History Variables}. In
  \bibinfo{booktitle}{\emph{NASA Formal Methods}} \emph{(\bibinfo{series}{NFM
  '18})}. \bibinfo{publisher}{Springer International Publishing},
  \bibinfo{address}{Cham, Switzerland}, \bibinfo{pages}{70--86}.
\newblock
\showISBNx{978-3-319-77935-5}
\urldef\tempurl%
\url{https://doi.org/10.1007/978-3-319-77935-5_5}
\showDOI{\tempurl}


\bibitem[\protect\citeauthoryear{Chand, Liu, and Stoller}{Chand
  et~al\mbox{.}}{2016}]%
        {chand2016formal}
\bibfield{author}{\bibinfo{person}{Saksham Chand}, \bibinfo{person}{Yanhong~A.
  Liu}, {and} \bibinfo{person}{Scott~D. Stoller}.}
  \bibinfo{year}{2016}\natexlab{}.
\newblock \showarticletitle{Formal Verification of Multi-Paxos for Distributed
  Consensus}. In \bibinfo{booktitle}{\emph{FM 2016: Formal Methods}}
  \emph{(\bibinfo{series}{FM '16})}. \bibinfo{publisher}{Springer International
  Publishing}, \bibinfo{address}{Cham, Switzerland}, \bibinfo{pages}{119--136}.
\newblock
\showISBNx{978-3-319-48989-6}
\urldef\tempurl%
\url{https://doi.org/10.1007/978-3-319-48989-6_8}
\showDOI{\tempurl}


\bibitem[\protect\citeauthoryear{Charron-Bost and Schiper}{Charron-Bost and
  Schiper}{2009}]%
        {charron2009heard}
\bibfield{author}{\bibinfo{person}{Bernadette Charron-Bost} {and}
  \bibinfo{person}{Andr{\'e} Schiper}.} \bibinfo{year}{2009}\natexlab{}.
\newblock \showarticletitle{The Heard-Of Model: Computing in Distributed
  Systems with Benign Faults}.
\newblock \bibinfo{journal}{\emph{Distrib. Comput.}} \bibinfo{volume}{22},
  \bibinfo{number}{1} (\bibinfo{date}{April} \bibinfo{year}{2009}),
  \bibinfo{pages}{49--71}.
\newblock
\showISSN{0178-2770}
\urldef\tempurl%
\url{https://doi.org/10.1007/s00446-009-0084-6}
\showDOI{\tempurl}


\bibitem[\protect\citeauthoryear{Chaudhuri, Doligez, Lamport, and
  Merz}{Chaudhuri et~al\mbox{.}}{2008}]%
        {chaudhuri2008tlaps}
\bibfield{author}{\bibinfo{person}{Kaustuv Chaudhuri}, \bibinfo{person}{Damien
  Doligez}, \bibinfo{person}{Leslie Lamport}, {and} \bibinfo{person}{Stephan
  Merz}.} \bibinfo{year}{2008}\natexlab{}.
\newblock \showarticletitle{A TLA+ Proof System}. In
  \bibinfo{booktitle}{\emph{Proceedings of the LPAR Workshops, CEUR Workshop}},
  Vol.~\bibinfo{volume}{418}. \bibinfo{publisher}{CEUR-WS.org},
  \bibinfo{pages}{17--37}.
\newblock
\urldef\tempurl%
\url{http://ceur-ws.org/Vol-418/paper2.pdf}
\showURL{%
\tempurl}


\bibitem[\protect\citeauthoryear{Chaudhuri, Doligez, Lamport, and
  Merz}{Chaudhuri et~al\mbox{.}}{2010}]%
        {chaudhuri2010tla+}
\bibfield{author}{\bibinfo{person}{Kaustuv Chaudhuri}, \bibinfo{person}{Damien
  Doligez}, \bibinfo{person}{Leslie Lamport}, {and} \bibinfo{person}{Stephan
  Merz}.} \bibinfo{year}{2010}\natexlab{}.
\newblock \showarticletitle{The TLA+ Proof System: Building a Heterogeneous
  Verification Platform}. In \bibinfo{booktitle}{\emph{Theoretical Aspects of
  Computing -- ICTAC 2010}} \emph{(\bibinfo{series}{ICTAC '10})}.
  \bibinfo{publisher}{Springer Berlin Heidelberg}, \bibinfo{address}{Berlin,
  Heidelberg}, \bibinfo{pages}{44--44}.
\newblock
\showISBNx{978-3-642-14808-8}
\urldef\tempurl%
\url{https://doi.org/10.1007/978-3-642-14808-8_3}
\showDOI{\tempurl}


\bibitem[\protect\citeauthoryear{Clarke}{Clarke}{1980}]%
        {clarke1980proving}
\bibfield{author}{\bibinfo{person}{Edmund~Melson Clarke}.}
  \bibinfo{year}{1980}\natexlab{}.
\newblock \showarticletitle{Proving Correctness of Coroutines Without History
  Variables}.
\newblock \bibinfo{journal}{\emph{Acta Inf.}} \bibinfo{volume}{13},
  \bibinfo{number}{2} (\bibinfo{date}{Feb.} \bibinfo{year}{1980}),
  \bibinfo{pages}{169--188}.
\newblock
\showISSN{0001-5903}
\urldef\tempurl%
\url{https://doi.org/10.1007/BF00263992}
\showDOI{\tempurl}


\bibitem[\protect\citeauthoryear{Clint}{Clint}{1973}]%
        {clint1973program}
\bibfield{author}{\bibinfo{person}{M. Clint}.} \bibinfo{year}{1973}\natexlab{}.
\newblock \showarticletitle{Program Proving: Coroutines}.
\newblock \bibinfo{journal}{\emph{Acta Inf.}} \bibinfo{volume}{2},
  \bibinfo{number}{1} (\bibinfo{date}{March} \bibinfo{year}{1973}),
  \bibinfo{pages}{50--63}.
\newblock
\showISSN{0001-5903}
\urldef\tempurl%
\url{https://doi.org/10.1007/BF00571463}
\showDOI{\tempurl}


\bibitem[\protect\citeauthoryear{Clint}{Clint}{1981}]%
        {clint1981use}
\bibfield{author}{\bibinfo{person}{Maurice Clint}.}
  \bibinfo{year}{1981}\natexlab{}.
\newblock \showarticletitle{On the Use of History Variables}.
\newblock \bibinfo{journal}{\emph{Acta Inf.}} \bibinfo{volume}{16},
  \bibinfo{number}{1} (\bibinfo{date}{Aug.} \bibinfo{year}{1981}),
  \bibinfo{pages}{15--30}.
\newblock
\showISSN{0001-5903}
\urldef\tempurl%
\url{https://doi.org/10.1007/BF00289587}
\showDOI{\tempurl}


\bibitem[\protect\citeauthoryear{Cousineau, Doligez, Lamport, Merz, Ricketts,
  and Vanzetto}{Cousineau et~al\mbox{.}}{2012}]%
        {cousineau2012tlaproofs}
\bibfield{author}{\bibinfo{person}{Denis Cousineau}, \bibinfo{person}{Damien
  Doligez}, \bibinfo{person}{Leslie Lamport}, \bibinfo{person}{Stephan Merz},
  \bibinfo{person}{Daniel Ricketts}, {and} \bibinfo{person}{Hern{\'a}n
  Vanzetto}.} \bibinfo{year}{2012}\natexlab{}.
\newblock \showarticletitle{TLA+ Proofs}. In \bibinfo{booktitle}{\emph{FM 2012:
  Formal Methods}}. \bibinfo{publisher}{Springer Berlin Heidelberg},
  \bibinfo{address}{Berlin, Heidelberg}, \bibinfo{pages}{147--154}.
\newblock
\showISBNx{978-3-642-32759-9}
\urldef\tempurl%
\url{https://doi.org/10.1007/978-3-642-32759-9_14}
\showDOI{\tempurl}


\bibitem[\protect\citeauthoryear{Dr\u{a}goi, Henzinger, and
  Zufferey}{Dr\u{a}goi et~al\mbox{.}}{2016}]%
        {druagoi2016psync}
\bibfield{author}{\bibinfo{person}{Cezara Dr\u{a}goi},
  \bibinfo{person}{Thomas~A. Henzinger}, {and} \bibinfo{person}{Damien
  Zufferey}.} \bibinfo{year}{2016}\natexlab{}.
\newblock \showarticletitle{PSync: A Partially Synchronous Language for
  Fault-tolerant Distributed Algorithms}.
\newblock \bibinfo{journal}{\emph{SIGPLAN Not.}} \bibinfo{volume}{51},
  \bibinfo{number}{1} (\bibinfo{date}{Jan.} \bibinfo{year}{2016}),
  \bibinfo{pages}{400--415}.
\newblock
\showISSN{0362-1340}
\urldef\tempurl%
\url{https://doi.org/10.1145/2914770.2837650}
\showDOI{\tempurl}


\bibitem[\protect\citeauthoryear{Gerla, Lee, Pau, and Lee}{Gerla
  et~al\mbox{.}}{2014}]%
        {gerla2014internet}
\bibfield{author}{\bibinfo{person}{Mario Gerla}, \bibinfo{person}{Eun-Kyu Lee},
  \bibinfo{person}{Giovanni Pau}, {and} \bibinfo{person}{Uichin Lee}.}
  \bibinfo{year}{2014}\natexlab{}.
\newblock \showarticletitle{Internet of vehicles: From intelligent grid to
  autonomous cars and vehicular clouds}. In \bibinfo{booktitle}{\emph{2014 IEEE
  World Forum on Internet of Things (WF-IoT)}} \emph{(\bibinfo{series}{WF-IoT
  '14})}. \bibinfo{publisher}{IEEE Press}, \bibinfo{address}{Piscataway, NJ,
  USA}, \bibinfo{pages}{241--246}.
\newblock
\urldef\tempurl%
\url{https://doi.org/10.1109/WF-IoT.2014.6803166}
\showDOI{\tempurl}


\bibitem[\protect\citeauthoryear{Gorbovitski}{Gorbovitski}{2011}]%
        {Gor11thesis}
\bibfield{author}{\bibinfo{person}{Michael Gorbovitski}.}
  \bibinfo{year}{2011}\natexlab{}.
\newblock \emph{\bibinfo{title}{A system for invariant-driven
  transformations}}.
\newblock \bibinfo{thesistype}{Ph.D. Dissertation}. \bibinfo{school}{Stony
  Brook University}, \bibinfo{address}{Stony Brook, NY, USA}.
\newblock Advisor(s) Liu, Yanhong A.
\newblock


\bibitem[\protect\citeauthoryear{Hawblitzel, Howell, Kapritsos, Lorch, Parno,
  Roberts, Setty, and Zill}{Hawblitzel et~al\mbox{.}}{2015}]%
        {hawblitzel2015ironfleet}
\bibfield{author}{\bibinfo{person}{Chris Hawblitzel}, \bibinfo{person}{Jon
  Howell}, \bibinfo{person}{Manos Kapritsos}, \bibinfo{person}{Jacob~R. Lorch},
  \bibinfo{person}{Bryan Parno}, \bibinfo{person}{Michael~L. Roberts},
  \bibinfo{person}{Srinath Setty}, {and} \bibinfo{person}{Brian Zill}.}
  \bibinfo{year}{2015}\natexlab{}.
\newblock \showarticletitle{IronFleet: Proving Practical Distributed Systems
  Correct}. In \bibinfo{booktitle}{\emph{Proceedings of the 25th Symposium on
  Operating Systems Principles}} \emph{(\bibinfo{series}{SOSP '15})}.
  \bibinfo{publisher}{ACM}, \bibinfo{address}{New York, NY, USA},
  \bibinfo{pages}{1--17}.
\newblock
\showISBNx{978-1-4503-3834-9}
\urldef\tempurl%
\url{https://doi.org/10.1145/2815400.2815428}
\showDOI{\tempurl}


\bibitem[\protect\citeauthoryear{K\"{u}fner, Nestmann, and Rickmann}{K\"{u}fner
  et~al\mbox{.}}{2012}]%
        {kufner2012formal}
\bibfield{author}{\bibinfo{person}{Philipp K\"{u}fner}, \bibinfo{person}{Uwe
  Nestmann}, {and} \bibinfo{person}{Christina Rickmann}.}
  \bibinfo{year}{2012}\natexlab{}.
\newblock \showarticletitle{Formal Verification of Distributed Algorithms: From
  Pseudo Code to Checked Proofs}. In \bibinfo{booktitle}{\emph{Proceedings of
  the 7th IFIP TC 1/WG 202 International Conference on Theoretical Computer
  Science}} \emph{(\bibinfo{series}{TCS '12})}.
  \bibinfo{publisher}{Springer-Verlag}, \bibinfo{address}{Berlin, Heidelberg},
  \bibinfo{pages}{209--224}.
\newblock
\showISBNx{978-3-642-33474-0}
\urldef\tempurl%
\url{https://doi.org/10.1007/978-3-642-33475-7_15}
\showDOI{\tempurl}


\bibitem[\protect\citeauthoryear{Lamport}{Lamport}{1978a}]%
        {lamport1978implementation}
\bibfield{author}{\bibinfo{person}{Leslie Lamport}.}
  \bibinfo{year}{1978}\natexlab{a}.
\newblock \showarticletitle{The implementation of reliable distributed
  multiprocess systems}.
\newblock \bibinfo{journal}{\emph{Computer Networks (1976)}}
  \bibinfo{volume}{2}, \bibinfo{number}{2} (\bibinfo{date}{May}
  \bibinfo{year}{1978}), \bibinfo{pages}{95--114}.
\newblock
\showISSN{0376-5075}
\urldef\tempurl%
\url{https://doi.org/10.1016/0376-5075(78)90045-4}
\showDOI{\tempurl}


\bibitem[\protect\citeauthoryear{Lamport}{Lamport}{1978b}]%
        {lamport1978time}
\bibfield{author}{\bibinfo{person}{Leslie Lamport}.}
  \bibinfo{year}{1978}\natexlab{b}.
\newblock \showarticletitle{Time, Clocks, and the Ordering of Events in a
  Distributed System}.
\newblock \bibinfo{journal}{\emph{Commun. ACM}} \bibinfo{volume}{21},
  \bibinfo{number}{7} (\bibinfo{date}{July} \bibinfo{year}{1978}),
  \bibinfo{pages}{558--565}.
\newblock
\showISSN{0001-0782}
\urldef\tempurl%
\url{https://doi.org/10.1145/359545.359563}
\showDOI{\tempurl}


\bibitem[\protect\citeauthoryear{Lamport}{Lamport}{1994}]%
        {lamport1994temporal}
\bibfield{author}{\bibinfo{person}{Leslie Lamport}.}
  \bibinfo{year}{1994}\natexlab{}.
\newblock \showarticletitle{The Temporal Logic of Actions}.
\newblock \bibinfo{journal}{\emph{ACM Trans. Program. Lang. Syst.}}
  \bibinfo{volume}{16}, \bibinfo{number}{3} (\bibinfo{date}{May}
  \bibinfo{year}{1994}), \bibinfo{pages}{872--923}.
\newblock
\showISSN{0164-0925}
\urldef\tempurl%
\url{https://doi.org/10.1145/177492.177726}
\showDOI{\tempurl}


\bibitem[\protect\citeauthoryear{Lamport}{Lamport}{2001}]%
        {lamport2001paxos}
\bibfield{author}{\bibinfo{person}{Leslie Lamport}.}
  \bibinfo{year}{2001}\natexlab{}.
\newblock \showarticletitle{Paxos made simple}.
\newblock \bibinfo{journal}{\emph{ACM SIGACT News}} \bibinfo{volume}{32},
  \bibinfo{number}{4} (\bibinfo{date}{Dec.} \bibinfo{year}{2001}),
  \bibinfo{pages}{51--58}.
\newblock
\showISSN{0163-5700}
\urldef\tempurl%
\url{https://doi.org/10.1145/568425.568433}
\showDOI{\tempurl}


\bibitem[\protect\citeauthoryear{Lamport}{Lamport}{2002}]%
        {lamport2002specifying}
\bibfield{author}{\bibinfo{person}{Leslie Lamport}.}
  \bibinfo{year}{2002}\natexlab{}.
\newblock \bibinfo{booktitle}{\emph{Specifying Systems: The TLA+ Language and
  Tools for Hardware and Software Engineers}}.
\newblock \bibinfo{publisher}{Addison-Wesley Longman Publishing Co., Inc.},
  \bibinfo{address}{Boston, MA, USA}.
\newblock
\showISBNx{032114306X}


\bibitem[\protect\citeauthoryear{Lamport}{Lamport}{2012}]%
        {lamport2012write}
\bibfield{author}{\bibinfo{person}{Leslie Lamport}.}
  \bibinfo{year}{2012}\natexlab{}.
\newblock \showarticletitle{How to write a 21st century proof}.
\newblock \bibinfo{journal}{\emph{Journal of Fixed Point Theory and
  Applications}} \bibinfo{volume}{11}, \bibinfo{number}{1}
  (\bibinfo{date}{March} \bibinfo{year}{2012}), \bibinfo{pages}{43--63}.
\newblock
\showISSN{1661-7746}
\urldef\tempurl%
\url{https://doi.org/10.1007/s11784-012-0071-6}
\showDOI{\tempurl}


\bibitem[\protect\citeauthoryear{Lamport}{Lamport}{2019}]%
        {lamport2017writings}
\bibfield{author}{\bibinfo{person}{Leslie Lamport}.}
  \bibinfo{year}{2019}\natexlab{}.
\newblock \bibinfo{title}{My Writings :: Proving the Correctness of
  Multiprocess Programs}.
\newblock
\newblock
\urldef\tempurl%
\url{https://lamport.azurewebsites.net/pubs/pubs.html#proving}
\showURL{%
Retrieved November 1, 2019 from \tempurl}


\bibitem[\protect\citeauthoryear{{Lamport} and {Merz}}{{Lamport} and
  {Merz}}{2017}]%
        {2017arXiv170305121L}
\bibfield{author}{\bibinfo{person}{Leslie {Lamport}} {and}
  \bibinfo{person}{Stephan {Merz}}.} \bibinfo{year}{2017}\natexlab{}.
\newblock \showarticletitle{Auxiliary Variables in TLA+}.
\newblock  (\bibinfo{year}{2017}).
\newblock
\showeprint{1703.05121}
\urldef\tempurl%
\url{https://arxiv.org/abs/1703.05121}
\showURL{%
\tempurl}


\bibitem[\protect\citeauthoryear{Lamport, Merz, and Doligez}{Lamport
  et~al\mbox{.}}{2014}]%
        {basicpaxos2014}
\bibfield{author}{\bibinfo{person}{Leslie Lamport}, \bibinfo{person}{Stephan
  Merz}, {and} \bibinfo{person}{Damien Doligez}.}
  \bibinfo{year}{2014}\natexlab{}.
\newblock \bibinfo{title}{Paxos.tla}.
\newblock
\newblock
\urldef\tempurl%
\url{https://github.com/tlaplus/v1-tlapm/blob/master/examples/paxos/Paxos.tla}
\showURL{%
Retrieved February 6, 2018 from \tempurl}


\bibitem[\protect\citeauthoryear{Liu}{Liu}{2013}]%
        {Liu13book}
\bibfield{author}{\bibinfo{person}{Yanhong~Annie Liu}.}
  \bibinfo{year}{2013}\natexlab{}.
\newblock \bibinfo{booktitle}{\emph{{Systematic Program Design: From Clarity To
  Efficiency}}}.
\newblock \bibinfo{publisher}{Cambridge University Press}.
\newblock
\showISBNx{9781107610798}


\bibitem[\protect\citeauthoryear{Liu, Brandvein, Stoller, and Lin}{Liu
  et~al\mbox{.}}{2016}]%
        {Liu+16IncOQ-PPDP}
\bibfield{author}{\bibinfo{person}{Yanhong~A. Liu}, \bibinfo{person}{Jon
  Brandvein}, \bibinfo{person}{Scott~D. Stoller}, {and} \bibinfo{person}{Bo
  Lin}.} \bibinfo{year}{2016}\natexlab{}.
\newblock \showarticletitle{Demand-driven Incremental Object Queries}. In
  \bibinfo{booktitle}{\emph{Proceedings of the 18th International Symposium on
  Principles and Practice of Declarative Programming}}
  \emph{(\bibinfo{series}{PPDP '16})}. \bibinfo{publisher}{ACM},
  \bibinfo{address}{New York, NY, USA}, \bibinfo{pages}{228--241}.
\newblock
\showISBNx{978-1-4503-4148-6}
\urldef\tempurl%
\url{https://doi.org/10.1145/2967973.2968610}
\showDOI{\tempurl}


\bibitem[\protect\citeauthoryear{Liu, Stoller, and Lin}{Liu
  et~al\mbox{.}}{2017}]%
        {liu2017clarity}
\bibfield{author}{\bibinfo{person}{Yanhong~A. Liu}, \bibinfo{person}{Scott~D.
  Stoller}, {and} \bibinfo{person}{Bo Lin}.} \bibinfo{year}{2017}\natexlab{}.
\newblock \showarticletitle{From Clarity to Efficiency for Distributed
  Algorithms}.
\newblock \bibinfo{journal}{\emph{ACM Trans. Program. Lang. Syst.}}
  \bibinfo{volume}{39}, \bibinfo{number}{3}, Article \bibinfo{articleno}{12}
  (\bibinfo{date}{May} \bibinfo{year}{2017}), \bibinfo{numpages}{41}~pages.
\newblock
\showISSN{0164-0925}
\urldef\tempurl%
\url{https://doi.org/10.1145/2994595}
\showDOI{\tempurl}


\bibitem[\protect\citeauthoryear{Liu, Stoller, Lin, and Gorbovitski}{Liu
  et~al\mbox{.}}{2012}]%
        {Liu+12DistPL-OOPSLA}
\bibfield{author}{\bibinfo{person}{Yanhong~A. Liu}, \bibinfo{person}{Scott~D.
  Stoller}, \bibinfo{person}{Bo Lin}, {and} \bibinfo{person}{Michael
  Gorbovitski}.} \bibinfo{year}{2012}\natexlab{}.
\newblock \showarticletitle{From Clarity to Efficiency for Distributed
  Algorithms}. In \bibinfo{booktitle}{\emph{Proceedings of the ACM
  International Conference on Object Oriented Programming Systems Languages and
  Applications}} \emph{(\bibinfo{series}{OOPSLA '12})}.
  \bibinfo{publisher}{ACM}, \bibinfo{address}{New York, NY, USA},
  \bibinfo{pages}{395--410}.
\newblock
\showISBNx{978-1-4503-1561-6}
\urldef\tempurl%
\url{https://doi.org/10.1145/2384616.2384645}
\showDOI{\tempurl}


\bibitem[\protect\citeauthoryear{Lynch and Tuttle}{Lynch and Tuttle}{1987}]%
        {lynch1987hierarchical}
\bibfield{author}{\bibinfo{person}{Nancy~A. Lynch} {and}
  \bibinfo{person}{Mark~R. Tuttle}.} \bibinfo{year}{1987}\natexlab{}.
\newblock \showarticletitle{Hierarchical Correctness Proofs for Distributed
  Algorithms}. In \bibinfo{booktitle}{\emph{Proceedings of the Sixth Annual ACM
  Symposium on Principles of Distributed Computing}}
  \emph{(\bibinfo{series}{PODC '87})}. \bibinfo{publisher}{ACM},
  \bibinfo{address}{New York, NY, USA}, \bibinfo{pages}{137--151}.
\newblock
\showISBNx{0-89791-239-X}
\urldef\tempurl%
\url{https://doi.org/10.1145/41840.41852}
\showDOI{\tempurl}


\bibitem[\protect\citeauthoryear{Merz}{Merz}{2003}]%
        {merz2003logic}
\bibfield{author}{\bibinfo{person}{Stephan Merz}.}
  \bibinfo{year}{2003}\natexlab{}.
\newblock \showarticletitle{On the Logic of TLA+}.
\newblock \bibinfo{journal}{\emph{Computing and Informatics}}
  \bibinfo{volume}{22}, \bibinfo{number}{3-4} (\bibinfo{year}{2003}),
  \bibinfo{pages}{351--379}.
\newblock
\showISSN{2585-8807}
\urldef\tempurl%
\url{http://www.cai.sk/ojs/index.php/cai/article/view/460/367}
\showURL{%
\tempurl}


\bibitem[\protect\citeauthoryear{Merz}{Merz}{2008}]%
        {merz2008spec}
\bibfield{author}{\bibinfo{person}{Stephan Merz}.}
  \bibinfo{year}{2008}\natexlab{}.
\newblock \bibinfo{booktitle}{\emph{The Specification Language TLA+}}.
\newblock \bibinfo{publisher}{Springer Berlin Heidelberg},
  \bibinfo{address}{Berlin, Heidelberg}, \bibinfo{pages}{401--451}.
\newblock
\showISBNx{978-3-540-74107-7}
\urldef\tempurl%
\url{https://doi.org/10.1007/978-3-540-74107-7_8}
\showDOI{\tempurl}


\bibitem[\protect\citeauthoryear{Merz and Vanzetto}{Merz and Vanzetto}{2012a}]%
        {merz2012automatic}
\bibfield{author}{\bibinfo{person}{Stephan Merz} {and}
  \bibinfo{person}{Hern\'{a}n Vanzetto}.} \bibinfo{year}{2012}\natexlab{a}.
\newblock \showarticletitle{Automatic Verification of TLA+ Proof Obligations
  with SMT Solvers}. In \bibinfo{booktitle}{\emph{Proceedings of the 18th
  International Conference on Logic for Programming, Artificial Intelligence,
  and Reasoning}} \emph{(\bibinfo{series}{LPAR'12})}.
  \bibinfo{publisher}{Springer-Verlag}, \bibinfo{address}{Berlin, Heidelberg},
  \bibinfo{pages}{289--303}.
\newblock
\showISBNx{978-3-642-28716-9}
\urldef\tempurl%
\url{https://doi.org/10.1007/978-3-642-28717-6_23}
\showDOI{\tempurl}


\bibitem[\protect\citeauthoryear{Merz and Vanzetto}{Merz and Vanzetto}{2012b}]%
        {merz2012harnessing}
\bibfield{author}{\bibinfo{person}{Stephan Merz} {and}
  \bibinfo{person}{Hern{\'a}n Vanzetto}.} \bibinfo{year}{2012}\natexlab{b}.
\newblock \showarticletitle{Harnessing SMT Solvers for TLA+ Proofs}. In
  \bibinfo{booktitle}{\emph{Proceedings of the 12th International Workshop on
  Automated Verification of Critical Systems}} \emph{(\bibinfo{series}{AVoCS
  '12})}, Vol.~\bibinfo{volume}{53}. \bibinfo{publisher}{European Association
  of Software Science and Technology}, \bibinfo{pages}{1--15}.
\newblock
\showISSN{1863-2122}
\urldef\tempurl%
\url{https://doi.org/10.14279/tuj.eceasst.53.766}
\showDOI{\tempurl}


\bibitem[\protect\citeauthoryear{Owicki and Gries}{Owicki and Gries}{1976}]%
        {owicki1976axiomatic}
\bibfield{author}{\bibinfo{person}{Susan Owicki} {and} \bibinfo{person}{David
  Gries}.} \bibinfo{year}{1976}\natexlab{}.
\newblock \showarticletitle{An Axiomatic Proof Technique for Parallel Programs
  I}.
\newblock \bibinfo{journal}{\emph{Acta Inf.}} \bibinfo{volume}{6},
  \bibinfo{number}{4} (\bibinfo{date}{Dec.} \bibinfo{year}{1976}),
  \bibinfo{pages}{319--340}.
\newblock
\showISSN{0001-5903}
\urldef\tempurl%
\url{https://doi.org/10.1007/BF00268134}
\showDOI{\tempurl}


\bibitem[\protect\citeauthoryear{Padon, Losa, Sagiv, and Shoham}{Padon
  et~al\mbox{.}}{2017}]%
        {padon2017paxos}
\bibfield{author}{\bibinfo{person}{Oded Padon}, \bibinfo{person}{Giuliano
  Losa}, \bibinfo{person}{Mooly Sagiv}, {and} \bibinfo{person}{Sharon Shoham}.}
  \bibinfo{year}{2017}\natexlab{}.
\newblock \showarticletitle{Paxos Made EPR: Decidable Reasoning About
  Distributed Protocols}.
\newblock \bibinfo{journal}{\emph{Proc. ACM Program. Lang.}}
  \bibinfo{volume}{1}, \bibinfo{number}{OOPSLA}, Article
  \bibinfo{articleno}{108} (\bibinfo{date}{Oct.} \bibinfo{year}{2017}),
  \bibinfo{numpages}{31}~pages.
\newblock
\showISSN{2475-1421}
\urldef\tempurl%
\url{https://doi.org/10.1145/3140568}
\showDOI{\tempurl}


\bibitem[\protect\citeauthoryear{Paige and Koenig}{Paige and Koenig}{1982}]%
        {PaiKoe82}
\bibfield{author}{\bibinfo{person}{Robert Paige} {and} \bibinfo{person}{Shaye
  Koenig}.} \bibinfo{year}{1982}\natexlab{}.
\newblock \showarticletitle{Finite Differencing of Computable Expressions}.
\newblock \bibinfo{journal}{\emph{ACM Trans. Program. Lang. Syst.}}
  \bibinfo{volume}{4}, \bibinfo{number}{3} (\bibinfo{date}{July}
  \bibinfo{year}{1982}), \bibinfo{pages}{402--454}.
\newblock
\showISSN{0164-0925}
\urldef\tempurl%
\url{https://doi.org/10.1145/357172.357177}
\showDOI{\tempurl}


\bibitem[\protect\citeauthoryear{research INRIA Joint~Centre}{research INRIA
  Joint~Centre}{2014}]%
        {tlaps2019provers}
\bibfield{author}{\bibinfo{person}{Microsoft research INRIA Joint~Centre}.}
  \bibinfo{year}{2014}\natexlab{}.
\newblock \bibinfo{title}{TLA+ Proof System, Tactics}.
\newblock
\newblock
\urldef\tempurl%
\url{https://tla.msr-inria.inria.fr/tlaps/content/Documentation/Tutorial/Tactics.html}
\showURL{%
Retrieved November 1, 2019 from \tempurl}


\bibitem[\protect\citeauthoryear{Rothamel and Liu}{Rothamel and Liu}{2008}]%
        {RotLiu08OSQ-GPCE}
\bibfield{author}{\bibinfo{person}{Tom Rothamel} {and}
  \bibinfo{person}{Yanhong~A. Liu}.} \bibinfo{year}{2008}\natexlab{}.
\newblock \showarticletitle{Generating Incremental Implementations of
  Object-set Queries}. In \bibinfo{booktitle}{\emph{Proceedings of the 7th
  International Conference on Generative Programming and Component
  Engineering}} \emph{(\bibinfo{series}{GPCE '08})}. \bibinfo{publisher}{ACM},
  \bibinfo{address}{New York, NY, USA}, \bibinfo{pages}{55--66}.
\newblock
\showISBNx{978-1-60558-267-2}
\urldef\tempurl%
\url{https://doi.org/10.1145/1449913.1449923}
\showDOI{\tempurl}


\bibitem[\protect\citeauthoryear{Schilling}{Schilling}{2017}]%
        {schilling2017perspectives}
\bibfield{author}{\bibinfo{person}{Klaus Schilling}.}
  \bibinfo{year}{2017}\natexlab{}.
\newblock \showarticletitle{Perspectives for miniaturized, distributed,
  networked cooperating systems for space exploration}.
\newblock \bibinfo{journal}{\emph{Robotics and Autonomous Systems}}
  \bibinfo{volume}{90}, \bibinfo{number}{1} (\bibinfo{date}{April}
  \bibinfo{year}{2017}), \bibinfo{pages}{118--124}.
\newblock
\showISSN{0921-8890}
\urldef\tempurl%
\url{https://doi.org/10.1016/j.robot.2016.10.007}
\showDOI{\tempurl}
\newblock
\shownote{Special Issue on New Research Frontiers for Intelligent Autonomous
  Systems.}


\bibitem[\protect\citeauthoryear{Sergey, Wilcox, and Tatlock}{Sergey
  et~al\mbox{.}}{2017}]%
        {sergeyprogramming}
\bibfield{author}{\bibinfo{person}{Ilya Sergey}, \bibinfo{person}{James~R.
  Wilcox}, {and} \bibinfo{person}{Zachary Tatlock}.}
  \bibinfo{year}{2017}\natexlab{}.
\newblock \showarticletitle{Programming and Proving with Distributed
  Protocols}.
\newblock \bibinfo{journal}{\emph{Proc. ACM Program. Lang.}}
  \bibinfo{volume}{2}, \bibinfo{number}{POPL}, Article \bibinfo{articleno}{28}
  (\bibinfo{date}{Dec.} \bibinfo{year}{2017}), \bibinfo{numpages}{30}~pages.
\newblock
\showISSN{2475-1421}
\urldef\tempurl%
\url{https://doi.org/10.1145/3158116}
\showDOI{\tempurl}


\bibitem[\protect\citeauthoryear{Tschorsch and Scheuermann}{Tschorsch and
  Scheuermann}{2016}]%
        {tschorsch2016bitcoin}
\bibfield{author}{\bibinfo{person}{Florian Tschorsch} {and}
  \bibinfo{person}{Bj{\o}rn Scheuermann}.} \bibinfo{year}{2016}\natexlab{}.
\newblock \showarticletitle{Bitcoin and Beyond: A Technical Survey on
  Decentralized Digital Currencies}.
\newblock \bibinfo{journal}{\emph{IEEE Communications Surveys Tutorials}}
  \bibinfo{volume}{18}, \bibinfo{number}{3} (\bibinfo{date}{March}
  \bibinfo{year}{2016}), \bibinfo{pages}{2084--2123}.
\newblock
\showISSN{1553-877X}
\urldef\tempurl%
\url{https://doi.org/10.1109/COMST.2016.2535718}
\showDOI{\tempurl}


\bibitem[\protect\citeauthoryear{Van~Renesse and Altinbuken}{Van~Renesse and
  Altinbuken}{2015}]%
        {van2015paxos}
\bibfield{author}{\bibinfo{person}{Robbert Van~Renesse} {and}
  \bibinfo{person}{Deniz Altinbuken}.} \bibinfo{year}{2015}\natexlab{}.
\newblock \showarticletitle{Paxos Made Moderately Complex}.
\newblock \bibinfo{journal}{\emph{ACM Comput. Surv.}} \bibinfo{volume}{47},
  \bibinfo{number}{3}, Article \bibinfo{articleno}{42} (\bibinfo{date}{Feb.}
  \bibinfo{year}{2015}), \bibinfo{numpages}{36}~pages.
\newblock
\showISSN{0360-0300}
\urldef\tempurl%
\url{https://doi.org/10.1145/2673577}
\showDOI{\tempurl}


\bibitem[\protect\citeauthoryear{Wilcox, Woos, Panchekha, Tatlock, Wang, Ernst,
  and Anderson}{Wilcox et~al\mbox{.}}{2015}]%
        {wilcox2015verdi}
\bibfield{author}{\bibinfo{person}{James~R. Wilcox}, \bibinfo{person}{Doug
  Woos}, \bibinfo{person}{Pavel Panchekha}, \bibinfo{person}{Zachary Tatlock},
  \bibinfo{person}{Xi Wang}, \bibinfo{person}{Michael~D. Ernst}, {and}
  \bibinfo{person}{Thomas Anderson}.} \bibinfo{year}{2015}\natexlab{}.
\newblock \showarticletitle{Verdi: A Framework for Implementing and Formally
  Verifying Distributed Systems}.
\newblock \bibinfo{journal}{\emph{SIGPLAN Not.}} \bibinfo{volume}{50},
  \bibinfo{number}{6} (\bibinfo{date}{June} \bibinfo{year}{2015}),
  \bibinfo{pages}{357--368}.
\newblock
\showISSN{0362-1340}
\urldef\tempurl%
\url{https://doi.org/10.1145/2813885.2737958}
\showDOI{\tempurl}


\bibitem[\protect\citeauthoryear{Zave}{Zave}{2012}]%
        {zave2012using}
\bibfield{author}{\bibinfo{person}{Pamela Zave}.}
  \bibinfo{year}{2012}\natexlab{}.
\newblock \showarticletitle{Using Lightweight Modeling to Understand Chord}.
\newblock \bibinfo{journal}{\emph{SIGCOMM Comput. Commun. Rev.}}
  \bibinfo{volume}{42}, \bibinfo{number}{2} (\bibinfo{date}{March}
  \bibinfo{year}{2012}), \bibinfo{pages}{49--57}.
\newblock
\showISSN{0146-4833}
\urldef\tempurl%
\url{https://doi.org/10.1145/2185376.2185383}
\showDOI{\tempurl}


\end{thebibliography}

\appendix
\onecolumn
{
\setlength{\baselineskip}{1.98ex}
\input{tlatex.sty}
\section{\texorpdfstring{\tlaplus{}}{TLA+} specification of Basic Paxos}
\label{appendix:spec}
\tlatex
\@x{}\moduleLeftDash\@xx{ {\MODULE} PaxosSpec}\moduleRightDash\@xx{}%
\begin{lcom}{0}%
\begin{cpar}{0}{F}{F}{0}{0}{}%
This is a specification in TLA\textsuperscript{+} 
of \ensuremath{Basic Paxos}.
\end{cpar}%
\end{lcom}%
\@x{ {\EXTENDS} Integers ,\, TLAPS ,\, NaturalsInduction}%
\@x{ {\CONSTANTS} {\mathcal{A}} ,\, {\mathcal{Q}} ,\, {\mathcal{V}}}%
\@y{\@s{0.0}%
 Sets of acceptors, quorums of acceptors, and values to propose
}%
\@pvspace{8.0pt}%
\@x{ {\VARIABLES} sent}%
\@y{\@s{0.0}%
 Set of sent messages
}%
\@pvspace{8.0pt}%
 \@x{ {\ASSUME} QuorumAssumption \.{\defeq} {\mathcal{Q}} \.{\subseteq} {\SUBSET}
 {\mathcal{A}} \.{\land} \A\, Q1 ,\, Q2 \.{\in} {\mathcal{Q}} \.{:} Q1 \.{\cap} Q2
 \.{\neq} {\emptyset}}%
\@pvspace{8.0pt}%
\@x{ {\mathcal{B}} \.{\defeq} {\mathds{N}}}%
\@y{\@s{0.0}%
 Set of ballots
}%
\@x{ vars \.{\defeq} {\langle} sent {\rangle}}%
\@x{ Send ( m ) \.{\defeq} sent \.{'} \.{=} sent \.{\cup} \{ m \}}%
\@x{ None \.{\defeq} {\CHOOSE} v \.{:} v \.{\notin} {\mathcal{V}}}%
\@pvspace{8.0pt}%
\begin{lcom}{0}%
\begin{cpar}{0}{F}{F}{0}{0}{}%
 Phase \ensuremath{1a}: A \textsf{1a} message with ballot \ensuremath{b} is sent by some proposer (to all processes). 
\end{cpar}%
\end{lcom}%
 \@x{ Phase1a ( b ) \.{\defeq} Send ( [ type \.{\mapsto}\@w{1a} ,\, bal
 \.{\mapsto} b ] )}%
\@pvspace{8.0pt}%
\begin{lcom}{0}%
\begin{cpar}{0}{F}{F}{0}{0}{}%
 Phase \ensuremath{1b}: For an acceptor \ensuremath{a}, if there is
 a \textsf{1a} message \ensuremath{m} with
 ballot \ensuremath{m.bal} that is
 higher than the highest it has seen, 
 \ensuremath{a} sends a \textsf{1b} message with \ensuremath{m.bal} alongwith the highest-numbered pair it has voted for.
\end{cpar}%
\end{lcom}%
 \@x{ 2bs ( a ) \.{\defeq} \. \{ m \.{\in} sent
 \.{:} m . type \.{=}\@w{2b} \.{\land} m . acc \.{=} a \} \@s{2.0}}%
 \@x{ max\_prop ( a ) \.{\defeq} \.{\IF} 2bs ( a ) \.{=} {\emptyset} \.{\THEN} \{ [ bal \.{\mapsto} \.{-} 1 ,\, val \.{\mapsto}
 None ] \}}%
 \@x{\@s{114} \.{\ELSE} \{ m \.{\in} 2bs
 \.{:} \A\, m2 \.{\in} 2bs \.{:} m . bal \.{\geq} m2 . bal \}}%
\@pvspace{8.0pt}%
 \@x{ Phase1b ( a ) \.{\defeq} \E\, m \.{\in} sent ,\, r \.{\in} max\_prop (
 a ) \.{:}}%
\@x{\@s{10.0} \.{\land} m . type \.{=}\@w{1a}}%
 \@x{\@s{10.0} \.{\land} \A\, m2 \.{\in} sent \.{:} m2 . type \.{\in}
 \{\@w{1b} ,\,\@w{2b} \} \.{\land} m2 . acc \.{=} a \.{\implies} m . bal
 \.{>} m2 . bal}%
 \@x{\@s{10.0} \.{\land} Send ( [ type \.{\mapsto}\@w{1b} ,\, bal \.{\mapsto}
 m . bal ,\, maxVBal \.{\mapsto} r . bal ,\, maxVal \.{\mapsto} r . val ,\,
 acc \.{\mapsto} a ] )}%
\@pvspace{8.0pt}%
\begin{lcom}{0}%
\begin{cpar}{0}{F}{F}{0}{0}{}%
 Phase \ensuremath{2a}: If there is no \textsf{2a} message in \ensuremath{sent} with ballot \ensuremath{b}, and 
 a quorum of acceptors has sent a set \ensuremath{S} of \textsf{1b} messages with ballot \ensuremath{b},
 a proposer sends a
 \textsf{2a} message 
 with ballot \ensuremath{b} and value \ensuremath{v}, where \ensuremath{v} is the value with the highest ballot in \ensuremath{S}, or is any value in \ensuremath{\mathcal{V}} if no acceptor that responded in \ensuremath{S} has voted for anything.
\end{cpar}%
\end{lcom}%
\@x{ Phase2a ( b ) \.{\defeq}}%
 \@x{\@s{4.0} \.{\land} {\nexists}\, m \.{\in} sent \.{:} ( m . type
 \.{=}\@w{2a} ) \.{\land} ( m . bal \.{=} b )}%
 \@x{\@s{4.0} \.{\land} \E\, v \.{\in} {\mathcal{V}} ,\, Q \.{\in} {\mathcal{Q}} ,\, S
 \.{\subseteq} \{ m \.{\in} sent \.{:} m . type \.{=}\@w{1b} \.{\land} m .
 bal \.{=} b \} \.{:}}%
 \@x{\@s{16.89} \.{\land} \A\, a \.{\in} Q \.{:} \E\, m \.{\in} S \.{:} m .
 acc \.{=} a}%
 \@x{\@s{16.89} \.{\land} \.{\lor} \A\, m \.{\in} S \.{:} m . maxVBal \.{=}
 \.{-} 1}%
\@x{\@s{25.78} \.{\lor} \E\, b2 \.{\in} 0 \.{\dotdot} ( b \.{-} 1 ) \.{:}}%
\@x{\@s{38.66} \.{\land} \A\, m \.{\in} S \.{:} m . maxVBal \.{\leq} b2}%
 \@x{\@s{38.66} \.{\land} \E\, m \.{\in} S \.{:} m . maxVBal \.{=} b2
 \.{\land} m . maxVal \.{=} v}%
 \@x{\@s{16.89} \.{\land} Send ( [ type \.{\mapsto}\@w{2a} ,\, bal \.{\mapsto}
 b ,\, val \.{\mapsto} v ] )}%
\@pvspace{8.0pt}%
\begin{lcom}{0}%
\begin{cpar}{0}{F}{F}{0}{0}{}%
 Phase \ensuremath{2b}: For an acceptor \ensuremath{a}, if there is a \textsf{2a} message \ensuremath{m} with ballot \ensuremath{m.bal}
 that is higher than or equal to the highest it has seen, \ensuremath{a} 
 sends a \textsf{2b} message with \ensuremath{m.bal} and \ensuremath{m.val}.
\end{cpar}%
\end{lcom}%
\@x{ Phase2b ( a ) \.{\defeq} \E\, m \.{\in} sent \.{:}}%
\@x{\@s{8.0} \.{\land} m . type \.{=}\@w{2a}}%
 \@x{\@s{8.0} \.{\land} \A\, m2 \.{\in} sent \.{:} m2 . type
 \.{\in} \{\@w{1b} ,\,\@w{2b} \} \.{\land} m2 . acc \.{=} a \.{\implies} m .
 bal \.{\geq} m2 . bal}%
 \@x{\@s{8.0} \.{\land} Send ( [ type \.{\mapsto}\@w{2b} ,\, bal
 \.{\mapsto} m . bal ,\, val \.{\mapsto} m . val ,\, acc \.{\mapsto} a ] )}%
\@pvspace{8.0pt}%
\@x{ Init\@s{3.3} \.{\defeq} sent \.{=} {\emptyset}}%
 \@x{ Next \.{\defeq} \.{\lor} \E\, b\@s{0.51} \.{\in} {\mathcal{B}} \.{:} Phase1a (
 b ) \.{\lor} Phase2a ( b )}%
 \@x{\@s{32.5} \.{\lor} \E\, a \.{\in} {\mathcal{A}} \.{:} Phase1b ( a ) \.{\lor}
 Phase2b ( a )}%
\@x{ Spec\@s{1.17} \.{\defeq} Init \.{\land} {\Box} [ Next ]_{ vars}}%
\@x{}\bottombar\@xx{}%

\pagebreak
\section{Agreement property to prove for Basic Paxos and invariants used in proof}
\label{appendix:prop}
\tlatex
\@x{}\moduleLeftDash\@xx{ {\MODULE} PaxosProp}\moduleRightDash\@xx{}%
\begin{lcom}{0}%
\begin{cpar}{0}{F}{F}{0}{0}{}%
\ensuremath{VotedForIn(a, v, b)} means that acceptor \ensuremath{a} has sent some \textsf{2b} message \ensuremath{m} with \ensuremath{m.val} equal to \ensuremath{v} and \ensuremath{m.bal} equal to \ensuremath{b}. This specifies that acceptor \ensuremath{a} has voted the pair \ensuremath{\langle v, b\rangle}.
\end{cpar}%
\end{lcom}%
 \@x{ VotedForIn ( a ,\, v ,\, b ) \.{\defeq} \E\, m \.{\in} sent \.{:} m .
 type \.{=}\@w{2b} \.{\land} m . val\@s{2.0} \.{=} v \.{\land} m .
 bal\@s{2.0} \.{=} b \.{\land} m . acc\@s{2.0} \.{=} a}%
\@pvspace{8.0pt}%
\begin{lcom}{0}%
\begin{cpar}{0}{F}{F}{0}{0}{}%
\ensuremath{ChosenIn(v, b)} means that every acceptor in some quorum \ensuremath{Q} has voted the pair \ensuremath{\langle v, b\rangle}.
\end{cpar}%
\end{lcom}%
 \@x{ ChosenIn ( v ,\, b ) \.{\defeq} \E\, Q \.{\in} {\mathcal{Q}} \.{:} \A\, a
 \.{\in} Q \.{:} VotedForIn ( a ,\, v ,\, b )}%
\@pvspace{8.0pt}%
\begin{lcom}{0}%
\begin{cpar}{0}{F}{F}{0}{0}{}%
\ensuremath{Chosen(v)} means that for some ballot \ensuremath{b}, \ensuremath{ChosenIn(v, b)} holds.
\end{cpar}%
\end{lcom}%
 \@x{ Chosen ( v ) \.{\defeq} \E\, b \.{\in} {\mathcal{B}} \.{:} ChosenIn ( v ,\, b
 )}%
\@pvspace{8.0pt}%
\begin{lcom}{0}%
\begin{cpar}{0}{F}{F}{0}{0}{}%
 \ensuremath{WontVoteIn(a, b)} means that acceptor \ensuremath{a} has seen a higher ballot than \ensuremath{b}, and did not and will not vote 
 any value with ballot \ensuremath{b}.
\end{cpar}%
\end{lcom}%
 \@x{ WontVoteIn ( a ,\, b ) \.{\defeq} \.{\land} \A\, v\@s{2.61} \.{\in}
 {\mathcal{V}} \.{:} {\lnot} VotedForIn ( a ,\, v ,\, b )}%
 \@x{\@s{81.5} \.{\land} \E\, m \.{\in} sent \.{:} m . type \.{\in} \{\@w{1b}
 ,\,\@w{2b} \} \.{\land} m . acc \.{=} a \.{\land} m . bal \.{>} b}%
\@pvspace{8.0pt}%
\begin{lcom}{0}%
\begin{cpar}{0}{F}{F}{0}{0}{}%
 \ensuremath{SafeAt(v, b)} means that no value except perhaps \ensuremath{v} has been or will be chosen in any ballot lower
 than \ensuremath{b}.
\end{cpar}%
\end{lcom}%
 \@x{ SafeAt ( v ,\, b ) \.{\defeq} \A\, b2 \.{\in} 0 \.{\dotdot} ( b \.{-} 1
 ) \.{:} \E\, Q \.{\in} {\mathcal{Q}} \.{:} \A\, a \.{\in} Q \.{:} VotedForIn ( a
 ,\, v ,\, b2 ) \.{\lor} WontVoteIn ( a ,\, b2 )}%
\@pvspace{8.0pt}%
\begin{lcom}{0}%
\begin{cpar}{0}{F}{F}{0}{0}{}%
\ensuremath{Messages} defines the set of valid messages. 
\ensuremath{TypeOK} defines invariants for the types of the variables. 
\end{cpar}%
\end{lcom}%
 \@x{ Messages \.{\defeq} [ type \.{:} \{\@w{1a} \} ,\, bal\@s{0.29} \.{:}
 {\mathcal{B}} ] \.{\cup}}%
 \@x{\@s{50} [ type \.{:} \{\@w{1b} \} ,\, bal \.{:} {\mathcal{B}} ,\, maxVBal
 \.{:} {\mathcal{B}} \.{\cup} \{ \.{-} 1 \} ,\, maxVal \.{:} {\mathcal{V}} \.{\cup} \{
 None \} ,\, acc \.{:} {\mathcal{A}} ] \.{\cup}}%
 \@x{\@s{50} [ type \.{:} \{\@w{2a} \} ,\, bal\@s{0.29} \.{:} {\mathcal{B}} ,\,
 val \.{:} {\mathcal{V}} ] \.{\cup}}%
 \@x{\@s{50} [ type \.{:} \{\@w{2b} \} ,\, bal \.{:} {\mathcal{B}} ,\, val \.{:}
 {\mathcal{V}} ,\, acc \.{:} {\mathcal{A}} ]}%
\@x{ TypeOK \.{\defeq} sent \.{\subseteq} Messages}%
\@pvspace{8.0pt}%
\begin{lcom}{0}%
\begin{cpar}{0}{F}{F}{0}{0}{}%
\ensuremath{MsgInv} defines properties satisfied by the contents of messages, for \textsf{1b}, \textsf{2a}, and \textsf{2b} messages.
\end{cpar}%
\end{lcom}%
\@x{ MsgInv \.{\defeq} \A\, m \.{\in} sent \.{:}}%
 \@x{\@s{8.0} \.{\land} m . type \.{=}\@w{1b} \.{\implies}
 \.{\land} VotedForIn ( m . acc ,\, m . maxVal ,\, m . maxVBal ) \.{\lor} m .
 maxVBal \.{=} \.{-} 1}%
 \@x{\@s{83} \.{\land} \A\, b \.{\in} m . maxVBal \.{+} 1 \.{\dotdot} m .
 bal \.{-} 1 \.{:} {\nexists}\, v \.{\in} {\mathcal{V}} \.{:} VotedForIn ( m . acc
 ,\, v ,\, b )}%
 \@x{\@s{8.0} \.{\land} m . type \.{=}\@w{2a}\@s{0.29} \.{\implies}
 \.{\land} SafeAt ( m . val ,\, m . bal )}%
 \@x{\@s{83} \.{\land} \A\, m2 \.{\in} sent \.{:} ( m2 . type \.{=}\@w{2a}
 \.{\land} m2 . bal \.{=} m . bal ) \.{\implies} m2 \.{=} m}%
 \@x{\@s{8.0} \.{\land} m . type \.{=}\@w{2b} \.{\implies} \E\, m2
 \.{\in} sent \.{:} m2 . type \.{=}\@w{2a} \.{\land} m2 . bal\@s{2.0} \.{=} m
 . bal \.{\land} m2 . val\@s{2.0} \.{=} m . val}%
\@pvspace{8.0pt}%
\begin{lcom}{0}%
\begin{cpar}{0}{F}{F}{0}{0}{}%
\ensuremath{Inv} is the complete inductive invariant.
\end{cpar}%
\end{lcom}%
\@x{ Inv \.{\defeq} TypeOK \.{\land} MsgInv}%
\@pvspace{8.0pt}%
\begin{lcom}{0}%
\begin{cpar}{0}{F}{F}{0}{0}{}%
\ensuremath{Agree} states that at most one value can be chosen.
\end{cpar}%
\end{lcom}%
 \@x{ Agree \.{\defeq} \A\, v1 ,\, v2 \.{\in} {\mathcal{V}} \.{:} Chosen ( v1 )
 \.{\land} Chosen ( v2 ) \.{\implies} v1 \.{=} v2}%
\@x{}\bottombar\@xx{}%

\pagebreak
\section{TLAPS checked proof of Basic Paxos}
\label{appendix:proof}
\tlatex
\@x{}\moduleLeftDash\@xx{ {\MODULE} PaxosProof}\moduleRightDash\@xx{}%
\begin{lcom}{0}%
\begin{cpar}{0}{F}{F}{0}{0}{}%
The following two lemmas are straightforward consequences of the predicates defined in Appendix~\ref{appendix:prop}.
\end{cpar}%
\end{lcom}%
 \@x{ {\LEMMA} VotedInv \.{\defeq} MsgInv \.{\land} TypeOK \.{\implies} \A\, a
 \.{\in} {\mathcal{A}} ,\, v \.{\in} {\mathcal{V}} ,\, b \.{\in} {\mathcal{B}} \.{:}
 VotedForIn ( a ,\, v ,\, b ) \.{\implies} SafeAt ( v ,\, b )}%
\@x{ {\BY} {\!\!} {\DEF} VotedForIn ,\, MsgInv ,\, Messages ,\, TypeOK}%
\@pvspace{8.0pt}%
 \@x{ {\LEMMA} VotedOnce \.{\defeq} MsgInv \.{\implies}\@s{2.0} \A\, a1 ,\, a2
 \.{\in} {\mathcal{A}} ,\, v1 ,\, v2 \.{\in} {\mathcal{V}} ,\, b \.{\in} {\mathcal{B}} \.{:}}%
 \@x{\@s{32.44} VotedForIn ( a1 ,\, v1 ,\, b ) \.{\land} VotedForIn ( a2 ,\,
 v2 ,\, b ) \.{\implies} v1 \.{=} v2}%
\@x{ {\BY} {\!\!} {\DEF} MsgInv ,\, VotedForIn}%
\begin{lcom}{0}%
\begin{cpar}{0}{F}{F}{0}{0}{}%
Lemma \ensuremath{SafeAtStable} asserts that, if \ensuremath{SafeAt(v, b)} holds, it holds in the next state as well.
\end{cpar}%
\end{lcom}%
 \@x{ {\LEMMA} SafeAtStable \.{\defeq} Inv \.{\land} Next \.{\implies} \A\, v
 \.{\in} {\mathcal{V}} ,\, b \.{\in} {\mathcal{B}} \.{:} SafeAt ( v ,\, b )
 \.{\implies} SafeAt ( v ,\, b ) \.{'}}%
 \@x{\@pfstepnum{1}{} .\@s{2.0} {\SUFFICES} {\ASSUME} Inv ,\, Next ,\, {\NEW} v
 \.{\in} {\mathcal{V}} ,\, {\NEW} b \.{\in} {\mathcal{B}} ,\, SafeAt ( v ,\, b )}%
\@x{\@s{55.37} {\PROVE} SafeAt ( v ,\, b ) \.{'}\@s{2.0} {\OBVIOUS}}%
\@x{\@pfstepnum{1}{} .\@s{2.0} {\USE} {\!\!} {\DEF} Send ,\, Inv ,\, {\mathcal{B}}}%
 \@x{\@pfstepnum{1}{1.}\  {\ASSUME} {\NEW} b2 \.{\in} {\mathcal{B}} ,\, Phase1a ( b2
 )\@s{2.0} {\PROVE}\@s{-4.1} SafeAt ( v ,\, b ) \.{'}}%
 \@x{\@s{20.1} {\BY}\@pfstepnum{1}{1} ,\, \texttt{SMT} {\DEF} SafeAt ,\, Phase1a ,\,
 VotedForIn ,\, WontVoteIn}%
 \@x{\@pfstepnum{1}{2.}\  {\ASSUME} {\NEW} a \.{\in} {\mathcal{A}} ,\, Phase1b ( a
 )\@s{2.0} {\PROVE}\@s{-4.1} SafeAt ( v ,\, b ) \.{'}}%
 \@x{\@s{20.1} {\BY}\@pfstepnum{1}{2} ,\, QuorumAssumption ,\, \texttt{SMTT(60)}
 {\DEF} TypeOK ,\, SafeAt ,\, WontVoteIn ,\, VotedForIn ,\, Phase1b}%
 \@x{\@pfstepnum{1}{3.}\  {\ASSUME} {\NEW} b2 \.{\in} {\mathcal{B}} ,\, Phase2a ( b2
 ) \@s{2.0} {\PROVE}\@s{-4.1} SafeAt ( v ,\, b ) \.{'}}%
 \@x{\@s{20.1} {\BY}\@pfstepnum{1}{3} ,\, QuorumAssumption ,\, \texttt{SMT} {\DEF}
 TypeOK ,\, SafeAt ,\, WontVoteIn ,\, VotedForIn ,\, Phase2a}%
 \@x{\@pfstepnum{1}{4.}\  {\ASSUME} {\NEW} a \.{\in} {\mathcal{A}} ,\, Phase2b ( a
 ) \@s{2.0} {\PROVE} SafeAt ( v ,\, b ) \.{'}}%
 \@x{\@s{4.0}\@pfstepnum{2}{1.}\  {\PICK} m \.{\in} sent \.{:} Phase2b ( a )
 {\bang} ( m ) \@s{2.0} {\BY}\@pfstepnum{1}{4}{\DEF} Phase2b}%
 \@x{\@s{4.0}\@pfstepnum{2}{2.}\  \A\, a2 \.{\in} {\mathcal{A}} ,\, b2 \.{\in}
 {\mathcal{B}} ,\, v2 \.{\in} {\mathcal{V}} \.{:} VotedForIn ( a2 ,\, v2 ,\, b2 )
 \.{\implies} VotedForIn ( a2 ,\, v2 ,\, b2 ) \.{'}}%
\@x{\@s{23.6} {\BY}\@pfstepnum{2}{1}{\DEF} TypeOK ,\, VotedForIn}%
 \@x{\@s{4.0}\@pfstepnum{2}{3.}\  {\ASSUME} {\NEW} a2 \.{\in} {\mathcal{A}} ,\,
 {\NEW} b2 \.{\in} {\mathcal{B}} ,\, WontVoteIn ( a2 ,\, b2 ) ,\, {\NEW} v2 \.{\in}
 {\mathcal{V}}}%
 \@x{\@s{24} {\PROVE} {\lnot} VotedForIn ( a2 ,\, v2 ,\, b2 )
 \.{'}}%
 \@x{\@s{8.0}\@pfstepnum{3}{1.}\  {\PICK} m1 \.{\in} sent \.{:} m1 . type
 \.{\in} \{\@w{1b} ,\,\@w{2b} \} \.{\land} m1 . acc \.{=} a2 \.{\land} m1 .
 bal \.{>} b2\@s{2} {\BY}\@pfstepnum{2}{3}{\DEF} WontVoteIn}%
 \@x{\@s{8.0}\@pfstepnum{3}{2.}\  a2 \.{=} a \.{\implies} b2 \.{\neq} m .
 bal\@s{2} {\BY}\@pfstepnum{2}{1} ,\,\@pfstepnum{2}{3}
 ,\,\@pfstepnum{3}{1} ,\, a2 \.{=} a \.{\implies} m . bal \.{\geq} m1 . bal
 {\DEF} TypeOK ,\, Messages}%
 \@x{\@s{8.0}\@pfstepnum{3}{3.}\  a2 \.{\neq} a \.{\implies} {\lnot}
 VotedForIn ( a2 ,\, v2 ,\, b2 ) \.{'}\@s{2} {\BY}\@pfstepnum{2}{1} ,\,\@pfstepnum{2}{3}{\DEF}
 WontVoteIn ,\, VotedForIn}%
 \@x{\@s{8.0}\@pfstepnum{3}{} .\@s{2} {\QED} {\BY}\@pfstepnum{2}{1}
 ,\,\@pfstepnum{2}{2} ,\,\@pfstepnum{2}{3} ,\,\@pfstepnum{3}{2}
 ,\,\@pfstepnum{3}{3} {\DEF} Phase2b ,\, VotedForIn ,\, WontVoteIn ,\, TypeOK ,\,
 Messages ,\, Send}%
 \@x{\@s{4.0}\@pfstepnum{2}{4.}\  \A\, a2 \.{\in} {\mathcal{A}} ,\, b2 \.{\in}
 {\mathcal{B}} \.{:} WontVoteIn ( a2 ,\, b2 ) \.{\implies} WontVoteIn ( a2 ,\, b2 )
 \.{'}\@s{2} {\BY}\@pfstepnum{2}{1} ,\,\@pfstepnum{2}{3}{\DEF}
 WontVoteIn ,\, Send}%
 \@x{\@s{4.0}\@pfstepnum{2}{} .\@s{2} {\QED} {\BY}\@pfstepnum{2}{2}
 ,\,\@pfstepnum{2}{4} ,\, QuorumAssumption {\DEF} SafeAt}%
 \@x{\@pfstepnum{1}{} .\@s{2} {\QED} {\BY}\@pfstepnum{1}{1} ,\,\@pfstepnum{1}{2}
 ,\,\@pfstepnum{1}{3} ,\,\@pfstepnum{1}{4}{\DEF} Next}%
\begin{lcom}{0}%
\begin{cpar}{0}{F}{F}{0}{0}{}%
\ensuremath{Invariant} asserts the temporal formula that if \ensuremath{Spec} holds then \ensuremath{Inv} always holds. 
\end{cpar}%
\end{lcom}%
\@x{ {\THEOREM} Invariant \.{\defeq} Spec \.{\implies} {\Box} Inv}%
\@x{\@pfstepnum{1}{} .\@s{2} {\USE} {\!\!} {\DEF} {\mathcal{B}} ,\, 2bs ,\, max\_prop}%
 \@x{\@pfstepnum{1}{1.}\  Init \.{\implies} Inv\@s{2} {\BY} {\!\!} {\DEF} Init ,\, Inv ,\,
 TypeOK ,\, MsgInv ,\, VotedForIn}%
 \@x{\@pfstepnum{1}{2.}\  Inv \.{\land} [ Next ]_{ vars} \.{\implies} Inv
 \.{'}}%
\@x{\@s{4.0}\@pfstepnum{2}{} .\@s{2} {\SUFFICES} {\ASSUME} Inv ,\, Next\@s{2} {\PROVE} Inv \.{'}}%
 \@x{\@s{19} {\BY} {\!\!} {\DEF} vars ,\, Inv ,\, TypeOK ,\, MsgInv ,\, VotedForIn
 ,\, SafeAt ,\, WontVoteIn}%
\@x{\@s{4.0}\@pfstepnum{2}{} .\@s{2} {\USE} {\!\!} {\DEF} Inv}%
\begin{lcom}{6.5}%
\begin{cpar}{0}{F}{F}{0}{0}{}%
\@pfstepnum{2}{1} proves \ensuremath{TypeOK'} for \ensuremath{Next}. 
Each of \@pfstepnum{3}{1-4} assumes the action of a phase and proves \ensuremath{TypeOK'} for that case.
\end{cpar}%
\end{lcom}%
\@x{\@s{4.0}\@pfstepnum{2}{1.}\  TypeOK \.{'}}%
 \@x{\@s{8.0}\@pfstepnum{3}{1.}\  {\ASSUME} {\NEW} b \.{\in} {\mathcal{B}} ,\,
 Phase1a ( b )\@s{2} {\PROVE}\@s{-4.1} TypeOK \.{'} \@s{2}{\BY}\@pfstepnum{3}{1}{\DEF} TypeOK
 ,\, Phase1a ,\, Send ,\, Messages}%
 \@x{\@s{8.0}\@pfstepnum{3}{2.}\  {\ASSUME} {\NEW} b \.{\in} {\mathcal{B}} ,\,
 Phase2a ( b )\@s{2} {\PROVE}\@s{-4.1} TypeOK \.{'}}%
 \@x{\@s{12.0}\@pfstepnum{4}{} .\ {\PICK} v \.{\in} {\mathcal{V}} \.{:} Send ( [ type
 \.{\mapsto}\@w{2a} ,\, bal \.{\mapsto} b ,\, val \.{\mapsto} v ] )
 \@s{2} {\BY}\@pfstepnum{3}{2}{\DEF} Phase2a}%
 \@x{\@s{12.0}\@pfstepnum{4}{} .\ {\QED} {\BY} {\!\!} {\DEF} TypeOK ,\, Send ,\,
 Messages}%
 \@x{\@s{8.0}\@pfstepnum{3}{3.}\  {\ASSUME} {\NEW} a \.{\in} {\mathcal{A}} ,\,
 Phase1b ( a ) \@s{2} {\PROVE}\@s{-4.1} TypeOK \.{'}}%
 \@x{\@s{12.0}\@pfstepnum{4}{} .\  {\PICK} m \.{\in} sent ,\, r \.{\in}
 max\_prop ( a ) \.{:} Phase1b ( a ) {\bang} ( m ,\, r )\@s{2}
 {\BY}\@pfstepnum{3}{3}{\DEF} Phase1b}%
 \@x{\@s{12.0}\@pfstepnum{4}{} .\  {\QED} \@s{2} {\BY} {\!\!} {\DEF} Send ,\, TypeOK ,\,
 Messages}%
 \@x{\@s{8.0}\@pfstepnum{3}{4.}\  {\ASSUME} {\NEW} a \.{\in} {\mathcal{A}} ,\,
 Phase2b ( a ) \@s{2}{\PROVE}\@s{-4.1} TypeOK \.{'}}%
 \@x{\@s{12.0}\@pfstepnum{4}{} .\  {\PICK} m \.{\in} sent \.{:} Phase2b ( a )
 {\bang} ( m ) \@s{2}{\BY}\@pfstepnum{3}{4}{\DEF} Phase2b}%
 \@x{\@s{12.0}\@pfstepnum{4}{} .\  {\QED} {\BY} {\!\!} {\DEF} Send ,\, TypeOK ,\,
 Messages}%
 \@x{\@s{8.0}\@pfstepnum{3}{} .\  {\QED} {\BY}\@pfstepnum{3}{1}
 ,\,\@pfstepnum{3}{2} ,\,\@pfstepnum{3}{3} ,\,\@pfstepnum{3}{4}{\DEF}
 Next}%
\begin{lcom}{6.5}%
\begin{cpar}{0}{F}{F}{0}{0}{}%
 \@pfstepnum{2}{2} proves \ensuremath{MsgInv'} for \ensuremath{Next}. 
 Each of \@pfstepnum{3}{1-4} assumes the action of a phase and proves \ensuremath{MsgInv'} for that case.
\end{cpar}\end{lcom}
\@x{\@s{4.0}\@pfstepnum{2}{2.}\  MsgInv \.{'}}%
 \@x{\@s{8.0}\@pfstepnum{3}{1.}\  {\ASSUME} {\NEW} b \.{\in} {\mathcal{B}} ,\,
 Phase1a ( b )\@s{2} {\PROVE}\@s{-4.1} MsgInv \.{'}}%
 \@x{\@s{12.0}\@pfstepnum{4}{1.}\  \A\, a2 ,\, v2 ,\, b2 \.{:} VotedForIn ( a2
 ,\, v2 ,\, b2 ) \.{'} \.{\equiv} VotedForIn ( a2 ,\, v2 ,\, b2 )
 \@s{2} {\BY}\@pfstepnum{3}{1}{\DEF} Send ,\, VotedForIn ,\, Phase1a}%
 \@x{\@s{12.0}\@pfstepnum{4}{} .\  {\QED} {\BY}\@pfstepnum{3}{1}
 ,\,\@pfstepnum{4}{1} ,\, QuorumAssumption ,\, SafeAtStable {\DEF} Phase1a
 ,\, MsgInv ,\, TypeOK ,\, Messages ,\, Send}%
 \@x{\@s{8.0}\@pfstepnum{3}{2.}\  {\ASSUME} {\NEW} a \.{\in} {\mathcal{A}} ,\,
 Phase1b ( a )\@s{2} {\PROVE}\@s{-4.1} MsgInv \.{'}}%
 \@x{\@s{12.0}\@pfstepnum{4}{} .\  {\PICK} m \.{\in} sent ,\, r \.{\in}
 max\_prop ( a ) \.{:} Phase1b ( a ) {\bang} ( m ,\, r )
 \@s{2}{\BY}\@pfstepnum{3}{2}{\DEF} Phase1b}%
 \@x{\@s{12.0}\@pfstepnum{4}{} .\  {\DEFINE} m2 \.{\defeq} [ type
 \.{\mapsto}\@w{1b} ,\, bal \.{\mapsto} m . bal ,\, maxVBal \.{\mapsto} r .
 bal ,\, maxVal \.{\mapsto} r . val ,\, acc \.{\mapsto} a ]}%
 \@x{\@s{12.0}\@pfstepnum{4}{1.}\  \A\, a2 ,\, v2 ,\, b2 \.{:} VotedForIn ( a2
 ,\, v2 ,\, b2 ) \.{'} \.{\equiv} VotedForIn ( a2 ,\, v2 ,\, b2 ) \@s{2} {\BY}
 {\DEF} Send ,\, VotedForIn}%
 \@x{\@s{12.0}\@pfstepnum{4}{2.}\  VotedForIn ( m2 . acc ,\, m2 . maxVal ,\,
 m2 . maxVBal ) \.{\lor} m2 . maxVBal \.{=} \.{-} 1}%
\@x{\@s{33.6} {\BY} {\!\!} {\DEF} TypeOK ,\, Messages ,\, VotedForIn}%
 \@x{\@s{12.0}\@pfstepnum{4}{3.}\  \A\, b \.{\in} ( r . bal \.{+} 1 )
 \.{\dotdot} ( m2 . bal \.{-} 1 ) \.{:} {\nexists}\, v \.{\in} {\mathcal{V}} \.{:}
 VotedForIn ( m2 . acc ,\, v ,\, b )}%
\@x{\@s{33.6} {\BY} {\!\!} {\DEF} TypeOK ,\, Messages ,\, VotedForIn ,\, Send}%
 \@x{\@s{12.0}\@pfstepnum{4}{} .\  {\QED} {\BY}\@pfstepnum{4}{1}
 ,\,\@pfstepnum{4}{2} ,\,\@pfstepnum{4}{3} ,\, SafeAtStable {\DEF} MsgInv ,\,
 TypeOK ,\, Messages ,\, Send}%
 \@x{\@s{8.0}\@pfstepnum{3}{3.}\  {\ASSUME} {\NEW} b \.{\in} {\mathcal{B}} ,\,
 Phase2a ( b )\@s{2} {\PROVE}\@s{-4.1} MsgInv \.{'}}%
 \@x{\@s{12.0}\@pfstepnum{4}{1.}\  {\nexists}\, m \.{\in} sent \.{:}
 ( m . type \.{=}\@w{2a} ) \.{\land} ( m . bal \.{=} b )
 \@s{2} {\BY}\@pfstepnum{3}{3}{\DEF} Phase2a}%
 \@x{\@s{12.0}\@pfstepnum{4}{2.}\  {\PICK} v \.{\in} {\mathcal{V}} ,\, Q \.{\in}
 {\mathcal{Q}} ,\, S \.{\subseteq} \{ m \.{\in} sent \.{:} m . type \.{=}\@w{1b}
 \.{\land} m . bal \.{=} b \} \.{:}}%
 \@x{\@s{37.1} \.{\land} \A\, a \.{\in} Q \.{:} \E\, m \.{\in} S \.{:} m .
 acc \.{=} a}%
 \@x{\@s{37.1} \.{\land} \.{\lor} \A\, m \.{\in} S \.{:} m . maxVBal \.{=}
 \.{-} 1}%
\@x{\@s{45.99} \.{\lor} \E\, b2 \.{\in} 0 \.{\dotdot} ( b \.{-} 1 ) \.{:}}%
\@x{\@s{58.89} \.{\land} \A\, m \.{\in} S \.{:} m . maxVBal \.{\leq} b2}%
 \@x{\@s{58.89} \.{\land} \E\, m \.{\in} S \.{:} m . maxVBal \.{=} b2
 \.{\land} m . maxVal \.{=} v}%
 \@x{\@s{37.1} \.{\land} Send ( [ type \.{\mapsto}\@w{2a} ,\, bal \.{\mapsto}
 b ,\, val \.{\mapsto} v ] )\@s{2.0} {\BY}\@pfstepnum{3}{3}{\DEF} Phase2a}%
 \@x{\@s{12.0}\@pfstepnum{4}{3.}\  sent \.{'} \.{=} sent \.{\cup} \{ [ type
 \.{\mapsto}\@w{2a} ,\, bal \.{\mapsto} b ,\, val \.{\mapsto} v ] \}
 \@s{2} {\BY}\@pfstepnum{4}{2}{\DEF} Send}%
 \@x{\@s{12.0}\@pfstepnum{4}{4.}\  \A\, a2 ,\, v2 ,\, b2 \.{:} VotedForIn ( a2
 ,\, v2 ,\, b2 ) \.{'} \.{\equiv} VotedForIn ( a2 ,\, v2 ,\, b2 )
 \@s{2} {\BY}\@pfstepnum{4}{3}{\DEF} VotedForIn}%
 \@x{\@s{12.0}\@pfstepnum{4}{5.}\  \A\, m ,\, m2 \.{\in} sent \.{'} \.{:} m .
 type \.{=}\@w{2a} \.{\land} m2 . type \.{=}\@w{2a} \.{\land} m2 . bal \.{=}
 m . bal \.{\implies} m2 \.{=} m}%
 \@x{\@s{33.6} {\BY}\@pfstepnum{4}{1} ,\,\@pfstepnum{4}{3} ,\, \texttt{Isa} {\DEF}
 MsgInv}%
\@x{\@s{12.0}\@pfstepnum{4}{6.}\  SafeAt ( v ,\, b )}%
 \@x{\@s{16.0}\@pfstepnum{5}{1.} {\CASE} \A\, m \.{\in} S \.{:} m . maxVBal
 \.{=} \.{-} 1 \@s{2} {\BY}\@pfstepnum{4}{2} ,\,\@pfstepnum{5}{1}{\DEF} TypeOK
 ,\, MsgInv ,\, SafeAt ,\, WontVoteIn}%
 \@x{\@s{16.0}\@pfstepnum{5}{2.}\  {\ASSUME} {\NEW} b2\@s{2.97} \.{\in} 0
 \.{\dotdot} ( b \.{-} 1 ) ,\, \A\, m \.{\in} S \.{:} m . maxVBal \.{\leq} b2
 ,\,}%
 \@x{\@s{71} {\NEW} m2 \.{\in} S ,\, m2 . maxVBal \.{=} b2 ,\, m2 . maxVal
 \.{=} v\@s{2} {\PROVE} SafeAt ( v ,\, b )}%
 \@x{\@s{20.0}\@pfstepnum{6}{} .\  {\SUFFICES} {\ASSUME} {\NEW} b3 \.{\in} 0
 \.{\dotdot} ( b \.{-} 1 )}%
 \@x{\@s{76.5} {\PROVE} \E\, Q2 \.{\in} {\mathcal{Q}} \.{:} \A\, a \.{\in}
 Q2 \.{:} VotedForIn ( a ,\, v ,\, b3 ) \.{\lor} WontVoteIn ( a ,\, b3 )
 \@s{2} {\BY} {\!\!} {\DEF} SafeAt}%
 \@x{\@s{20.0}\@pfstepnum{6}{1.} {\CASE} b3 \.{\in} 0 \.{\dotdot} ( b2 \.{-} 1
 ) \@s{2} {\BY}\@pfstepnum{5}{2} ,\,\@pfstepnum{6}{1} ,\, VotedInv {\DEF} SafeAt ,\,
 MsgInv ,\, TypeOK ,\, Messages}%
\@x{\@s{20.0}\@pfstepnum{6}{2.} {\CASE} b3 \.{=} b2}%
 \@x{\@s{24.0}\@pfstepnum{7}{1.}\  VotedForIn ( m2 . acc ,\, v ,\, b2 )
 \@s{2} {\BY}\@pfstepnum{5}{2}{\DEF} MsgInv}%
 \@x{\@s{24.0}\@pfstepnum{7}{2.}\  \A\, a2 \.{\in} Q ,\, w \.{\in} {\mathcal{V}}
 \.{:} VotedForIn ( a2 ,\, w ,\, b2 ) \.{\implies} w \.{=} v}%
 \@x{\@s{45.6} {\BY}\@pfstepnum{7}{1} ,\, VotedOnce ,\, QuorumAssumption
 {\DEF} TypeOK ,\, Messages}%
 \@x{\@s{24.0}\@pfstepnum{7}{} .\  {\QED} {\BY}\@pfstepnum{4}{2}
 ,\,\@pfstepnum{6}{2} ,\,\@pfstepnum{7}{2}{\DEF} WontVoteIn}%
 \@x{\@s{20.0}\@pfstepnum{6}{3.} {\CASE} b3 \.{\in} ( b2 \.{+} 1 ) \.{\dotdot}
 ( b \.{-} 1 )\@s{2} {\BY}\@pfstepnum{4}{2} ,\,\@pfstepnum{5}{2}
 ,\,\@pfstepnum{6}{3}{\DEF} MsgInv ,\, TypeOK ,\, Messages ,\, WontVoteIn}%
 \@x{\@s{20.0}\@pfstepnum{6}{} .\  {\QED}\@s{2.0} {\BY}\@pfstepnum{6}{1}
 ,\,\@pfstepnum{6}{2} ,\,\@pfstepnum{6}{3}\ }%
 \@x{\@s{16.0}\@pfstepnum{5}{} .\  {\QED}\@s{2.0} {\BY}\@pfstepnum{4}{2}
 ,\,\@pfstepnum{5}{1} ,\,\@pfstepnum{5}{2}\ }%
 \@x{\@s{12.0}\@pfstepnum{4}{7.}\  ( \A\, m \.{\in} sent \.{:} m . type
 \.{=}\@w{2a} \.{\implies} SafeAt ( m . val ,\, m . bal ) ) \.{'}}%
 \@x{\@s{33.6} {\BY}\@pfstepnum{4}{3} ,\,\@pfstepnum{4}{6} ,\, SafeAtStable
 {\DEF} MsgInv ,\, TypeOK ,\, Messages}%
 \@x{\@s{12.0}\@pfstepnum{4}{} .\  {\QED} {\BY}\@pfstepnum{4}{3}
 ,\,\@pfstepnum{4}{4} ,\,\@pfstepnum{4}{5} ,\,\@pfstepnum{4}{7} ,\, \A\, m
 \.{\in} sent \.{'} \.{\,\backslash\,} sent \.{:} m . type \.{\neq}\@w{1b} {\DEF} MsgInv ,\, TypeOK ,\, Messages}%
 \@x{\@s{8.0}\@pfstepnum{3}{4.}\  {\ASSUME} {\NEW} a \.{\in} {\mathcal{A}} ,\,
 Phase2b ( a )\@s{2} {\PROVE}\@s{-4.1} MsgInv \.{'}}%
 \@x{\@s{12.0}\@pfstepnum{4}{} .\  {\PICK} m \.{\in} sent \.{:} Phase2b ( a )
 {\bang} ( m )\@s{2} {\BY}\@pfstepnum{3}{4}{\DEF} Phase2b}%
 \@x{\@s{12.0}\@pfstepnum{4}{1.}\  \A\, a2 ,\, v2 ,\, b2 \.{:} VotedForIn ( a2
 ,\, v2 ,\, b2 ) \.{\implies} VotedForIn ( a2 ,\, v2 ,\, b2 ) \.{'}\@s{2} {\BY}
 {\DEF} VotedForIn ,\, Send}%
 \@x{\@s{12.0}\@pfstepnum{4}{2.}\  \A\, m2 \.{\in} sent \.{:} m2 . type
 \.{=}\@w{1b} \.{\implies} \A\, v \.{\in} {\mathcal{V}} ,\, b2 \.{\in} ( m2 .
 maxVBal \.{+} 1 ) \.{\dotdot} ( m2 . bal \.{-} 1 ) \.{:}}%
 \@x{\@s{54.04} {\lnot} VotedForIn ( m2 . acc ,\, v ,\, b2 ) \.{\implies}
 {\lnot} VotedForIn ( m2 . acc ,\, v ,\, b2 ) \.{'}}%
 \@x{\@s{33.6} {\BY} {\!\!} {\DEF} Send ,\, VotedForIn ,\, MsgInv ,\, TypeOK ,\,
 Messages}%
 \@x{\@s{12.0}\@pfstepnum{4}{} .\  {\QED} {\BY}\@pfstepnum{2}{1}
 ,\,\@pfstepnum{4}{1} ,\,\@pfstepnum{4}{2} ,\, SafeAtStable {\DEF} MsgInv ,\,
 Send ,\, TypeOK ,\, Messages}%
 \@x{\@s{8.0}\@pfstepnum{3}{} .\  {\QED} {\BY}\@pfstepnum{3}{1}
 ,\,\@pfstepnum{3}{2} ,\,\@pfstepnum{3}{3} ,\,\@pfstepnum{3}{4}{\DEF}
 Next}%
 \@x{\@s{4.0}\@pfstepnum{2}{} .\  {\QED} {\BY}\@pfstepnum{2}{1}
 ,\,\@pfstepnum{2}{2}{\DEF} Inv}%
 \@x{\@pfstepnum{1}{} .\  {\QED} {\BY}\@pfstepnum{1}{1} ,\,\@pfstepnum{1}{2} ,\,
 \texttt{PTL} {\DEF} Spec}%
\begin{lcom}{0}%
\begin{cpar}{0}{F}{F}{0}{0}{}%
\ensuremath{Agreement} asserts that \ensuremath{Spec} implies that \ensuremath{Agree} always holds.
\end{cpar}%
\end{lcom}%
\@x{ {\THEOREM} Agreement \.{\defeq} Spec \.{\implies} {\Box} Agree}%
\@x{\@pfstepnum{1}{} .\  {\USE} {\!\!} {\DEF} {\mathcal{B}}}%
\@x{\@pfstepnum{1}{1.}\  Inv \.{\implies} Agree}%
 \@x{\@s{4.0}\@pfstepnum{2}{} .\  {\SUFFICES} {\ASSUME} Inv ,\, {\NEW} v1
 \.{\in} {\mathcal{V}} ,\,\@s{2.0} {\NEW} v2 \.{\in} {\mathcal{V}} ,\, {\NEW} b1 \.{\in}
 {\mathcal{B}} ,\, {\NEW} b2 \.{\in} {\mathcal{B}} ,\,}%
 \@x{\@s{93.5} ChosenIn ( v1 ,\, b1 ) ,\, ChosenIn ( v2 ,\, b2 ) ,\, b1
 \.{\leq} b2}%
 \@x{\@s{60} {\PROVE} v1 \.{=} v2\@s{4} {\BY} {\!\!} {\DEF} Agree ,\,
 Chosen}%
 \@x{\@s{4.0}\@pfstepnum{2}{1.} {\CASE} b1 \.{=} b2\@s{2} {\BY}\@pfstepnum{2}{1} ,\,
 VotedOnce ,\, QuorumAssumption ,\, \texttt{SMTT(100)} {\DEF} ChosenIn ,\, Inv}%
\@x{\@s{4.0}\@pfstepnum{2}{2.} {\CASE} b1 \.{<} b2}%
 \@x{\@s{8.0}\@pfstepnum{3}{1.}\  SafeAt ( v2 ,\, b2 )\@s{2} {\BY} VotedInv ,\,
 QuorumAssumption {\DEF} ChosenIn ,\, Inv}%
 \@x{\@s{8.0}\@pfstepnum{3}{2.}\  {\PICK} Q1 \.{\in} {\mathcal{Q}} \.{:} \A\, a
 \.{\in} Q1 \.{:} VotedForIn ( a ,\, v2 ,\, b1 ) \.{\lor} WontVoteIn ( a ,\,
 b1 )\@s{2} {\BY}\@pfstepnum{2}{2} ,\,\@pfstepnum{3}{1}{\DEF} SafeAt}%
 \@x{\@s{8.0}\@pfstepnum{3}{3.}\  {\PICK} Q2 \.{\in} {\mathcal{Q}} \.{:} \A\, a
 \.{\in} Q2 \.{:} VotedForIn ( a ,\, v1 ,\, b1 )\@s{2} {\BY} {\!\!} {\DEF} ChosenIn}%
 \@x{\@s{8.0}\@pfstepnum{3}{4.}\  {\QED} {\BY}\@pfstepnum{3}{2}
 ,\,\@pfstepnum{3}{3} ,\, QuorumAssumption ,\, VotedOnce ,\, \texttt{Z3} {\DEF}
 WontVoteIn ,\, Inv}%
 \@x{\@s{4.0}\@pfstepnum{2}{} .\  {\QED} {\BY}\@pfstepnum{2}{1}
 ,\,\@pfstepnum{2}{2}\ }%
\@x{\@pfstepnum{1}{} .\  {\QED} {\BY} \@pfstepnum{1}{1} ,\, Invariant ,\, \texttt{PTL}}%
\@x{}\bottombar\@xx{}%

\pagebreak
\section{\texorpdfstring{\tlaplus{}}{TLA+} specification of Multi-Paxos with Preemption}
\label{appendix:mppspec}
\tlatex
\@x{}\moduleLeftDash\@xx{ {\MODULE} MultiPaxosPreemptionSpec}\moduleRightDash\@xx{}%
\begin{lcom}{0}%
\begin{cpar}{0}{F}{F}{0}{0}{}%
This is a specification in TLA\textsuperscript{+} 
of Multi-Paxos with Preemption.
\end{cpar}%
\end{lcom}%
\@x{ {\EXTENDS} Integers ,\, TLAPS}%
\@x{ {\CONSTANTS} {\mathcal{P}} ,\, {\mathcal{A}} ,\, {\mathcal{Q}} ,\, {\mathcal{V}} \@s{17.75}}%
\@y{\@s{0.0}%
 Sets of proposers, acceptors, quorums of acceptors, and values to propose
}%
\@pvspace{8.0pt}%
\@x{ {\VARIABLES} sent}%
\@y{\@s{0.0}%
 Set of sent messages
}%
\@pvspace{8.0pt}%
 \@x{ {\ASSUME} QuorumAssumption \.{\defeq} {\mathcal{Q}} \.{\subseteq} {\SUBSET}
 {\mathcal{A}} \.{\land} \A\, Q1 ,\, Q2 \.{\in} {\mathcal{Q}} \.{:} Q1 \.{\cap} Q2
 \.{\neq} {\emptyset}}%
\@pvspace{8.0pt}%
\@x{ {\mathcal{B}} \.{\defeq} {\mathds{N}}}%
\@y{\@s{0.0}%
 Set of ballots
}%
\@xx{}%
\@x{ {\mathcal{S}} \.{\defeq} {\mathds{N}}}%
\@y{\@s{0.0}%
 Set of slots
}%
\@x{ vars \.{\defeq} {\langle} sent {\rangle}}%
\@x{ Send ( m ) \.{\defeq} sent \.{'} \.{=} sent \.{\cup} m}%
\@x{ None \.{\defeq} {\CHOOSE} v \.{:} v \.{\notin} {\mathcal{V}}}%
\@pvspace{8.0pt}%
\begin{lcom}{0}%
\begin{cpar}{0}{F}{F}{0}{0}{}%
 Phase \ensuremath{1a}: A \textsf{1a} message with ballot \ensuremath{b} is sent by some proposer (to all processes). 
\end{cpar}%
\end{lcom}%
\@x{ Phase1a ( p ) \.{\defeq} \E\, b \.{\in} \mathcal{B} \.{:}}%
 \@x{\@s{4.0} \.{\land} \.{\lor} {\nexists}\, m \.{\in} sent \.{:} m . type
 \.{=}\@w{preempt} \.{\land} m . to \.{=} p}%
 \@x{\@s{12.89} \.{\lor} \E\, m \.{\in} sent \.{:} \.{\land} m . type
 \.{=}\@w{preempt} \.{\land} m . to \.{=} p \.{\land} b \.{>} m . bal}%
 \@x{\@s{66.9} \.{\land} \A\, m2 \.{\in} sent \.{:} m2 . type
 \.{=}\@w{1a} \.{\land} m2 . from \.{=} p \.{\implies} m . bal \.{>}
 m2 . bal}%
 \@x{\@s{4.0} \.{\land} Send ( \{ [ type \.{\mapsto}\@w{1a} ,\, from
 \.{\mapsto} p ,\, bal \.{\mapsto} b ] \} )}%
\@pvspace{8.0pt}%
\begin{lcom}{0}%
\begin{cpar}{0}{F}{F}{0}{0}{}%
 Phase \ensuremath{1b}: For an acceptor \ensuremath{a}, if there is
 a \textsf{1a} message \ensuremath{m} with
 ballot \ensuremath{m.bal} that is
 higher than the highest it has seen, 
 \ensuremath{a} sends a \textsf{1b} message with \ensuremath{m.bal} alongwith the highest-numbered pair it has voted for.
\end{cpar}%
\end{lcom}%
 \@x{ sent1b2b ( a ) \.{\defeq} \{ m \.{\in} sent \.{:} m . type \.{\in}
 \{\@w{1b} ,\,\@w{2b} \} \.{\land} m . from \.{=} a \}}%
\@x{ PartialBmax ( T ) \.{\defeq} \{ t \.{\in} T \.{:} \A\, t1 \.{\in} T \.{:} t . slot \.{=} t1 .
 slot \.{\implies} t . bal \.{\geq} t1 . bal \}}%
 \@x{ voteds ( a ) \.{\defeq} \{ [ bal \.{\mapsto} m . bal ,\, slot
 \.{\mapsto} m . slot ,\, val \.{\mapsto} m . val ] \.{:} m \.{\in} \{ m
 \.{\in} sent \.{:} m . type \.{=}\@w{2b} \.{\land} m . from \.{=} a \} \}}%
\@pvspace{8.0pt}%
\@x{ Phase1b ( a ) \.{\defeq} \E\, m \.{\in} sent \.{:}}%
\@x{\@s{4.0} \.{\land} m . type \.{=}\@w{1a}}%
 \@x{\@s{4.0} \.{\land} \A\, m2 \.{\in} sent1b2b ( a ) \.{:} m . bal \.{>} m2
 . bal}%
 \@x{\@s{4.0} \.{\land} Send ( \{ [ type \.{\mapsto}\@w{1b} ,\, from
 \.{\mapsto} a ,\, bal \.{\mapsto} m . bal ,\, voted \.{\mapsto} PartialBmax
 ( voteds ( a ) ) ] \} )}%
\@pvspace{8.0pt}%
\begin{lcom}{0}%
\begin{cpar}{0}{F}{F}{0}{0}{}%
 Phase \ensuremath{2a}: If there is no \textsf{2a} message in \ensuremath{sent} with ballot \ensuremath{b}, and 
 a quorum of acceptors has sent a set \ensuremath{S} of \textsf{1b} messages with ballot \ensuremath{b},
 a proposer sends a
 \textsf{2a} message 
 with ballot \ensuremath{b} and value \ensuremath{v}, where \ensuremath{v} is the value with the highest ballot in \ensuremath{S}, or is any value in \ensuremath{\mathcal{V}} if no acceptor that responded in \ensuremath{S} has voted for anything.
\end{cpar}%
\end{lcom}%
\@x{ Bmax ( T ) \.{\defeq} \{ [ slot \.{\mapsto} t . slot ,\, val \.{\mapsto} t . val ]
 \.{:} t \.{\in} PartialBmax ( T ) \}}%
\@x{ FreeSlots ( T ) \.{\defeq} \{ s \.{\in} Slots \.{:} {\nexists}\, t \.{\in} T \.{:} t . slot
 \.{=} s \}}%
\@x{ NewProposals ( T ) \.{\defeq} {\CHOOSE} D \.{\subseteq} [ slot \.{:} FreeSlots ( T ) ,\,
 val \.{:} \mathcal{V} ] \.{:} \A\, d1 ,\, d2 \.{\in} D \.{:} d1 . slot \.{=} d2 . slot
 \.{\implies} d1 \.{=} d2 }%
\@x{ ProposeDecrees ( T ) \.{\defeq} Bmax ( T ) \.{\cup} NewProposals ( T )}%
 \@x{ VS ( S ,\, Q ) \.{\defeq} {\UNION} \{ m . voted \.{:} m \.{\in} \{ m
 \.{\in} S \.{:} m . from \.{\in} Q \} \}}%
\@pvspace{8.0pt}%
\@x{ Phase2a ( p ) \.{\defeq} \E\, b \.{\in} \mathcal{B} \.{:}}%
 \@x{\@s{4.0} \.{\land} {\nexists}\, m \.{\in} sent \.{:} ( m . type
 \.{=}\@w{2a} ) \.{\land} ( m . bal \.{=} b )}%
 \@x{\@s{4.0} \.{\land} \E\, Q \.{\in} \mathcal{Q} ,\, S \.{\in} {\SUBSET} \{ m
 \.{\in} sent \.{:} ( m . type \.{=}\@w{1b} ) \.{\land} ( m . bal \.{=} b )
 \} \.{:}}%
 \@x{\@s{16.89} \.{\land} \A\, a \.{\in} Q \.{:} \E\, m \.{\in} S \.{:} m .
 from \.{=} a}%
 \@x{\@s{16.89} \.{\land} Send ( \{ [ type \.{\mapsto}\@w{2a} ,\, from
 \.{\mapsto} p ,\, bal \.{\mapsto} b ,\, decrees \.{\mapsto} ProposeDecrees (
 VS ( S ,\, Q ) ) ] \} )}%
\@pvspace{8.0pt}%
\begin{lcom}{0}%
\begin{cpar}{0}{F}{F}{0}{0}{}%
 Phase \ensuremath{2b}: For an acceptor \ensuremath{a}, if there is a \textsf{2a} message \ensuremath{m} with ballot \ensuremath{m.bal}
 that is higher than or equal to the highest it has seen, \ensuremath{a} 
 sends a \textsf{2b} message with \ensuremath{m.bal} and \ensuremath{m.val}.
\end{cpar}
\end{lcom}%
\@x{ Phase2b ( a ) \.{\defeq} \E\, m \.{\in} sent \.{:}}%
\@x{\@s{4.0} \.{\land} m . type \.{=}\@w{2a}}%
 \@x{\@s{4.0} \.{\land} \A\, m2 \.{\in} sent1b2b ( a ) \.{:} m . bal \.{\geq}
 m2 . bal}%
 \@x{\@s{4.0} \.{\land} Send ( \{ [ type \.{\mapsto}\@w{2b} ,\, from
 \.{\mapsto} a ,\, bal \.{\mapsto} m . bal ,\, slot \.{\mapsto} d . slot ,\,
 val \.{\mapsto} d . val ] \.{:} d \.{\in} m . decrees \} )}%
\@pvspace{8.0pt}%
 \@x{ Preempt ( a ) \.{\defeq} \E\, m \.{\in} sent ,\, m2 \.{\in} sent1b2b ( a
 ) \.{:}}%
\@x{\@s{4.0} \.{\land} m . type \.{\in} \{\@w{1a} ,\,\@w{2a} \}}%
\@x{\@s{4.0} \.{\land} m2 . bal\@s{0.27} \.{>} m . bal}%
 \@x{\@s{4.0} \.{\land} \A\, m3 \.{\in} sent1b2b ( a ) \.{:} m2 . bal \.{\geq}
 m3 . bal}%
 \@x{\@s{4.0} \.{\land} Send ( \{ [ type \.{\mapsto}\@w{preempt} ,\, to
 \.{\mapsto} m . from ,\, bal \.{\mapsto} m2 . bal ] \} )}%
\@pvspace{8.0pt}%
\@x{ Init\@s{3.3} \.{\defeq} sent \.{=} {\emptyset}}%
 \@x{ Next \.{\defeq} \.{\lor} \E\, b\@s{0.51} \.{\in} {\mathcal{B}} \.{:} Phase1a (
 b ) \.{\lor} Phase2a ( b )}%
 \@x{\@s{32.5} \.{\lor} \E\, a \.{\in} {\mathcal{A}} \.{:} Phase1b ( a ) \.{\lor}
 Phase2b ( a ) \.{\lor} Preempt ( a )}%
\@x{ Spec\@s{1.17} \.{\defeq} Init \.{\land} {\Box} [ Next ]_{ vars}}%
\@x{}\bottombar\@xx{}%
}

\pagebreak
\section{Comparison of Multi-Paxos Invariants}
\label{appendix:mpinvcomp}

\figref{fig-hist-mpinv} compares the message invariants used by Chand et al.'s~\cite{chand2016formal} and our proof. For brevity we change variable names for Chand et al.'s proof---(1) $acceptorVoted$ to $aVoted$, (2) $acceptorMaxBal$ to $aBal$, and (3) $MaxVotedBallotInSlot$ to $MaxBalInSlot$. The invariants we derive for Multi-Paxos and Multi-Paxos with Preemption are the same.

\begin{figure*}[ht!]
    \centering
    \small
    \noindent
    \begin{tabular}{|@{~}p{0.1\textwidth}@{~}|>{\raggedright\arraybackslash}p{0.5\textwidth}@{~}|@{~}p{0.35\textwidth}@{~}|}
         \hline
         & Chand et al.'s proof~\cite{chand2016formal} & Our proof\\
         \hline
         \multirow{4}{*}{\shortstack[c]{Type\\Invariants}} & (I16) $sent \subseteq Messages$ & $sent \subseteq Messages$\\
         & (I17) $aVoted\! \in\! [\Acceptors\! ->\! \SUBSET\! [bal\! :\! \Ballots, slot\! :\! \Slots, val\! :\!\Values]]$ & \\
         & (I18) $aBal \in [\Acceptors -> \Ballots \cup \{-1\}]$ & \\
         \hline
         \multirow{10}{*}{\shortstack[c]{Process\\Invariants\\\\\\$\A\! a\! \in\! \mathcal{A}$}} & (I19) $\A r \in aVoted[a] : aBal[a] \geq r.bal$ & \\
         & (I20) $aBal[a] = -1 => aVoted[a] = \emptyset$ & \\
         & (I21) $\A r \in aVoted[a] : VotedForIn(a, r.bal, r.slot, r.val)$ &\\
         & (I22) $\A b \in \Ballots, s \in \Slots, v \in \Values :$ &\\
         & $\quad c > MaxBalInSlot(aVoted[a], s) =>$ &\\
         & $\quad ~VotedForIn(a, b, s, v)$ &\\
         & (I23) $\A b \in \Ballots, s \in \Slots, v \in \Values :
       VotedForIn(a, b, s, v) =>$ &\\
        & $\quad \E r \in aVoted[a] : r.slot = s /\ r.bal \geq b$ & \\
         
         \hline
         \multirow{14}{*}{\shortstack[c]{Message\\Invariants\\\\\\$\A\!m\!\in\! sent$}} & (I24)  $m.type \!=\! \msgtype{2b} \!=>\! m.bal\! \leq\! aBal[m.from]$ & \\
         & (I25) $m.type \!=\! \msgtype{1b} \!=>\! m.bal\! \leq\! aBal[m.from]$ & \\
         \cline{2-3}
         
         & & (I26) $m.type = \msgtype{1b} => \A s \in \Slots,$\\
         & & $b \in 0..m.bal-1, v \in \Values:$\\
         & & $\; VotedForIn(m.from, b, s, v) =>$\\
         & & $\;\; \E r\! \in\! m.voted\!:\! r.slot\! =\! s /\ r.bal\! \geq\! b$\\
         \cline{2-3}
        
         & (I27) $m.type = \msgtype{1b} => \A b \in \Ballots, s \in \Slots, v \in \Values:$ & $m.type = \msgtype{1b} => \A r \in m.voted:$\\
         & $\quad b \in MaxBalInSlot(m.voted, s)+1..m.bal-1 =>$ & $\; \A b \in r.bal+1..m.bal-1, v \in \Values:$\\
         & $\quad ~ VotedForIn(m.from, b, s, v)$ & $\;\; ~VotedForIn(m.from, b, r.slot, v)$\\
         \cline{2-3}        
        
         & \multicolumn{2}{p{0.78\textwidth}|}{(I28) $m.type = \msgtype{1b} \Rightarrow \A r \in m.voted : VotedForIn(m.from, r.bal, r.slot, r.val)$}\\
         & \multicolumn{2}{p{0.78\textwidth}|}{(I29) $m.type = \msgtype{2a} \Rightarrow \A d \in m.decrees : SafeAt(d.bal, d.slot, d.val)$}\\
         & \multicolumn{2}{p{0.78\textwidth}|}{(I30) $\A d1,d2 \in m.decrees : d1.slot = d2.slot \Rightarrow d1 = d2$}\\
         & \multicolumn{2}{p{0.78\textwidth}|}{(I31) $m.type = \msgtype{2a} \Rightarrow \A m2 \in sent : m2.type = \msgtype{2a} \land m2.bal = m.bal \Rightarrow m2 = m$}\\
         & \multicolumn{2}{p{0.78\textwidth}|}{(I32) $m.type = \msgtype{2b} \Rightarrow \E m2 \in sent : m2.type = \msgtype{2a} \land m2.bal = m.bal\ \land$}\\
         & \multicolumn{2}{p{0.78\textwidth}|}{$\quad  \E d \in m2.decrees: d.slot = m.slot \land d.val = m.val$}\\
         \hline
    \end{tabular}
    \caption{Comparison of invariants for Multi-Paxos. Our proof does not need (I17)-(I25), and needs only (I16), an additional (I26), a simpler (I27), and (I28)-(I32).}
    \label{fig-hist-mpinv}
\end{figure*}

\section{Phase 2a and unique ballots}
\label{appendix-p2a}

The first conjunct of $Phase2a$, shown in \figref{fig-hist-spec-p2a}, states that no \msg{2a} message has been sent with the same ballot on which $Phase2a$ is being executed. To understand why the algorithm is unsafe without this conjunct, consider a system with 3 processes, $\{P1, P2, P3\}$, where each process is both an acceptor and a proposer. Assume that the first conjunct of $Phase2a$ is omitted. 
Assume that each majority is considered a quorum. Consider the following run:
\begin{enumerate}

    \item $P1$ executes Phase 1a on some ballot $b$ and sends a \msg{1a} message.
    \item $P1$ and $P2$ receive the \msg{1a} message sent in step (1), execute Phase 1b, and send \msg{1b} messages with $maxVBal$ as $-1$ and $maxVal$ as $\bot$.
    \item $P1$ receives the two \msg{1b} messages sent in step (2), executes Phase 2a on ballot $b$, and send a \msg{2a} with proposed value $v1$. Note that because $maxVBal$ of both \msg{1b} messages was $-1$, $v1$ is a random value.
    \item $P1$ and $P2$ receive the \msg{2a} message sent in step (3), execute Phase 2b, and send \msg{2b} messages with ballot $b$ and value $v1$. With two such \msg{2b} messages, $Chosen(v1)$ holds.
    
    \item $P3$ receives the \msg{1a} message sent in step (1), executes Phase 1b, and sends a \msg{1b} message with $maxVBal$ as $-1$ and $maxVal$ as $\bot$.
    \item $P1$ receives the \msg{1b} message sent in step (5), uses also a previously received \msg{1b} message sent in step (2), executes Phase 2a again on ballot $b$ (allowed after removing the first conjunct in $Phase2a$),
    and send a \msg{2a} with proposed value $v2$.  Again, because $maxVBal$ of both \msg{1b} messages was $-1$, $v2$ is a random value, which can be different from $v1$.
    \item $P1$ and $P3$ receive the \msg{2a} message sent in step (6), execute Phase 2b, and send \msg{2b} messages with ballot $b$ and value $v2$. With two such \msg{2b} messages, $Chosen(v2)$ holds.
    
\end{enumerate}
Because $v1$ and $v2$ may be two different random values, $Agreement$ does not hold.  That is, the algorithm without the first conjunct in $Phase2a$ is unsafe.

\end{document}